\documentclass[preprint, aps, prd]{revtex4-1}

\def\pd{\partial}
\def\mc{\mathcal}
\def\ul{\underline}

\usepackage[dvips]{graphicx}
\usepackage{amssymb}
\usepackage{amssymb,amsmath}
\usepackage{graphicx}
\usepackage{dsfont}
\usepackage{caption}
\usepackage{subcaption}
\usepackage{mathtools}
\usepackage{verbatim}
\usepackage{graphicx}
\usepackage{multirow}
\usepackage[outline]{contour}
\usepackage{xcolor,colortbl}
\makeatother
\begin{document}
\title{Wrapped D4-branes from maximal 6D gauged supergravity}

\author{Patharadanai Nuchino} \email[REVTeX Support:
]{danai.nuchino@hotmail.com and parinya.ka@hotmail.com} 
\affiliation{Department of Physics, Faculty of Science, Ramkhamhaeng University, 282 Ramkhamhaeng Road, Bang Kapi, Bangkok 10240, Thailand}
 \author{Parinya Karndumri}
\affiliation{String Theory and
Supergravity Group, Department of Physics, Faculty of Science,
Chulalongkorn University, 254 Phayathai Road, Pathumwan, Bangkok
10330, Thailand}

\date{\today}
\begin{abstract}
We study a large class of domain wall solutions with $Mkw_3\times \Sigma^2$ and $Mkw_2\times \Sigma^3$ slices from maximal gauged supergravity in six dimensions. $\Sigma^2$ and $\Sigma^3$ are given by a Riemann surface and a $3$-manifold with constant curvature while $Mkw_{3,2}$ denotes three/two-dimensional Minkowski space. We consider the maximal gauged supergravity with $CSO(p,q,5-p-q)$ and $CSO(p,q,4-p-q)\ltimes \mathbb{R}^4$ gauge groups arising from an $S^1$ reduction of seven-dimensional maximal gauged supergravity with $CSO(p,q,5-p-q)$ and $CSO(p,q,4-p-q)$ gauge groups. The two types of gauge group can be embedded in type IIA theory via consistent truncations on $H^{p,q}\times \mathbb{R}^{5-p-q}$ and $H^{p,q}\times\mathbb{R}^{4-p-q}\times S^1$, respectively. By performing topological twists on $\Sigma^2$ and $\Sigma^3$, we find a number of solutions interpolating between locally flat domain walls in the UV and curved domain walls with $Mkw_3\times \Sigma^2$ and $Mkw_2\times \Sigma^3$ slices in the IR. Many solutions admit physical IR singularities and can be interpreted as holographic RG flows across dimensions from five-dimensional field theories to three- and two-dimensional non-conformal field theories in the IR. Upon uplifted to type IIA theory, we expect the solutions to describe brane configurations involving D4-branes wrapped on $\Sigma^2$ and $\Sigma^3$. 
\end{abstract}
\maketitle

\tableofcontents
%%%%%%%%%%%%%%%%%%%%%%%%%%%%%%%%%%%%%%%%%%%%%%%%%%%%%%%%%%%%%%%%%%%%%%%%%%%%%%%%%%%%%%%%%%%%
\section{Introduction}
For almost three decades, the AdS/CFT correspondence \cite{maldacena,Gubser_AdS_CFT,Witten_AdS_CFT} has led to new insight into various aspects of strongly-coupled superconformal field theories. In particular, the structure and dynamics of higher-dimensional superconformal field theories can be explored by considering twisted compactifications to the well-understood superconformal field theories in lower dimensions. The holographic study along this line involves constructing solutions to certain gauged supergravity of interest with a compact internal manifold $\mc{M}$ and performing a topological twist along $\mc{M}$ to preserve some amount of the original supersymmetry. Solutions of this type have been studied in a number of previous works, see  \cite{Maldacena_nogo}-\cite{ISO3_defect} for an incomplete list. 
\\
\indent On the other hand, a similar study in the case of non-conformal field theories has been less explored although the generalization to non-conformal field theories known as the DW/QFT correspondence \cite{DW_QFT1,DW_QFT2,DW_QFT3} has been proposed shortly after of the AdS/CFT correspondence. The holographic solutions in this case would correspond to non-conformal branes wrapped on a compact manifold $\mc{M}$ as given in \cite{pure_N1,N2_SYM_Gauntlett,N2_SYM_Zaffaroni} in which the construction of gravity duals of $N=1$ and $N=2$ super Yang-Mills theories in four dimensions has been performed by wrapping NS5/D5-branes on a two-sphere $S^2$. Other solutions describing D-branes wrapped on $S^2$ can be found in \cite{fractional_brane,SQCD_like,D6_wrapped}, see also \cite{D6_wrapped_spindle,non-confornal_brane_spindle,non-conformal_brane_Minwoo} for more recent results including a generalization to the case of $\mc{M}$ being two-dimensional orbifolds. 
\\
\indent In this paper, we will study supersymmetric solutions interpolating between locally flat domain walls and $Mkw_{3}\times \Sigma^2$- and $Mkw_{2}\times \Sigma^3$-sliced domain walls from the maximal gauged supergravity in six dimensions \cite{6D_Max_Gauging}. In these solutions, $\Sigma^2$ and $\Sigma^3$ denote a Riemann surface and a three-manifold with constant curvature while $Mkw_{3}$ and $Mkw_{2}$ are Minkowski spaces in three and two dimensions. These solutions should holographically describe RG flows across dimensions from non-conformal field theories in five dimensions to three- and two-dimensional field theories in the IR. The locally flat domain wall solutions dual to the maximal super Yang-Mills theories in five dimensions have been studied recently in \cite{6D_DW_I,6D_DW_II,6D_DW_III} with various types of gauge groups. In this work, we will consider two classes of gauge groups given by $CSO(p,q,5-p-q)$ and $CSO(p,q,4-p-q)\ltimes \mathbb{R}^4$. Both of these gauge groups can be obtained from an $S^1$ reduction of the maximal gauged supergravity in seven dimensions \cite{7D_Max_Gauging} with $CSO(p,q,5-p-q)$ and $CSO(p,q,4-p-q)$ gauge groups, respectively. 
\\
\indent It has been shown in \cite{Henning_KK} that the $CSO(p,q,5-p-q)$ gauged supergravity in seven dimensions can be embedded in eleven dimensions via a consistent truncation on $H^{p,q}\times\mathbb{R}^{5-p-q}$ which can be regarded as a product of a $(p+q-1)$-dimensional hyperboloid and a $(5-p-q)$-torus after some identification. Therefore, in this case, the six-dimensional $CSO(p,q,5-p-q)$ gauged supergravity can arise from a truncation of eleven-dimensional supergravity on $H^{p,q}\times\mathbb{R}^{5-p-q}\times S^1$. Alternatively, since an $S^1$ reduction of eleven-dimensional supergravity gives rise to type IIA theory, we can consider this six-dimensional gauged supergravity as type IIA theory on $H^{p,q}\times\mathbb{R}^{5-p-q}$. A particular case of $SO(5)$ gauge group corresponding to a truncation on $S^4$ with $p=5$ and $q=0$ has been explicitly constructed in \cite{Pope_typeII_S3_S4}. For other gauge groups, the complete truncations have not been worked out to date.
\\
\indent On the other hand, as shown in \cite{Malek_IIA_IIB}, the maximal seven-dimensional gauged supergravity with $CSO(p,q,4-p-q)$ gauge group can be obtained from a truncation of type IIA and type IIB theories on $H^{p,q}\times\mathbb{R}^{4-p-q}$. A further reduction on $S^1$ gives the maximal $CSO(p,q,4-p-q)\ltimes \mathbb{R}^4$ gauged supergravity in six dimensions \cite{6D_Max_Gauging}. Accordingly, the two classes of gauge groups under consideration here could be embedded in type IIA theory. Therefore, the resulting solutions, upon uplifted to type IIA theory, would describe D4-branes wrapped on $\Sigma^2$ and $\Sigma^3$ leading to super Yang-Mills theories in three and two dimensions, respectively. For $SO(5)$ gauge group, the solution describing a near horizon geometry of D4-branes wrapped on a Riemann surface has been studied recently in \cite{non-confornal_brane_spindle}. In this work, we will consider more general gauge groups and study the full holographic RG flows interpolating between non-conformal field theories in five and three/two dimensions  
\\
\indent Holographic solutions involving dual non-conformal field theories, obtained from gauged supergravity, are usually singular. Some singularities are simply an indication that the description in terms of gauged supergravity theories is invalid. However, there are also singularities that lead to some pathological geometries or unacceptable aspects of the dual field theories. The former could lead to physically acceptable holographic descriptions of the dual field theories under consideration while the latter must be discarded. In this paper, we will use two criteria proposed in \cite{Maldacena_nogo} and \cite{Gubser_Sing} to determine whether IR singularities appearing in the solutions are physically acceptable or not. Unphysical singularities could lead to nonsensible results in the dual field theories such as a negative vacuum expectation value for a positive definite operator or a description of the nonexistence Higgs branch.
\\
\indent The paper is organized as follows. In section \ref{6DN=(2,2)gSUGRA}, we give a very brief review of the maximal $N=(2,2)$ gauged supergravity in six dimensions. The embedding of $CSO(p,q,5-p-q)$ and $CSO(p,q,4-p-q)\ltimes \mathbb{R}^4$ gauge groups in $GL(5)\subset SO(5, 5)$ is also given. Supersymmetric solutions describing D4-branes wrapped on $\Sigma^2$ and $\Sigma^3$ from these two classes of gauge groups are respectively given in sections \ref{15_Sec} and \ref{40_Sec}. Conclusions and discussions are given in section \ref{conclusion_sec}. For convenience, we also collect some useful formulae and relations used throughout the paper in appendix \ref{AppA}. The effective potentials arising from twisted compactifications of six-dimensional gauged supergravity on $\Sigma^2$ and $\Sigma^3$ are given in appendix \ref{AppB}
%%%%%%%%%%%%%%%%%%%%%%%%%%%%%%%%%%%%%%%%%%%%%%%%%%%%%%%%%%%%%%%%%%%%%%%%%%%%%%%%%%%%%%%%%%%%%%%%%%%%%%%%%%%%%%%%%%%%%%%%%%%%%%%%%%%%%%%%%
\section{Six-dimensional $N=(2,2)$ gauged supergravity with $CSO(p,q,5-p-q)$ and $CSO(p,q,4-p-q)\ltimes\mathbb{R}^4$ gauge groups}\label{6DN=(2,2)gSUGRA}
We first give a brief review of six-dimensional $N=(2,2)$ gauged supergravity constructed in \cite{6D_Max_Gauging} by using the embedding tensor formalism. We mainly consider $CSO(p,q,5-p-q)$ and $CSO(p,q,4-p-q)\ltimes\mathbb{R}^4$ gauge groups and relevant formulae for finding supersymmetric solutions.
\\
\indent The bosonic field content of the theory consists of the vielbein $e^{\hat{\mu}}_\mu$, two-form potentials $B_{\mu\nu, M}=(B_{\mu\nu, m}, {B_{\mu\nu}}^m)$, three-forms $C_{\mu\nu\rho, A}$, vector fields ${A_\mu}^A$, and scalars ${V_A}^{\alpha\dot{\alpha}}$ that parametrize the coset space $SO(5,5)/\left(SO(5)\times SO(5)\right)$. We will use the convention that curved and flat space-time indices are denoted by $\mu$, $\nu$ and $\hat{\mu}$, $\hat{\nu}$, respectively. Lower and upper $m,n,\ldots=1,\ldots,5$ indices respectively label fundamental and anti-fundamental representations of $GL(5)\subset SO(5,5)$. The index $A=1,2,\ldots, 16$ describes Majorana-Weyl spinors of the $SO(5,5)$ duality symmetry, and the indices $\alpha=1,2,3,4$ and $\dot{\alpha}=\dot{1},\dot{2},\dot{3},\dot{4}$ are respectively two sets of $SO(5)$ spinor indices in the $SO(5)\times SO(5)$ local symmetry. For spinor fields, there are gravitini $\psi_{+\mu\alpha}$ and $\psi_{-\mu\dot{\alpha}}$, and the spin-$\frac{1}{2}$ fields $\chi_{+a\dot{\alpha}}$ and $\chi_{-\dot{a}\alpha}$ where $a,=1,\ldots,5$ and $\dot{a}=\dot{1},\ldots,\dot{5}$ are respectively two sets of $SO(5)$ vector indices in $SO(5)\times SO(5)$. Indices $M,N,\ldots =1,2,\ldots, 10$ denote $SO(5,5)$ vector representation. We use $\pm$ to indicate space-time chiralities of the spinors. 
\\
\indent We also note that in ungauged supergravity, only the electric two-forms $B_{\mu\nu, m}$ appear in the Lagrangian while the magnetic duals ${B_{\mu\nu}}^m$ are introduced on-shell. Therefore, only the $GL(5)$ subgroup of the full global $SO(5,5)$ symmetry is a manifest off-shell symmetry. When the gaugings involve a symmetry that is not a subgroup of the off-shell $GL(5)$ symmetry, the magnetic two-forms are introduced in the gauged Lagrangian via the topological term. The electric and magnetic two-forms are combined into a vector representation of the global symmetry group $SO(5,5)$ denoted by $B_{\mu\nu, M}=(B_{\mu\nu, m}, {B_{\mu\nu}}^m)$.

\subsection{The maximal gauged supergravity in six dimensions}
The most general gaugings can be described by the embedding tensor ${\Theta_A}^{MN}$ which implements the minimal coupling of various fields via the covariant derivative
\begin{equation}\label{gauge_covariant_derivative}
D_\mu=\partial_\mu-g{A_\mu}^A\ {\Theta_A}^{MN}\boldsymbol{t}_{MN}\, .
\end{equation}
It should also be pointed out that the gauge coupling constant $g$ can also be absorbed in the definition of the embedding tensor. The gauge generators $X_A={\Theta_A}^{MN}\boldsymbol{t}_{MN}$ of a gauge group $G_0\subset SO(5,5)$ are given by particular linear combinations of the $SO(5,5)$ generators $\boldsymbol{t}_{MN}$. 
\\
\indent Supersymmetry requires the embedding tensor to transform as $\mathbf{144}_c$ representation of $SO(5,5)$. Accordingly, ${\Theta_A}^{MN}$ can be written in terms of a vector-spinor representation $\theta^{AM}$ of $SO(5,5)$ as 
\begin{equation}
{\Theta_A}^{MN}\ =\ -\theta^{B[M}(\Gamma^{N]})_{BA}\ \equiv \ \left(\Gamma^{[M}\theta^{N]}\right)_A\, .
\end{equation}
$\theta^{AM}$ is subject to the constraint
\begin{equation}\label{MainLC}
(\Gamma_M)_{AB}\,\theta^{BM}\ =\ 0
\end{equation}
where $(\Gamma_M)_{AB}$ are symmetric $SO(5,5)$ gamma matrices. 
\\
\indent With the $SO(5,5)$ generators in vector and spinor representations given by
\begin{equation}
{(\boldsymbol{t}_{MN})_P}^Q=4\eta_{P[M}\delta^Q_{N]}\qquad\text{and}\qquad {(\boldsymbol{t}_{MN})_A}^B\ =\ {(\Gamma_{MN})_A}^B,
\end{equation}
the corresponding gauge generators take the forms
\begin{eqnarray}\label{DefGaugeGen}
& &{(X_A)_M}^N=2\left(\Gamma_{M}\theta^{N}\right)_A+2\left(\Gamma^{N}\theta_{M}\right)_A\nonumber \\
\textrm{and}\qquad & &{(X_A)_B}^C= \left(\Gamma^{M}\theta^{N}\right)_A{(\Gamma_{MN})_B}^C\, .
\end{eqnarray}
The $SO(5,5)$ invariant tensor is given by
\begin{equation}\label{off-diag-eta}
\eta_{MN}\ =\ \eta^{MN}\ =\ \begin{pmatrix} 	0 & \mathds{1}_5 \\
							\mathds{1}_5 & 0    \end{pmatrix}\, .
\end{equation}
Consistency requires the generators $X_A$ to form a closed subalgebra of $SO(5,5)$
\begin{equation}
\left[X_A,X_B\right]\ = \ -{(X_A)_B}^C\,X_C\, .
\end{equation} 
Therefore, the embedding tensor needs to satisfy the quadratic constraints of the form
\begin{equation}\label{QC}
\theta^{AM}\theta^{BN}\eta_{MN}\ =\ 0\qquad \textrm{and}\qquad \theta^{AM}\theta^{B[N}(\Gamma^{P]})_{AB}=0\, .
\end{equation}
\indent By decomposing $\theta^{AM}=(\theta^{Am},{\theta^{A}}_m)$ under $GL(5)\subset SO(5,5)$, gaugings triggered by $\theta^{Am}$ are electric in the sense that only electric two-forms participate in the gauged theory. On the other hand, in magnetic gaugings triggered by ${\theta^A}_m$, it is necessary to introduce magnetic two-forms together with additional three-form tensor fields. This leads to a hierarchy of non-abelian vectors and tensor fields of various ranks.
\\
\indent For all solutions considered in this paper, all of the two-form and three-form fields can be consistently truncated out. Accordingly, we will set all these fields to zero from now on for brevity. In addition, the ansatze for the gauge fields relevant for performing topological twists lead to identically vanishing topological terms. With all these, the bosonic Lagrangian of the six-dimensional $N=(2,2)$ gauged supergravity can be written as
\begin{equation}\label{6D_Lar}
e^{-1}\mathcal{L}=\frac{1}{4}R-\frac{1}{16}{P_{\mu}}^{a\dot{a}}{P^\mu}_{a\dot{a}}-\mathbf{V}-\frac{1}{4}M_{AB}{\mathcal{H}_{\mu\nu}}^{A}\mathcal{H}^{\mu\nu, B}\, .
\end{equation}
With vanishing two-form fields, the covariant gauge field strength tensors are simply given by the usual non-abelian gauge field strengths
\begin{equation}\label{Ful2Form}
{\mathcal{H}_{\mu\nu}}^A=2\partial_{[\mu}{A_{\nu]}}^{A}+g{X_{[BC]}}^{A}{A_{\mu}}^{B}{A_{\nu}}^{C}\, .
\end{equation}
The scalar matrix $M_{AB}$ is defined by 
\begin{equation}
M_{AB}=\Omega_{\alpha\beta}\Omega_{\dot{\alpha}\dot{\beta}}{V_A}^{\alpha\dot{\alpha}}{V_B}^{\beta\dot{\beta}}
\end{equation}
with $\Omega_{\alpha\beta}$ and $\Omega_{\dot{\alpha}\dot{\beta}}$ being the $USp(4)\sim SO(5)$ symplectic forms. Together with the complex counjugate $\Omega^{\alpha\beta}=(\Omega_{\alpha\beta})^*$ and $\Omega^{\dot{\alpha}\dot{\beta}}=(\Omega_{\dot{\alpha}\dot{\beta}})^*$, these symplectic forms can be used to raise and lower $\alpha$ and $\dot{\alpha}$ indices. 
\\
\indent The scalar fields are described by the $16\times16$ coset representative ${V_A}^{\alpha\dot{\alpha}}$ in chiral spinor representation of $SO(5,5)$. The inverse of the coset representative will be denoted by ${V^A}_{\alpha\dot{\alpha}}$. The kinetic term for scalar fields is given in terms of the scalar vielbein
\begin{equation}
{P_{\mu}}^{a\dot{a}}=\frac{1}{4}{(\gamma^a)}_{\alpha\beta}{(\gamma^{\dot{a}})}_{\dot{\alpha}\dot{\beta}}V^{A\alpha\dot{\alpha}}D_\mu {V_{A}}^{\beta\dot{\beta}}\label{PDef}
\end{equation}
with ${(\gamma^a)}_{\alpha\beta}=\frac{1}{4}\Omega_{\beta\gamma}{(\gamma^a)_{\alpha\dot{\alpha}}}^{\gamma\dot{\alpha}}$, ${(\gamma^{\dot{a}})}_{\dot{\alpha}\dot{\beta}}=\frac{1}{4}\Omega_{\dot{\beta}\dot{\gamma}}{(\gamma^{\dot{a}})_{\alpha\dot{\alpha}}}^{\alpha\dot{\gamma}}$, and $V^{A\alpha\dot{\alpha}}=\Omega^{\alpha\beta}\Omega^{\dot{\alpha}\dot{\beta}}{V^A}_{\beta\dot{\beta}}$. 
\\
\indent The scalar potential can be written in terms of the embedding tensor and the coset representative as
\begin{equation}\label{scalarPot1}
\mathbf{V}=\frac{g^2}{2}\theta^{AM}\theta^{BN}{\mathcal{V}_M}^a{\mathcal{V}_N}^b\left[{V_A}^{\alpha\dot{\alpha}}{(\gamma_a)_\alpha}^\beta{(\gamma_b)_\beta}^\gamma V_{B\gamma\dot{\alpha}}\right].
\end{equation}
The coset representative in vector representation ${\mathcal{V}_M}^{\underline{A}}=({\mathcal{V}_M}^a,{\mathcal{V}_M}^{\dot{a}})$ is related to ${V_A}^{\alpha\dot{\alpha}}$ by the following relations
\begin{eqnarray}
{\mathcal{V}_M}^a&=&\frac{1}{16}V^{A\alpha\dot{\alpha}}(\Gamma_M)_{AB}{(\gamma^a)_{\alpha\dot{\alpha}}}^{\beta\dot{\beta}}{V^B}_{\beta\dot{\beta}}\label{VVrel1}\\
\textrm{and}\qquad {\mathcal{V}_M}^{\dot{a}}&=&-\frac{1}{16}V^{A\alpha\dot{\alpha}}(\Gamma_M)_{AB}{(\gamma^{\dot{a}})_{\alpha\dot{\alpha}}}^{\beta\dot{\beta}}{V^B}_{\beta\dot{\beta}}\, . \label{VVrel2}
\end{eqnarray}
In these relations, ${(\Gamma_{\ul{A}})_{\alpha\dot{\alpha}}}^{\beta\dot{\beta}}=({(\gamma_a)_{\alpha\dot{\alpha}}}^{\beta\dot{\beta}},{(\gamma_{\dot{a}})_{\alpha\dot{\alpha}}}^{\beta\dot{\beta}})$ are $SO(5,5)$ gamma matrices in a basis with diagonal $\eta_{\ul{A}\ul{B}}$.
\\
\indent We finally note the fermionic supersymmetry transformations of fermionic fields
\begin{eqnarray}
\delta\psi_{+\mu\alpha}&=& D_\mu\epsilon_{+\alpha}+\frac{g}{4}\hat{\gamma}_\mu {T_\alpha}^{\dot{\beta}}\epsilon_{-\dot{\beta}}+\frac{1}{8}({\hat{\gamma}_\mu}^{\ \nu\rho}-6\delta^\nu_\mu\hat{\gamma}^\rho){\mathcal{H}_{\nu\rho}}^A\Omega_{\alpha\beta}{V_A}^{\beta\dot{\beta}}\epsilon_{-\dot{\beta}},\label{1stSUSY}\\
\delta\psi_{-\mu\dot{\alpha}}&=& D_\mu\epsilon_{-\dot{\alpha}}-\frac{g}{4}\hat{\gamma}_\mu {T^{\beta}}_{\dot{\alpha}}\epsilon_{+\beta}+\frac{1}{8}({\hat{\gamma}_\mu}^{\ \nu\rho}-6\delta^\nu_\mu\hat{\gamma}^\rho){\mathcal{H}_{\nu\rho}}^A\Omega_{\dot{\alpha}\dot{\beta}}{V_A}^{\beta\dot{\beta}}\epsilon_{+\beta},\label{2ndSUSY}\\
\delta\chi_{+a\dot{\alpha}}&=&\frac{1}{4}{P^\mu}_{a\dot{a}}\hat{\gamma}_\mu{(\gamma^{\dot{a}})_{\dot{\alpha}}}^{\dot{\beta}}\epsilon_{-\dot{\beta}}+2g{(T_{a})^\beta}_{\dot{\alpha}}\epsilon_{+\beta}-\frac{g}{2}{T^{\alpha}}_{\dot{\alpha}}{(\gamma_a)_\alpha}^\beta\epsilon_{+\beta}\nonumber\\&&+\frac{1}{4}{\mathcal{H}_{\mu\nu}}^A\hat{\gamma}^{\mu\nu}\Omega_{\dot{\alpha}\dot{\beta}}{V_A}^{\alpha\dot{\beta}}{(\gamma_a)_\alpha}^\beta\epsilon_{+\beta},\label{3rdSUSY}\\
\delta\chi_{-\dot{a}\alpha}&=&\frac{1}{4}{P^\mu}_{a\dot{a}}\hat{\gamma}_\mu{(\gamma^a)_\alpha}^\beta\epsilon_{+\beta}+2g{(T_{\dot{a}})_{\alpha}}^{\dot{\beta}}\epsilon_{-\dot{\beta}}+\frac{g}{2}{T_{\alpha}}^{\dot{\alpha}}{(\gamma_{\dot{a}})_{\dot{\alpha}}}^{\dot{\beta}}\epsilon_{-\dot{\beta}}\nonumber\\&&+\frac{1}{4}{\mathcal{H}_{\mu\nu}}^A\hat{\gamma}^{\mu\nu}\Omega_{\alpha\beta}{V_A}^{\beta\dot{\alpha}}{(\gamma_{\dot{a}})_{\dot{\alpha}}}^{\dot{\beta}}\epsilon_{-\dot{\beta}}\, .\label{4thtSUSY}
\end{eqnarray}
The covariant derivatives of supersymmetry parameters, $\epsilon_{+\alpha}$ and $\epsilon_{-\dot{\alpha}}$, are defined by
\begin{eqnarray}
D_\mu\epsilon_{+\alpha}&=& \partial_\mu\epsilon_{+\alpha}+\frac{1}{4}{\omega_\mu}^{\nu\rho}\hat{\gamma}_{\nu\rho}\epsilon_{+\alpha}+\frac{1}{4}{Q_\mu}^{ab}{(\gamma_{ab})_\alpha}^\beta\epsilon_{+\beta},\label{CoDivEp+}\\
D_\mu\epsilon_{-\dot{\alpha}}&=& \partial_\mu\epsilon_{-\dot{\alpha}}+\frac{1}{4}{\omega_\mu}^{\nu\rho}\hat{\gamma}_{\nu\rho}\epsilon_{-\dot{\alpha}}+\frac{1}{4}{Q_\mu}^{\dot{a}\dot{b}}{(\gamma_{\dot{a}\dot{b}})_{\dot{\alpha}}}^{\dot{\beta}}\epsilon_{-\dot{\beta}}\label{CoDivEp-}
\end{eqnarray}
with space-time gamma matrices denoted by $\hat{\gamma}_\mu=e_\mu^{\hat{\mu}}\hat{\gamma}_{\hat{\mu}}$. The $SO(5)\times SO(5)$ composite connections, $Q_\mu^{ab}$ and $Q_\mu^{\dot{a}\dot{b}}$, are given by
\begin{eqnarray}
{Q_{\mu}}^{ab}&=&\frac{1}{8}{(\gamma^{ab})}_{\alpha\beta}\Omega_{\dot{\alpha}\dot{\beta}} V^{A\alpha\dot{\alpha}}D_\mu {V_A}^{\beta\dot{\beta}},\label{QuDef}\\
{Q_{\mu}}^{\dot{a}\dot{b}}&=&\frac{1}{8}\Omega_{\alpha\beta}{(\gamma^{\dot{a}\dot{b}})}_{\dot{\alpha}\dot{\beta}}V^{A\alpha\dot{\alpha}}D_\mu {V_A}^{\beta\dot{\beta}}.\label{QdDef}
\end{eqnarray}
The T-tensors appearing in the fermionic supersymmetry transformations are defined as
\begin{equation}\label{TTenDef}
(T^a)^{\alpha\dot{\alpha}}={\mathcal{V}_M}^a\theta^{AM}{V_A}^{\alpha\dot{\alpha}}\qquad \textrm{and} \qquad (T^{\dot{a}})^{\alpha\dot{\alpha}}=-{\mathcal{V}_M}^{\dot{a}}\theta^{AM}{V_A}^{\alpha\dot{\alpha}}
\end{equation}
with
\begin{equation}
T^{\alpha\dot{\alpha}}\equiv (T^a)^{\beta\dot{\alpha}}{(\gamma_a)_\beta}^\alpha=-(T^{\dot{a}})^{\alpha\dot{\beta}}{(\gamma_{\dot{a}})_{\dot{\beta}}}^{\dot{\alpha}}.
\end{equation}
In terms of these T-tensors, the scalar potential can alternatively be written as
\begin{equation}\label{scalarPot2}
\mathbf{V}=g^2\left[(T^a)^{\alpha\dot{\alpha}}(T_a)_{\alpha\dot{\alpha}}-\frac{1}{2}T^{\alpha\dot{\alpha}}T_{\alpha\dot{\alpha}}\right].
\end{equation}

\subsection{Gaugings of $CSO(p,q,5-p-q)$ and $CSO(p,q,4-p-q)\ltimes\mathbb{R}^4$ groups}\label{gauging_Sec}
We now consider the embedding of $CSO(p,q,5-p-q)$ and $CSO(p,q,4-p-q)\ltimes\mathbb{R}^4$ gauge groups in $GL(5)\subset SO(5,5)$. Under $GL(5)$, the embedding tensor $\theta^{AM}$ in $\mathbf{144}_c$ representation of $SO(5,5)$ decomposes as
\begin{equation}\label{mainthetaDec}
\mathbf{144}_c\ \rightarrow\ \overline{\mathbf{5}}^{+3}\,\oplus\,\mathbf{5}^{+7}\,\oplus\,\mathbf{10}^{-1}\,\oplus\,\mathbf{15}^{-1}\,\oplus\,\mathbf{24}^{-5}\,\oplus\,\overline{\mathbf{40}}^{-1}\,\oplus\,\overline{\mathbf{45}}^{+3}\, .
\end{equation}
Since the $SO(5,5)$ spinor representation decomposes as $\mathbf{16}_s\ \rightarrow\ \overline{\mathbf{5}}^{+3}\,\oplus\,\mathbf{10}^{-1}\,\oplus\,\mathbf{1}^{-5}$, the gauge generators can be written as
\begin{equation}\label{GaugeGenSplit}
X_A\ =\ \mathbb{T}_{Am}X^m+\mathbb{T}_{A}^{mn}X_{mn}+\mathbb{T}_{A\ast}X_\ast\, .
\end{equation}
In this equation, $X^m$, $X_{mn}$, and $X_\ast$ denote respectively $\overline{\mathbf{5}}^{+3}$, $\mathbf{10}^{-1}$, and $\mathbf{1}^{-5}$ representations of $GL(5)$. The explicit form of matrices $\mathbb{T}_{Am}$, $\mathbb{T}_{A}^{mn}$, and $\mathbb{T}_{A\ast}$ can be found in appendix \ref{AppA}.
\\
\indent The gauge covariant derivative can now be rewritten as
\begin{equation}
D_\mu=\nabla_\mu-gA_{\mu,m} X^m-g{A_\mu}^{mn} X_{mn}-gA_\mu^{\ast} X_\ast
\end{equation}
in which the vector fields in $\mathbf{16}_c$ decompose as
\begin{equation}
{A_\mu}^A\ =\ \mathbb{T}^{Am}A_{\mu,m}+\mathbb{T}^{A}_{mn}{A_\mu}^{mn}+\mathbb{T}^{A}_{\ast}A_\mu^{\ast}\, .
\end{equation}
Matrices $\mathbb{T}^{Am}$, $\mathbb{T}^{A}_{mn}$, and $\mathbb{T}^{A}_{\ast}$ are respectively inverse matrices of $\mathbb{T}_{Am}$, $\mathbb{T}_{A}^{mn}$, and $\mathbb{T}_{A\ast}$ and are given by complex conjugations. 
\\
\indent By the decomposition of $SO(5,5)$ vector representation under $GL(5)$, $\mathbf{10}\ \rightarrow\ \mathbf{5}^{+2}\,\oplus\,\overline{\mathbf{5}}^{-2}$, we can write the two components $\theta^{Am}$ and ${\theta^{A}}_m$ of the embedding tensor as
\begin{eqnarray}
\theta^{Am}&:&\qquad \overline{\mathbf{5}}^{+3}\,\oplus\,\mathbf{10}^{-1}\,\oplus\,\mathbf{24}^{-5}\,\oplus\,\overline{\mathbf{40}}^{-1},\label{splitthetaDec1}\\ 
{\theta^{A}}_m&:&\qquad \overline{\mathbf{5}}^{+3}\,\oplus\,\mathbf{5}^{+7}\,\oplus\,\mathbf{10}^{-1}\,\oplus\,\mathbf{15}^{-1}\,\oplus\,\overline{\mathbf{45}}^{+3}\, .\label{splitthetaDec2}
\end{eqnarray}
The gaugings obtained by these components have been extensively studied in \cite{6D_DW_I} and \cite{6D_DW_II}. In this paper, we will only consider gaugings arising from $\mathbf{15}^{-1}$ and $\overline{\mathbf{40}}^{-1}$ representations.
\\
\indent The gaugings from the component $\mathbf{15}^{-1}$ are described by a symmetric tensor $Y_{mn}=Y_{(mn)}$ and lead to $CSO(p,q,5-p-q)$ gauge groups. Gaugings in $\overline{\mathbf{40}}^{-1}$ representation are characterized by a tensor $U^{np,m}=U^{[np],m}$ satisfying $U^{[np,m]}=0$ with the corresponding gauge groups given by $CSO(p,q,4-p-q)\ltimes\mathbb{R}^4$. The factor $\mathbb{R}^4$ is associated with the shift symmetries on the scalar fields. Both of these gauge groups can be obtained from a dimensional reduction on a circle of the maximal gauged supergravity in seven dimensions with $CSO(p,q,5-p-q)$ and $CSO(p,q,4-p-q)$ gauge groups, respectively.   
\\
\indent We end this section by giving a decomposition of scalar fields under $GL(5)$. The $25$ non-compact generators of $SO(5,5)$ are decomposed under $GL(5)$ as
\begin{equation}\label{15repscalarDEC}
\underbrace{\mathbf{1}+\mathbf{14}}_{\hat{\boldsymbol{t}}^+_{a\dot{b}}}\,+\underbrace{\mathbf{10}}_{\boldsymbol{s}_{mn}}\, .
\end{equation}
The generators $\hat{\boldsymbol{t}}^+_{a\dot{b}}$ correspond to non-compact generators of $GL(5)$ given by
\begin{equation}
\hat{\boldsymbol{t}}^+_{a\dot{b}}=\frac{1}{2}\left({\mathbb{M}_{a}}^M{\mathbb{M}_{\dot{b}}}^N+{\mathbb{M}_{\dot{b}}}^M{\mathbb{M}_{a}}^N\right)\boldsymbol{t}_{MN}
\end{equation}
where ${\mathbb{M}_{\underline{A}}}^M=({\mathbb{M}_{a}}^M,{\mathbb{M}_{\dot{a}}}^M)$ is a transformation matrix between diagonal and off-diagonal bases $\eta_{MN}$ and $\eta_{\underline{A}\underline{B}}$. The explicit form of this matrix can also be found in appendix \ref{AppA}. 
\\
\indent In accordance with \eqref{15repscalarDEC}, we will denote the $25$ scalar fields as 
\begin{equation}
\Phi^I=\{\varphi,\phi_1,\ldots,\phi_{14},\varsigma_1,\ldots,\varsigma_{10}\}\label{6D_scalar}
\end{equation}
with $I=1,\ldots,25$. The scalar $\varphi$ is the dilaton corresponding to the $SO(1,1)$ generator 
\begin{equation}\label{SO(1,1)Gen}
\boldsymbol{d}=\hat{\boldsymbol{t}}^+_{1\dot{1}}+\hat{\boldsymbol{t}}^+_{2\dot{2}}+\hat{\boldsymbol{t}}^+_{3\dot{3}}
+\hat{\boldsymbol{t}}^+_{4\dot{4}}+\hat{\boldsymbol{t}}^+_{5\dot{5}}\, .
\end{equation}
This $SO(1,1)\sim\mathbb{R}^+$ is identified with the $SO(1,1)$ factor of $GL(5)\sim SL(5)\times SO(1,1)$. The remaining fourteen linear combinations of $\hat{\boldsymbol{t}}^+_{a\dot{b}}$ generators correspond to the fourteen scalar fields $\{\phi_1,\ldots,\phi_{14}\}$ in the $SL(5)/SO(5)$ coset. The latter is the scalar manifold of the maximal gauged supergravity in seven dimensions. Finally, the ten scalars $\{\varsigma_1,\ldots,\varsigma_{10}\}$ correspond to the shift generators $\boldsymbol{s}_{mn}$ and will be called shift scalars. These scalars arise from the $S^1$ reduction of seven-dimensional vector fields.

\subsection{Physically acceptable singularities}
As mentioned in the introduction, hologrpahic solutions involving non-conformal field theories are always singular. However, some singularities can lead to physically acceptable solutions. In particular, the presence of these singularities indicates that the computation within the framework of gauged supergravity theories has limited viability. Before giving explicit solutions describing wrapped D4-branes, we will give a brief review of some criteria to separate physically acceptable singularities from a large number of all possible singular solutions. In this work, we use two criteria given in \cite{Gubser_Sing} and \cite{Maldacena_nogo} to which we will refer to as Gubser and Maldacena-Nunez criteria, respectively.   
\subsubsection{Gubser criterion}
A supergravity solution in $d+1$ dimensions that holographically describes $d$-dimensional Poincare invariant field theory takes the form of
\begin{equation}
ds^2=e^{A(r)}dx^2_{1,d-1}+dr^2\, .\label{RG_flow_sol}
\end{equation}
$dx^2_{1,d-1}$ is the metric on $d$-dimensional Minkowski space. This type of solutions can be obtained from a supergravity coupled to scalar fields. If $A(r)\sim r$ as $r\rightarrow \infty$, the geometry is asymptotically $AdS_{d+1}$, and the dual field theory admits a UV conformal fixed point. In a particular case in which $A(r)\sim r$ as $r\rightarrow \pm \infty$, the solution describes an RG flow from the UV fixed point to another conformal fixed point in the IR. If there do not exist any IR fixed points, the solution generically flows to a singularity for a finite value of $r$ with $A(r)\rightarrow -\infty$. If this singularity is physically acceptable, the solution describes a holographic RG flow to a non-conformal phase of the dual field theory. Gubser criterion can be used to determine whether the singularity is physical or not.
\\
\indent According to \cite{Gubser_Sing}, a physical singularity should be obtained from a limit of a family of black hole solutions with the singularity behide the horizons. Since the black holes with non-zero temperature are dual to field theories with finite temperature, this implies that the solution \eqref{RG_flow_sol} admits a finite temperature generalization of the form
\begin{equation}
ds^2=e^{2A(r)}\left(-h(r)dt^2+dx^2_{d-1}\right)+\frac{dr^2}{h(r)}\, .
\end{equation}   
The horizon is located at $r_0$ with $h(r_0)=0$, and the limit $h(r)\rightarrow 1$ recovers the Poincare invariant solution \eqref{RG_flow_sol}. With the assumption that $A'(r)$ is positive at some point outside the horizon, it can be shown that the finite temperature is possible only for solutions with the scalar potential $V<0$ near the horizon. For asymptotically $AdS_{d+1}$ space-times, if the null energy condition is imposed on the Lagrangian of matter fields, the bound on the scalar potential is given by $V\leq V_{\textrm{UV}}$ with $V_{\textrm{UV}}$ being the value of the potential at the UV fixed point. Therefore, physically acceptable IR singularities should lead to the scalar potential that is bounded above near the singularity.
\\
\indent In the present case, the singularities appear in the IR with lower dimensionality. In particular, the singular solutions (if physical) describe conformal phases of the dual field theories in $d-2$ or $d-3$ dimensions. To apply Gubser criterion in the solutions considered in this paper, we need to integrate out all the fields on $\Sigma^2$ or $\Sigma^3$ and obtain effective $(d-1)$- or $(d-2)$-dimensional Lagrangian of the form
\begin{equation}
\mc{L}_{d-\delta}=\sqrt{-g_{d-\delta}}\left(R+K-V^{(d-\delta)}_{\textrm{eff}}\right)
\end{equation}
with $\delta=1,2$. $K$ is the kinetic term for scalar fields, and $V^{(d-\delta)}_{\textrm{eff}}$ denotes the effective scalar potential in $d-\delta$ dimensions. The Gubser criterion then implies that for a physically acceptable singularity, $V^{(d-\delta)}_{\textrm{eff}}$ must be bounded above. The explicit form of $V^{(d-\delta)}_{\textrm{eff}}$ relevant for the analysis of this paper is given in appendix \ref{AppB} 

\subsubsection{Maldacena-Nunez criterion}
Unlike the Gubser criterion, the Maldacena-Nunez criterion is used to determine whether a given singularity is physical or not within the framework of ten- or eleven-dimensional supergravities. This is only possible for holographic solutions of gauged supergravity theories with known higher-dimensional origins. In this case, the lower-dimensional solutions can be uplifted to ten or eleven dimensions.
\\
\indent The Maldacena-Nunez criterion requires that in order to interpret a singular supergravity solution as a holographic description of an IR region of the dual field thoery, the component $\hat{g}_{00}$ of the ten- or eleven-dimensional metric should be bounded above as the singularity is approached. This implies that fixed proper energy excitations in supergravity correspond to lower and lower energy excitations in the dual field theory from the point of view of the coordinate time identified as the field theory time. In this criterion, we should emphasize that $\hat{g}_{00}$ is the component of the metric in Einstein frame. In addition, the behavior of $\hat{g}_{00}$ is related to the norm of the time-like Killing vector $\frac{\pd }{\pd t}$, so this criterion is applicable to dual field theories with a time translational symmetry. In summary, the Maldacena-Nunez criterion requires that the component $\hat{g}_{00}$ of ten- or eleven-dimensional metric should not diverge as the singularity is approached. 
%%%%%%%%%%%%%%%%%%%%%%%%%%%%%%%%%%%%%%%%%%%%%%%%%%%%%%%%%%%%%%%%%%%%%%%
\section{Wrapped D4-branes from $CSO(p,q,5-p-q)$ gauged supergravity}\label{15_Sec}
For gaugings in $\mathbf{15} ^{-1}$ representation, the embedding tensor is given by
\begin{equation}\label{15_rep_theta}
\theta^{Am}=0\qquad\text{and}\qquad{\theta^{A}}_m\ = \ \mathbb{T}^{An}Y_{nm}\,
\end{equation}
where
\begin{equation}\label{diagYmn}
Y_{mn}\ =\ \text{diag}(\underbrace{1,..,1}_p,\underbrace{-1,..,-1}_q,\underbrace{0,..,0}_r)
\end{equation}
for $p+q+r=5$. The only non-vanishing gauge generators are given by 
\begin{equation}
X_{mn}=\mathbb{T}^{A}_{mn}X_A=2Y_{p[m}{\boldsymbol{t}^p}_{n]}
\end{equation}
with ${\boldsymbol{t}^m}_{n}$ being $GL(5)$ generators. Therefore, the whole gauge group is entirely embedded in $GL(5)$. Accordingly, only ten gauge fields ${A_\mu}^{mn}={A_\mu}^{[mn]}$ participate in the gauging via the covariant derivatives 
\begin{equation}
D_\mu=\nabla_\mu-g{A_\mu}^{mn} X_{mn}\, .
\end{equation}
The gauge generators satisfy the commutation relations
\begin{equation}
[X_{mn},X_{pq}]={(X_{mn})_{pq}}^{rs}X_{rs}
\end{equation}
with ${(X_{mn})_{pq}}^{rs}=2{(X_{mn})_{[p}}^{[r}\delta_{q]}^{s]}$. The corresponding gauge group is given by
\begin{equation}
G_0\ =\ CSO(p,q,r)\ =\ SO(p,q) \ltimes \mathbb{R}^{(p+q)r}.
\end{equation}
\indent For vanishing two-form and three-form fields, the gauge covariant field strengths reduce to the usual non-abelian gauge field strengths  
\begin{eqnarray}
{\mathcal{H}_{\mu\nu}}^{mn}=\mathbb{T}_{A}^{mn}{\mathcal{H}_{\mu\nu}}^{A}={F_{\mu\nu}}^{mn}
\end{eqnarray}
with
\begin{equation}
{F_{\mu\nu}}^{mn}=\mathbb{T}_{A}^{mn}{F_{\mu\nu}}^{A}=2\partial_{[\mu}{A_{\nu]}}^{mn}+g{(X_{pq})_{rs}}^{mn}{A_{\mu}}^{pq}{A_{\nu}}^{rs}\, .\label{10_rep_F_def}
\end{equation} 
%%%%%%%%%%%%%%%%%%%%%%%%%%%%%%%%
\subsection{D4-branes wrapped on a Riemann surface}\label{15_Sig2_section}
We first look for solutions describing D4-branes wrapped on a two-dimensional Riemann surface $\Sigma^2_k$. The constant $k$ describes the curvature of $\Sigma^2_k$ with values $k=1,0,-1$ corresponding to a two-dimensional sphere $S^2$, a flat space $\mathbb{R}^2$, or a hyperbolic space $H^2$, respectively.
\\
\indent The metric ansatz for the six-dimensional space-time is given by
\begin{equation}\label{Sig_2_6D_metric}
ds_6^2=e^{2U(r)}\eta_{\bar{\mu}\bar{\nu}}dx^{\bar{\mu}} dx^{\bar{\nu}}+dr^2+e^{2V(r)}ds^2_{\Sigma^{2}_{k}}
\end{equation}
in which $\bar{\mu},\bar{\nu}=0,\ldots ,2$ are indices of a three-dimensional Minkowski space-time $Mkw_3$. The two functions $U(r)$ and $V(r)$ depend only on the radial coordinate $r$. The explicit form of the metric on $\Sigma^2_k$ can be written as
\begin{equation}\label{Sigma2metric}
 ds^2_{\Sigma^2_k}=d\theta^2+f_k(\theta)^2d\zeta^2
\end{equation}
with the function $f_k(\theta)$ defined by
\begin{equation}\label{fFn}
f_k(\theta)=\begin{cases}
                        	\sin{\theta}, \ \  \quad k=+1 \\
                       	\theta, \ \ \ \ \ \ \quad k=0\\
			\sinh{\theta}, \quad k=-1
                    \end{cases}.
\end{equation}
\indent With the following choice of vielbein
\begin{equation}
e^{\hat{\bar{\mu}}}=e^{U}dx^{\bar{\mu}}, \qquad e^{\hat{r}}=dr, \qquad e^{\hat{\theta}}= e^{V}d\theta, \qquad e^{\hat{\zeta}}=e^{V}f_{k}(\theta)d\zeta,\label{AdS5xSigma2bein}
\end{equation}
we find the following non-vanishing components of the spin connection
\begin{equation}\label{AdS4xSigma3SpinCon}
{\omega_{\hat{\bar{\mu}}}}^{\hat{\bar{\nu}}\hat{r}}=U'\delta_{\hat{\bar{\mu}}}^{\hat{\bar{\nu}}}, \qquad {\omega_{\hat{i}}}^{\hat{j}\hat{r}}= V'\delta_{\hat{i}}^{\hat{j}},\qquad
{\omega_{\hat{\zeta}}}^{\hat{\zeta}\hat{\theta}}=e^{-V}\frac{f'_{k}(\theta)}{f_{k}(\theta)}\, .
\end{equation}
The index $\hat{i}=\hat{\theta}, \hat{\zeta}$ is a flat index on $\Sigma^2_{k}$, and $f'_k(\theta)=\frac{df_k(\theta)}{d\theta}$. The $r$-derivatives will be denoted by $'$ while a $'$ on any function with an explicit argument refers to the derivative of the function with respect to that argument. 
\\
\indent To preserve some amount of supersymmetry, we perform a topological twist by turning on some gauge fields to cancel the spin connection ${\omega_{\hat{\zeta}}}^{\hat{\zeta}\hat{\theta}}$ on $\Sigma^2_k$. In the present case, we achieve this by turning on $SO(2)\times SO(2)$ gauge fields associated with $X_{12}$ and $X_{34}$ generators of the form
\begin{equation}\label{15_SO(2)_SO(2)_SIg2_A}
{A_{\hat{\zeta}}}^{12}=e^{-V}\frac{p_{1}}{4k}\frac{f'_{k}(\theta)}{f_{k}(\theta)}\qquad \textrm{and}\qquad
{A_{\hat{\zeta}}}^{34}=e^{-V}\frac{p_{2}}{4k}\frac{f'_{k}(\theta)}{f_{k}(\theta)}
\end{equation} 
where $p_1$ and $p_2$ are constant magnetic charges. The procedure of performing this topological twist requires the gauge group to contain $SO(2)\times SO(2)$ subgroup. All of these gauge groups are characterized by the embedding tensor of the form
\begin{equation}\label{SO(2)xSO(2)Y}
Y_{mn}=\textrm{diag}(1,1,\kappa,\kappa,\lambda)
\end{equation} 
for $\lambda=0,\pm 1$ and $\kappa=\pm 1$. These gauge groups are explicitly given by $SO(5)$ ($\kappa=\lambda=1$), $SO(4,1)$ ($\kappa=-\lambda=1$), $SO(3,2)$ ($\kappa=-\lambda=-1$), $CSO(4,0,1)$ ($\kappa=1,\lambda=0$), and $CSO(2,2,1)$ ($\kappa=-1,\lambda=0$).
\\
\indent There are five scalar fields that are invariant under $SO(2)\times SO(2)$ generated by $X_{12}$ and $X_{34}$, see more detail in \cite{6D_DW_I}. These consist of the dilaton $\varphi(r)$ and other four scalars corresponding to the non-compact generators
\begin{equation}
\mathcal{Y}_1=\hat{\boldsymbol{t}}^+_{1\dot{1}}+\hat{\boldsymbol{t}}^+_{2\dot{2}}-2\,\hat{\boldsymbol{t}}^+_{5\dot{5}},\qquad\mathcal{Y}_2=\hat{\boldsymbol{t}}^+_{3\dot{3}}+\hat{\boldsymbol{t}}^+_{4\dot{4}}-2\,\hat{\boldsymbol{t}}^+_{5\dot{5}},\qquad\mathcal{Y}_3=\boldsymbol{s}_{12},\qquad\mathcal{Y}_4=\boldsymbol{s}_{34}\, .
\end{equation}
With the coset representative
\begin{equation}\label{fullSO(2)xSO(2)singlet_coset}
V=e^{\varphi\boldsymbol{d}+\phi_1\mathcal{Y}_1+\phi_2\mathcal{Y}_2+\varsigma_1\mathcal{Y}_3+\varsigma_2\mathcal{Y}_4},
\end{equation}
the scalar potential is given by
\begin{equation}\label{YSO(2)xSO(2)PotSub}
\mathbf{V}=-\frac{g^2}{4}e^{2\left(\varphi-2(\phi_1+\phi_2)\right)}\left[4\kappa(2+\lambda e^{12\phi_1+8\phi_2})+\lambda e^{8\phi_1+12\phi_2}(4-\lambda e^{12\phi_1+8\phi_2})\right].
\end{equation}
The vielbein on the scalar manifold and the $SO(5)\times SO(5)$ composite connections are given respectively by
\begin{equation}
{P_{r}}^{a\dot{a}}=\frac{1}{4}{(\gamma^a)}_{\alpha\beta}{(\gamma^{\dot{a}})}_{\dot{\alpha}\dot{\beta}}V^{A\alpha\dot{\alpha}}\partial_r {V_{A}}^{\beta\dot{\beta}},\label{Scalar_viel_without_A}\\
\end{equation}
and 
\begin{eqnarray}
{Q_{r}}^{12}&=&-{Q_{r}}^{\dot{1}\dot{2}}\ =\ \varsigma'_1-4 \varsigma_1 (\varphi+\phi_1),\label{Q1_r_Sig2}\\
{Q_{r}}^{34}&=&-{Q_{r}}^{\dot{3}\dot{4}}\ =\ \varsigma'_2-4 \varsigma_2 (\varphi+\phi_2),\label{Q2_r_Sig2}\\
{Q_{\hat{\zeta}}}^{12}&=&{Q_{\hat{\zeta}}}^{\dot{1}\dot{2}}\ =\ -e^{-V}\frac{gp_{1}}{k}\frac{f'_{k}(\theta)}{f_{k}(\theta)},\label{Q1_Sig2}\\
{Q_{\hat{\zeta}}}^{34}&=&{Q_{\hat{\zeta}}}^{\dot{3}\dot{4}}\ =\ -\kappa e^{-V}\frac{gp_{2}}{k}\frac{f'_{k}(\theta)}{f_{k}(\theta)}\, .\label{Q2_Sig2}
\end{eqnarray}
\indent The topological twist can be achieved after imposing the twist condition 
\begin{equation}\label{15_SO(2)xSO(2)_twist_con}
g(p_1+\kappa p_2)=k
\end{equation}
together with the following projection conditions
\begin{equation}\label{SO(2)xSO(2)Projcon+}
\hat{\gamma}_{\hat{\zeta}\hat{\theta}}\epsilon_{+\alpha}={(\gamma_{12})_\alpha}^\beta\epsilon_{+\beta}={(\gamma_{34})_\alpha}^\beta\epsilon_{+\beta}
\end{equation}
and
\begin{equation}\label{SO(2)xSO(2)Projcon-}
\hat{\gamma}_{\hat{\zeta}\hat{\theta}}\epsilon_{-\dot{\alpha}}={(\gamma_{\dot{1}\dot{2}})_{\dot{\alpha}}}^{\dot{\beta}}
\epsilon_{-\dot{\beta}}={(\gamma_{\dot{3}\dot{4}})_{\dot{\alpha}}}^{\dot{\beta}}\epsilon_{-\dot{\beta}}\, .
\end{equation}
To obtain the corresponding BPS equations for scalars which depend on the radial coordinate $r$, we need to impose an additional projector of the form
\begin{equation}\label{15_DW_Proj}
\hat{\gamma}_r\epsilon_\pm=\epsilon_\mp\, .
\end{equation}
We will also take the supersymmetry parameters to depend only on $r$. With all these projectors, the unbroken supersymmetry has four supercharges. Each chirality of $\epsilon_\pm$ consists of four unbroken supercharges under the projectors \eqref{SO(2)xSO(2)Projcon+} and \eqref{SO(2)xSO(2)Projcon-}, and the two chiralities are related by the $\gamma_r$-projector \eqref{15_DW_Proj}. 
\\
\indent The non-vanishing components of the two-form field strengths are given by 
\begin{equation}
{\mathcal{H}_{\hat{\theta}\hat{\zeta}}}^{12}=-\frac{1}{4}p_1e^{-2V}\quad\text{ and }\quad {\mathcal{H}_{\hat{\theta}\hat{\zeta}}}^{34}=-\frac{1}{4}p_2e^{-2V}\, .
\end{equation} 
These identically satisfy the corresponding Bianchi identities. However, in order to consistently truncate out the two-form and three-form fields, we require the following term to vanish 
\begin{equation}
g{\theta^{A}}_mM_{AB}\mathbb{T}^{B}_{np}{\mathcal{H}_{\hat{\theta}\hat{\zeta}}}^{np}=-\frac{g\lambda}{\sqrt{2}}e^{-2(V-\varphi-2\phi_1-2\phi_2)}(p_2e^{4\phi_2}\varsigma_1+p_1e^{4\phi_1}\varsigma_2)\delta_m^5\, .\label{Sig_2_prob_terms}
\end{equation}
For $\lambda\neq0$, this leads to the following  constraint
\begin{equation}\label{2_4_duality_con}
p_2e^{4\phi_2}\varsigma_1+p_1e^{4\phi_1}\varsigma_2=0\, .
\end{equation}
It should be pointed out that this condition does not appear in $CSO(4,0,1)$ and $CSO(2,2,1)$ gauge groups with $\lambda=0$. This is due to the fact that the term given in \eqref{Sig_2_prob_terms} vanishes identically for $\lambda=0$.
\\
\indent Substituting the above ansatz into the conditions $\delta \psi_{+\bar{\mu}\alpha}=0$ and $\delta \psi_{-\bar{\mu}\dot{\alpha}}=0$ and imposing the projectors \eqref{SO(2)xSO(2)Projcon+}, \eqref{SO(2)xSO(2)Projcon-}, and \eqref{15_DW_Proj}, we find the following BPS equations for $U(r)$ and an algebraic constraint of the form
\begin{equation}\label{15_Sig2_U_flow}
U'=\frac{ge^{\varphi}}{4\sqrt{2}}(2e^{-4\phi_1}+2\kappa e^{-4\phi_2}+\lambda e^{8(\phi_1+\phi_2)})-\frac{e^{-2V-\varphi}}{8\sqrt{2}}(p_1e^{4\phi_1}+p_2e^{4\phi_2})
\end{equation}
and 
\begin{equation}\label{15_Sig2_U_Im}
0=(4 g e^{2\varphi-4\phi_1}-p_2 e^{4 \phi_2-2V})\varsigma_1+(4 g\kappa e^{2\varphi-4\phi_2}-p_1 e^{4 \phi_1-2V})\varsigma_2.
\end{equation}
These equations come from real and imaginary parts of $\delta \psi_{+\bar{\mu}\alpha}=0$ and $\delta \psi_{-\bar{\mu}\dot{\alpha}}=0$ conditions, respectively.
\\
\indent Similarly, the conditions $\delta \psi_{+i\alpha}=0$ and $\delta \psi_{-i\dot{\alpha}}=0$ result in the BPS equation for the warp factor $V(r)$
\begin{equation}\label{15_Sig2_V_flow}
V'=\frac{ge^{\varphi}}{4\sqrt{2}}(2e^{-4\phi_1}+2\kappa e^{-4\phi_2}+\lambda e^{8(\phi_1+\phi_2)})+\frac{3e^{-2V-\varphi}}{8\sqrt{2}}(p_1e^{4\phi_1}+p_2e^{4\phi_2})
\end{equation}
and another algebraic condition between the two shift scalars
\begin{equation}\label{15_Sig2_V_Im}
0=(4 g e^{2\varphi-4\phi_1}+3p_2 e^{4 \phi_2-2V})\varsigma_1+(4 g\kappa e^{2\varphi-4\phi_2}+3p_1 e^{4 \phi_1-2V})\varsigma_2.
\end{equation}
It can be verified that the difference between \eqref{15_Sig2_U_Im} and \eqref{15_Sig2_V_Im} precisely gives the constraint \eqref{2_4_duality_con}. In particular, using \eqref{2_4_duality_con} in \eqref{15_Sig2_U_Im} and \eqref{15_Sig2_V_Im} leads to the following two simple conditions
\begin{equation}
\left(1-\frac{\kappa p_2}{p_1}\right)\varsigma_1=0\qquad \textrm{and}\qquad \left(\kappa-\frac{p_1}{p_2}\right)\varsigma_2=0\, .
\end{equation}
We immediately see that there are two possibilities namely vanishing shift scalars $\varsigma_1=\varsigma_2=0$ or $p_2=p_1/\kappa$ with non-vanishing shift scalars.
\\
\indent Repeating the above analysis to the conditions $\delta \psi_{+r\alpha}=0$ and $\delta\psi_{-r\dot{\alpha}}=0$ as well as using the previously obtained results from $\delta \psi_{\pm \bar{\mu}\alpha}=0$ conditions, we find
\begin{eqnarray}
0&=& \partial_r\epsilon_{+\alpha}+\frac{1}{2}\left[{Q_r}^{12}{(\gamma_{12})_\alpha}^\beta+{Q_r}^{34}{(\gamma_{34})_\alpha}^\beta\right]\epsilon_{+\beta}+\frac{1}{2}U'\epsilon_{+\alpha},\label{15_Sig2_radial_eq1}\\
0&=& \partial_r\epsilon_{-\dot{\alpha}}+\frac{1}{2}\left[{Q_r}^{\dot{1}\dot{2}}{(\gamma_{\dot{1}\dot{2}})_{\dot{\alpha}}}^{\dot{\beta}}+{Q_r}^{\dot{3}\dot{4}}{(\gamma_{\dot{3}\dot{4}})_{\dot{\alpha}}}^{\dot{\beta}}\right]\epsilon_{-\dot{\beta}}-\frac{1}{2}U'\epsilon_{-\dot{\alpha}}\, .\qquad\label{15_Sig2_radial_eq2}
\end{eqnarray}
These equations can be solved by the Killing spinors of the form
\begin{equation}\label{15_sig2_full_Killing}
\epsilon_+(r)=e^{\frac{1}{2}\left[U(r)+ W(r)\gamma_{12}\right]}\epsilon_+^{0}\qquad\text{ and }\qquad\epsilon_-(r)=e^{\frac{1}{2}\left[U(r)- W(r)\gamma_{\dot{1}\dot{2}}\right]}\epsilon_-^{0}
\end{equation}
where $W(r)$ is an $r$-dependent function satisfying
\begin{equation}\label{15_Sig2_W_Im}
W'={Q_r}^{12}+{Q_r}^{34}=\varsigma'_1+\varsigma'_2-4\varphi(\varsigma_1+\varsigma_2)-4 \varsigma_1\phi_1-4 \varsigma_2\phi_2\, .
\end{equation}
$\epsilon_\pm^{0}$ are constant symplectic-Majorana-Weyl spinors satisfying projectors \eqref{SO(2)xSO(2)Projcon+}, \eqref{SO(2)xSO(2)Projcon-}, and \eqref{15_DW_Proj}. Finally, using the projector \eqref{15_DW_Proj} in the $\delta\chi_{+a\dot{\alpha}}=0$ and $\delta\chi_{-\dot{a}\alpha}=0$ conditions will lead to the BPS equations for scalar fields. 
\\
\indent We end the analysis of the BPS equations by giving some comments on the unbroken supersymmetry preserved by the solutions to these BPS equations. The solutions with $Mkw_3\times \mathbb{R}^2$ slices, or $k=0$, preserve $16$ supercharges while solutions with $Mkw_3\times S^2$ and $Mkw_3\times H^2$ slices preserve only $4$ supercharges. This is due to the fact that no spin connection on $\mathbb{R}^2$ needs to be canceled by performing the $SO(2)\times SO(2)$ twist. In particular, for the case of $k=0$, the projectors involving $\hat{\gamma}_{\hat{\zeta}\hat{\theta}}$ are not needed. This case is then equivalent to untwisted solutions of flat domain walls. For $k\neq 0$, the projectors \eqref{SO(2)xSO(2)Projcon+} and \eqref{SO(2)xSO(2)Projcon-} reduce the number of supercharges in both $\epsilon_+$ and $\epsilon_-$ to four. These two sets of four supercharges with opposite chiralities are then related to each other by the $\gamma_r$-projector \eqref{15_DW_Proj} resulting in four unbroken supercharges. This corresponds to $N=2$ supersymmetry in three dimensions. 
%%%%%%%%%%%%%%%%%%%%%%%%%%%%%%%%%%%%%%%%%%%%%%%%%%%%%%%%%%%%%
\subsubsection{Solutions with $\varsigma_1=\varsigma_2=0$}\label{15_M2_SO(2)xSO(2)_sec}
We begin with solutions with all shift scalars vanishing. For $\varsigma_1=\varsigma_2=0$, the function $W(r)$ is a constant which can be chosen to be zero. The Killing spinors then take a simple form 
\begin{equation}
\epsilon_\pm=e^{\frac{U(r)}{2}}\epsilon_{\pm}^{0}\, .\label{DW_Killing_spinor}	
\end{equation}
The resulting BPS equations take the form 
\begin{eqnarray}
\varphi'&=&-\frac{ge^{\varphi}}{20\sqrt{2}}(2e^{-4\phi_1}+2\kappa e^{-4\phi_2}+\lambda e^{8(\phi_1+\phi_2)})+\frac{e^{-2V-\varphi}}{40\sqrt{2}}(p_1e^{4\phi_1}+p_2e^{4\phi_2}),\label{15_Sig2_varphi_flow}\qquad\\
\phi'_1&=&\frac{ge^{\varphi}}{5\sqrt{2}}(3e^{-4\phi_1}-2\kappa e^{-4\phi_2}-\lambda e^{8(\phi_1+\phi_2)})-\frac{e^{-2V-\varphi}}{20\sqrt{2}}(3p_1e^{4\phi_1}-2p_2e^{4\phi_2}),\label{15_Sig2_phi1_flow}\\
\phi'_2&=&\frac{ge^{\varphi}}{5\sqrt{2}}(3\kappa e^{-4\phi_2}-2e^{-4\phi_1}-\lambda e^{8(\phi_1+\phi_2)})+\frac{e^{-2V-\varphi}}{20\sqrt{2}}(2p_1e^{4\phi_1}-3p_2e^{4\phi_2}),\\
U'&=&\frac{ge^{\varphi}}{4\sqrt{2}}(2e^{-4\phi_1}+2\kappa e^{-4\phi_2}+\lambda e^{8(\phi_1+\phi_2)})-\frac{e^{-2V-\varphi}}{8\sqrt{2}}(p_1e^{4\phi_1}+p_2e^{4\phi_2}),
\end{eqnarray}
%\\
\begin{eqnarray}
V'&=&\frac{ge^{\varphi}}{4\sqrt{2}}(2e^{-4\phi_1}+2\kappa e^{-4\phi_2}+\lambda e^{8(\phi_1+\phi_2)})+\frac{3e^{-2V-\varphi}}{8\sqrt{2}}(p_1e^{4\phi_1}+p_2e^{4\phi_2}).\label{15_Sig2_phi2_flow}
\end{eqnarray}
It can also be verified that these equations satisfy all bosonic field equations derived from the Lagrangian \eqref{6D_Lar}. We note that these equations do not admit a fixed point solution of the form $AdS_4\times\Sigma^2_k$. This solution requires $\varphi'=\phi_1'=\phi_2'=V'=0$ and $U'=\frac{1}{L_{\textrm{AdS}_4}}$ which are not satisfied by the BPS equations given above. Therefore, this implies that there do not exist any conformal fixed points in three-dimensional field theory arising from D4-branes wrapped on a Riemann surface with $SO(2)\times SO(2)$ twist. 
\\
\indent We are not able to analytically solve the above BPS equations, so we will look for possible numerical solutions. We are interested in solutions that are asymptotic to locally flat domain walls considered in \cite{6D_DW_I}. These solutions are holographically dual to maximal super Yang-Mills (SYM) theory in five dimensions. The full solutions interpolating between these flat domain walls and other singular geometries in the IR would describe three-dimensional super Yang-Mills theory in three dimensions arising from twisted compactifications of the maximal five-dimensional SYM on a Riemann surface.  
\\
\indent In the limit with the contribution of the gauge fields being small compared to the scalar potential, we expect the asymptotic behaviors of the solutions to be given by flat domain walls. Since the explicit form of these behaviors depends on the gauge groups under consideration, we will discuss each gauge group separately. 

\begin{figure}[h!]
  \centering
    \includegraphics[width=0.8\linewidth]{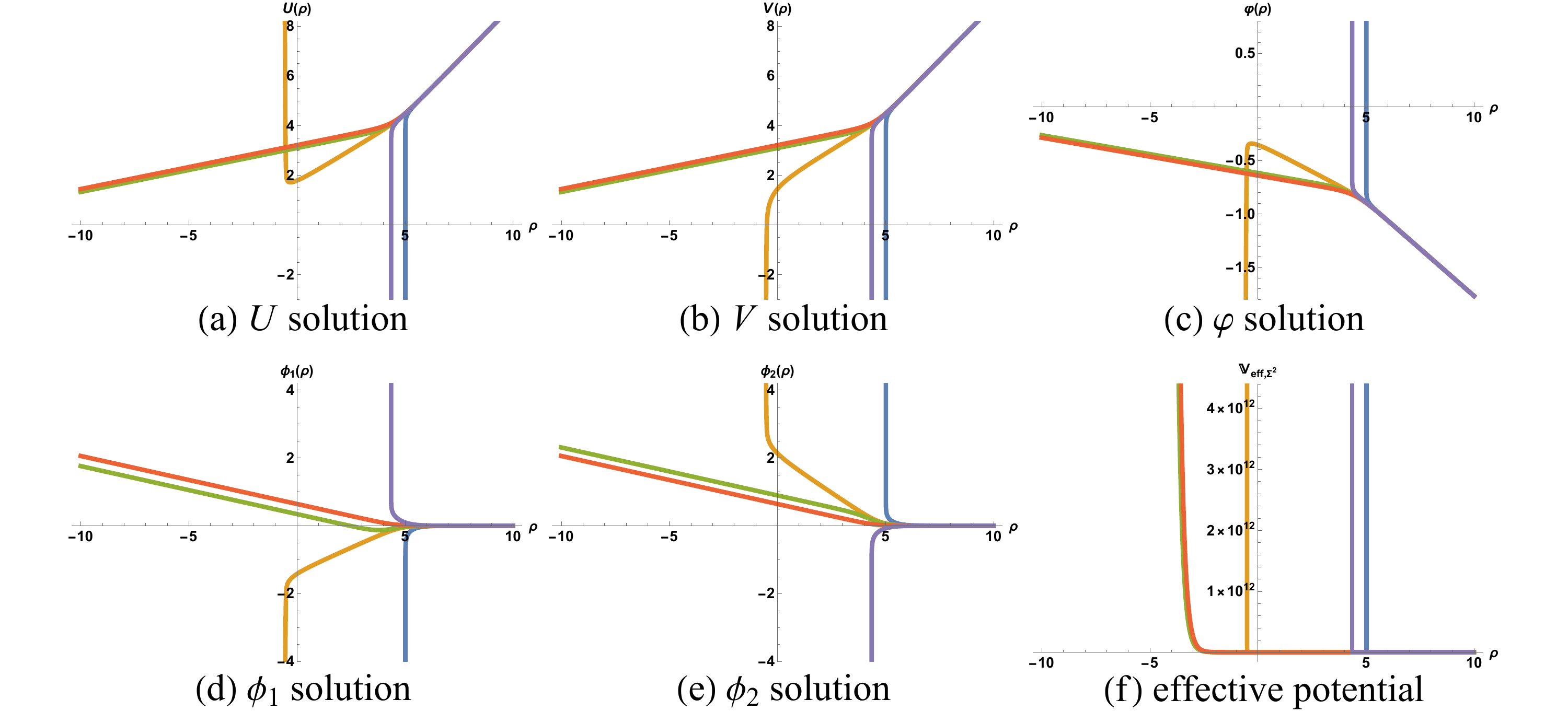}
\caption{Interpolating solutions between the locally $SO(5)$ flat domain wall as $\rho\rightarrow+\infty$ and $Mkw_3\times S^2$-sliced curved domain walls for $SO(2)\times SO(2)$ twist in $SO(5)$ gauge group. The blue, orange, green, red, and purple curves refer to $p_1=-2, 0, 0.1, 0.5, 1.3$, respectively.}
\label{15_S2_special_SO(2)xSO(2)_SO(5)gg_flows}
\end{figure}

\begin{figure}[h!]
  \centering
    \includegraphics[width=0.8\linewidth]{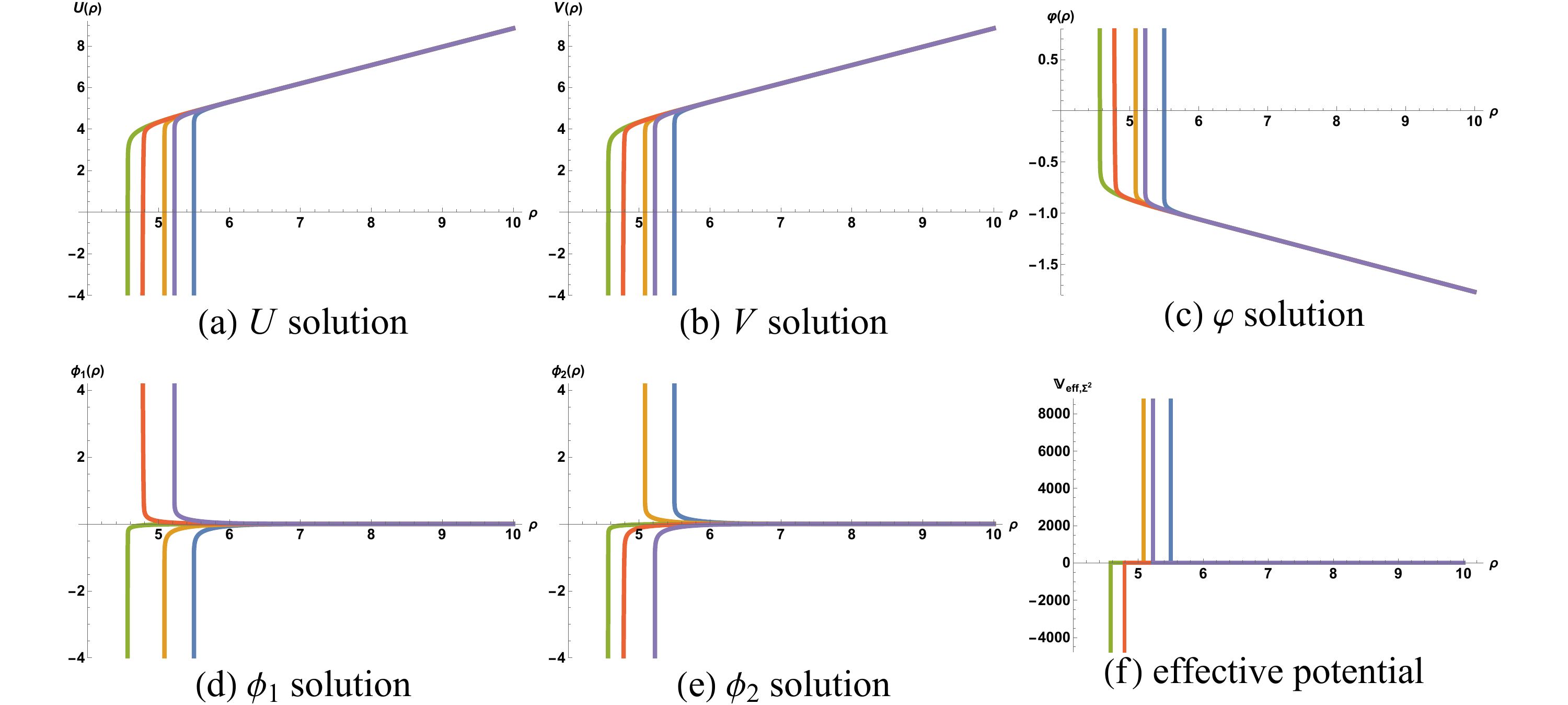}
\caption{Interpolating solutions between the locally $SO(5)$ flat domain wall as $\rho\rightarrow+\infty$ and $Mkw_3\times H^2$-sliced curved domain walls for $SO(2)\times SO(2)$ twist in $SO(5)$ gauge group. The blue, orange, green, red, and purple curves refer to $p_1=-8, -2.4, -0.5, 0, 2.6$, respectively.}
\label{15_H2_special_SO(2)xSO(2)_SO(5)gg_flows}
\end{figure}

\begin{figure}[h!]
  \centering
    \includegraphics[width=0.8\linewidth]{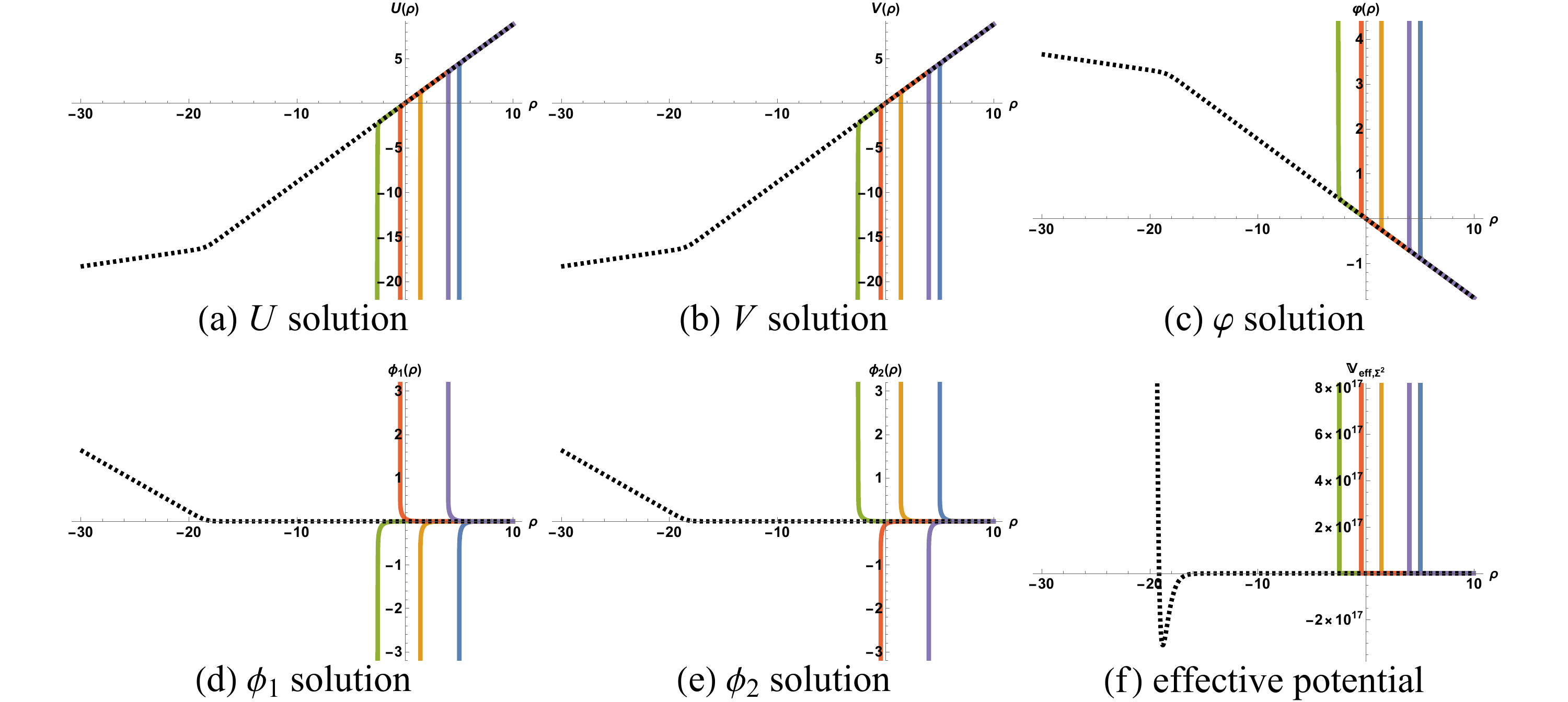}
\caption{Interpolating solutions between the locally $SO(5)$ flat domain wall as $\rho\rightarrow+\infty$ and $Mkw_3\times \mathbb{R}^2$-sliced curved domain walls for $SO(2)\times SO(2)$ twist in $SO(5)$ gauge group. The blue, orange, green, red, and purple curves refer to $p_1=-2, -2\times10^{-5}, -2\times10^{-9}, 10^{-7}, 0.1$, respectively. The dashed curve is the $SO(2)\times SO(2)$ flat domain wall obtained from setting $p_1=0$ which also implies $p_2=k=0$.} 
\label{15_R2_special_SO(2)xSO(2)_SO(5)gg_flows}
\end{figure}

\paragraph{Solutions from $SO(5)$, $SO(4,1)$, and $SO(3,2)$ gauge groups}
We begin with solutions in the case of semisimple gauge groups $SO(5)$, $SO(4,1)$, and $SO(3,2)$ with $\lambda\neq 0$. We first consider the simplest case of $\phi_1=\phi_2=0$. By changing to a new coordinate $\rho$ defined by $\frac{d\rho}{dr}=e^{\varphi+8(\phi_1+\phi_2)}$, we require the solutions to be asymptotically of the form 
\begin{equation}\label{SO(5)_flat_DW_asym}
U\sim V\sim\frac{5g \lambda \rho}{4\sqrt{2}},\qquad \varphi\sim-\frac{g \lambda \rho}{4\sqrt{2}},\qquad \phi_1\sim\phi_2\sim0
\end{equation}
for $\rho\rightarrow +\infty$. It can also be checked that with $\kappa=\lambda=1$ this solves the BPS equations in which the terms involving $e^{-2V-\varphi+4 \phi_1}$ and $e^{-2V-\varphi+4 \phi_2}$ become small compared to the remaining terms. This asymptotic behavior is the $SO(5)$ symmetric domain wall given in \cite{6D_DW_I}. Therefore, only solutions in the case of $SO(5)$ gauge group can have this asymptotic behavior. 
\\
\indent We now look for numerical solutions of the full BPS equations starting from the asymptotic behavior \eqref{SO(5)_flat_DW_asym}. Examples of these solutions are given in figures \ref{15_S2_special_SO(2)xSO(2)_SO(5)gg_flows}, \ref{15_H2_special_SO(2)xSO(2)_SO(5)gg_flows}, and \ref{15_R2_special_SO(2)xSO(2)_SO(5)gg_flows} for $k=+1$, $k=-1$, and $k=0$, respectively. In the case of $k=0$, for the sake of comparison, we have also included the flat domain wall solution with $p_1=p_2=k=0$ represented by the dashed curve in figure \ref{15_R2_special_SO(2)xSO(2)_SO(5)gg_flows}. This corresponds to the untwisted solution. In all of these solutions, we have set $g=1$ with different values of the independent parameter $p_1$. The values of $p_2$ are fixed by the twist condition \eqref{15_SO(2)xSO(2)_twist_con}. We see that all of these solutions are singular at a particular value of the coordinate $\rho$. 
\vfil
\vfil\pagebreak
\vfil
\vfil
\vfil\pagebreak

\begin{figure}[h!]
  \centering
    \includegraphics[width=0.92\linewidth]{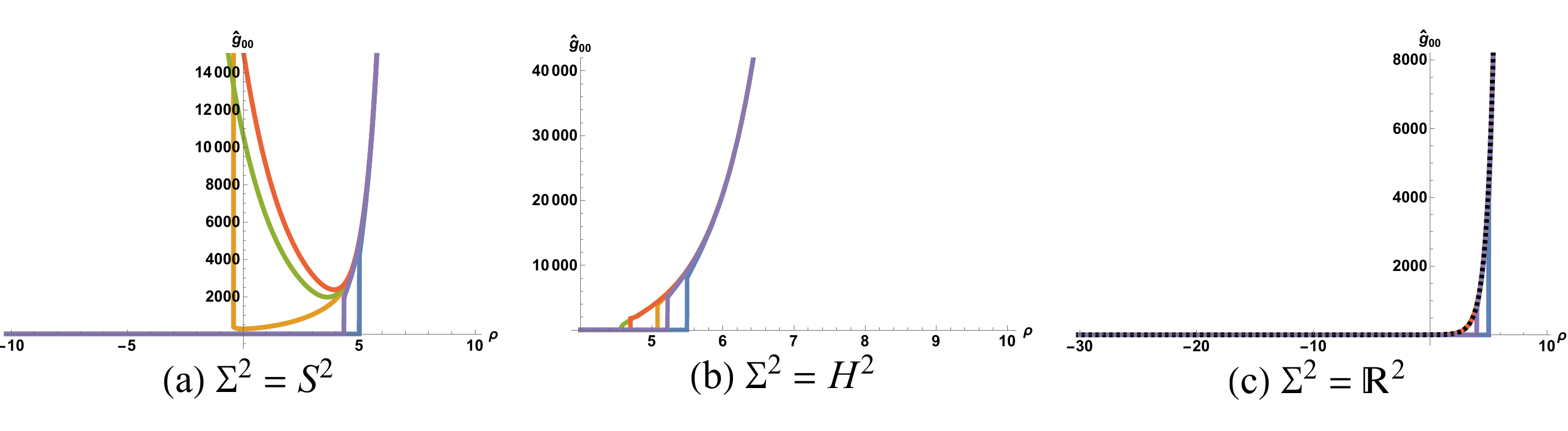}
\caption{The behavior of $\hat{g}_{00}$ along the RG flows given in figures \ref{15_S2_special_SO(2)xSO(2)_SO(5)gg_flows}, \ref{15_H2_special_SO(2)xSO(2)_SO(5)gg_flows}, and \ref{15_R2_special_SO(2)xSO(2)_SO(5)gg_flows}, respectively for $\Sigma^2=S^2, H^2, \mathbb{R}^2$ in $SO(5)$ gauge group. }
\label{g00fig1to3}
\end{figure}

Following \cite{Maldacena_nogo}, we can compute the effective scalar potential in four dimensions arising from a compactification of the maximal six-dimensional gauged supergravity on $\Sigma^2$. The result is given by   
\begin{equation}
\mathbf{V}_{\text{eff},\Sigma^2}=e^{-2V}\left(\mathbf{V}-\frac{k}{2}e^{-2V}+\frac{1}{2}M_{AB}{\mathcal{H}_{\theta\zeta}}^{A}\mathcal{H}^{\theta\zeta,B}\right)
\end{equation}
see appendix \ref{AppB} for more details. We can check whether the singularities appearing in the solutions are physically acceptable or not by using the criterion given in \cite{Gubser_Sing}. According to this criterion, the effective potential must be bounded from above near the singularities. The effective potential is bounded from above near the IR singularities for $Mkw_3\times H^2$-sliced curved domain walls with $-0.5\leq p_1\leq 0$ (the green and red curves in figure \ref{15_H2_special_SO(2)xSO(2)_SO(5)gg_flows}). On the other hand, the effective potential for $Mkw_3\times S^2$- and $Mkw_3\times \mathbb{R}^2$-sliced domain walls is not bounded from above near the IR singularities for any values of $p_1$.
\\
\indent In addition, we can also check the singularities by considering the $\hat{g}_{00}$ component of the uplifted metric to ten or eleven dimensions. As in \cite{7D_Max_twist}, the eleven-dimensional metric component $\hat{g}^{(11)}_{00}$ is given via a consistent truncation on $H^{p,q}\times \mathbb{R}^{5-p-q}$ by 
\begin{equation}
\hat{g}^{(11)}_{00}=\Delta^{\frac{1}{3}}g^{(7)}_{00}
\end{equation}
with $g^{(7)}_{00}$ being the $(00)$-component of the seven-dimensional metric. The warp factor is defined by
\begin{equation}
\Delta=\mc{M}^{ab}\eta_{ac}\eta_{bd}\mu^c\mu^d
\end{equation}
with $\eta_{ab}$, $a,b=1,2,\ldots, 5$ being the $CSO(p,q,5-p-q)$ invariant tensor. $\mc{M}^{ab}$ is the inverse of the $SL(5)$ scalar matrix defined in terms of the $SL(5)/SO(5)$ coset representative as
\begin{equation}
\mc{M}_{ab}={\mc{V}_a}^{\underline{a}}{\mc{V}_b}^{\underline{a}}\, .
\end{equation}
The coordinates $\mu^a$ satisfy $\mu^a\mu^b\eta_{ab}=1$. We can then embed the six-dimensional metric component $g_{00}$ in seven dimensions via a reduction on $S^1$ by the relation
\begin{equation}
g^{(7)}_{00}=e^{2\varphi}g_{00}
\end{equation}
with $\varphi$ being the aforementioned $GL(5)$ dilaton.  
\\
\indent Following \cite{Pope_typeII_S3_S4}, we can perform a dimensional reduction of the resulting eleven-dimensional metric on $S^1$ and find the $(00)$-component of the ten-dimensional metric of the form
\begin{equation}
\hat{g}_{00}=\Omega^{\frac{1}{8}}\Delta^{\frac{1}{3}}e^{2\varphi}g_{00}
\end{equation}
with
\begin{eqnarray}
\Omega=\Delta^{\frac{1}{3}}e^{-8\varphi}+\Delta^{-\frac{2}{3}}m_{ab}\chi^{ac}\chi^{bd}\eta_{ce}\eta_{df}\mu^e\mu^f\, .
\end{eqnarray}
We also note that the factor $\Omega$ is related to the type IIA dilaton via
\begin{equation}
e^{\phi}=\Omega^{\frac{3}{4}}\, .
\end{equation}
The axionic scalars $\chi^{ab}=-\chi^{ba}$ arise from the dimensional reduction of the $CSO(p,q,5-p-q)$ gauge fields $A^{ab}_\mu$ in seven dimensions, see \cite{Pope_typeII_S3_S4}. These scalars transform in the $SL(5)$ representation $\mathbf{10}$ and can be identified with the shift scalars $\varsigma_1,\ldots,\varsigma_{10}$ appearing in \eqref{6D_scalar}.
\\
\indent For the present case of $SO(5)$ gauge group, we have $\eta_{ab}=\delta_{ab}$ and vanishing shift scalars. Therefore, in this case, $\Omega=\Delta^{\frac{1}{3}}e^{-8\varphi}$, and the $(00)$-component of the type IIA metric is then given by
\begin{equation}
\hat{g}_{00}=\Delta^{\frac{3}{8}}e^{\varphi+2U}\, .
\end{equation} 
According to the criterion of \cite{Maldacena_nogo}, $|\hat{g}_{00}|$ should not diverge in order for the IR singularity to be physically acceptable. The behavior of $\hat{g}_{00}$ for the solutions shown in figures \ref{15_S2_special_SO(2)xSO(2)_SO(5)gg_flows}, \ref{15_H2_special_SO(2)xSO(2)_SO(5)gg_flows}, and \ref{15_R2_special_SO(2)xSO(2)_SO(5)gg_flows} are given in figure \ref{g00fig1to3}. From this figure, we readily see that, upon uplifted to type IIA theory, all the IR singularities of the solutions with $Mkw_3\times H^2$ and $Mkw_3\times \mathbb{R}^2$ slices are physically acceptable with $\hat{g}_{00}\rightarrow 0$ near the singularities. On the other hand, only the $Mkw_3\times S^2$-sliced solutions represented by the blue and purple curves contain physically acceptable singularities.   

\begin{figure}[h!]
  \centering
    \includegraphics[width=\linewidth]{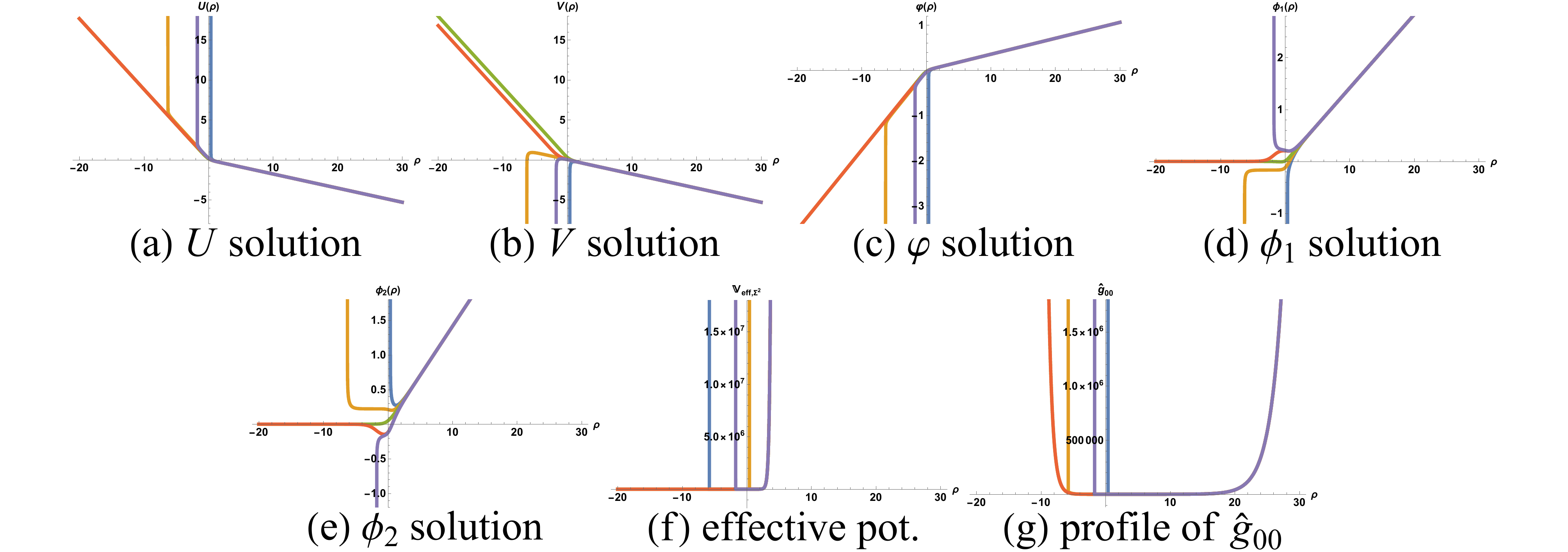}
\caption{Interpolating solutions between the locally $SO(2)\times SO(2)$ flat domain wall as $\rho\rightarrow+\infty$ and $Mkw_3\times S^2$-sliced curved domain walls for $SO(2)\times SO(2)$ twist in $SO(5)$ gauge group. The blue, orange, green, red, and purple curves refer to $p_1=-5, -2.94, -0.90, 1.90, 1.91$, respectively.}
\label{15_S2_SO(2)xSO(2)_SO(5)gg_flows}
\end{figure}

\begin{figure}[h!]
  \centering
    \includegraphics[width=\linewidth]{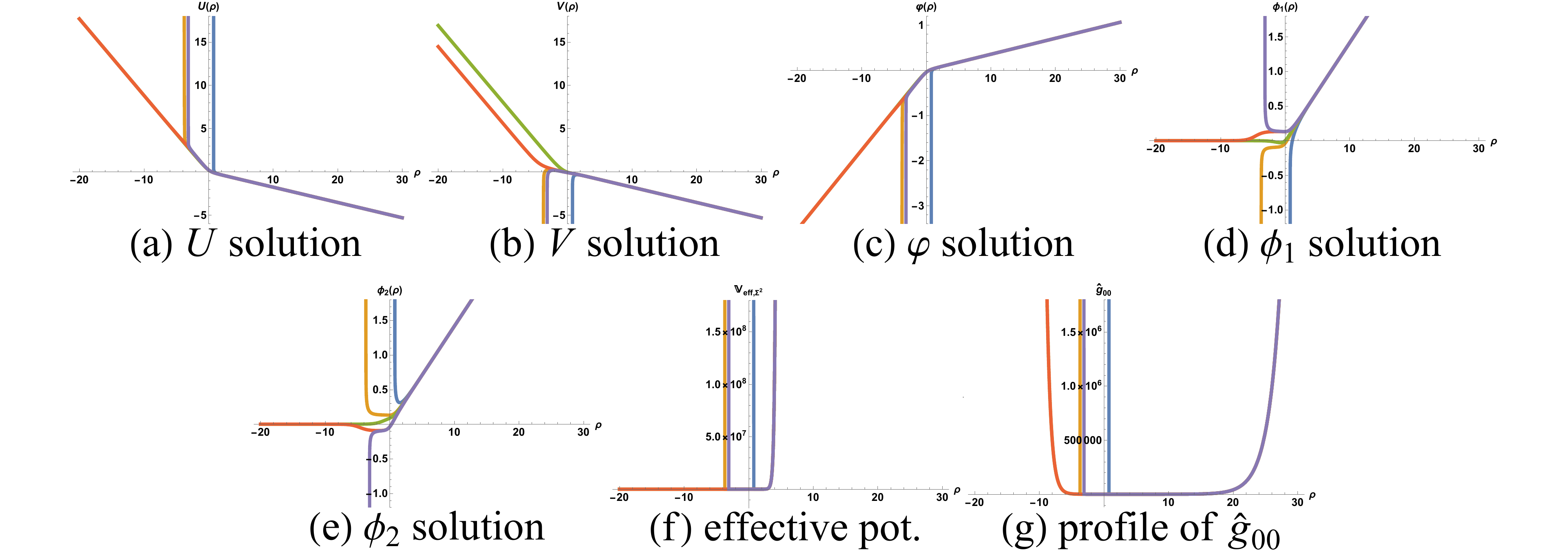}
\caption{Interpolating solutions between the locally $SO(2)\times SO(2)$ flat domain wall as $\rho\rightarrow+\infty$ and $Mkw_3\times H^2$-sliced curved domain walls for $SO(2)\times SO(2)$ twist in $SO(5)$ gauge group. The blue, orange, green, red, and purple curves refer to $p_1=-5.40, -0.43, 0, 1.42, 1.43$, respectively.}
\label{15_H2_SO(2)xSO(2)_SO(5)gg_flows}
\end{figure}

\begin{figure}[h!]
  \centering
    \includegraphics[width=\linewidth]{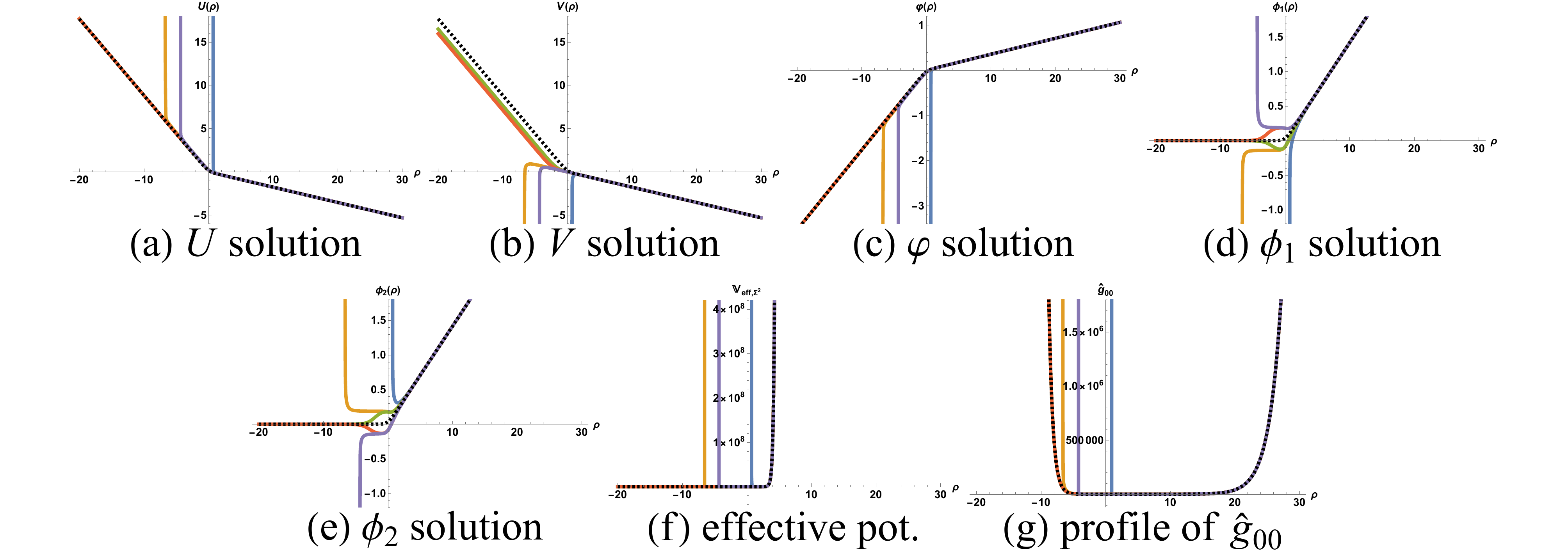}
\caption{Interpolating solutions interpolating between the locally $SO(2)\times SO(2)$ flat domain wall as $\rho\rightarrow+\infty$ and $Mkw_3\times \mathbb{R}^2$-sliced curved domain walls for $SO(2)\times SO(2)$ twist in $SO(5)$ gauge group. The blue, orange, green, red, and purple curves refer to $p_1=-6, -1.74, -1.70, 1.73, 1.74$, respectively. The dashed curve is the $SO(2)\times SO(2)$ flat domain wall obtained from $p_1=0$ which implies $p_2=k=0$.} 
\label{15_R2_SO(2)xSO(2)_SO(5)gg_flows}
\end{figure}

\begin{figure}[h!]
  \centering
    \includegraphics[width=\linewidth]{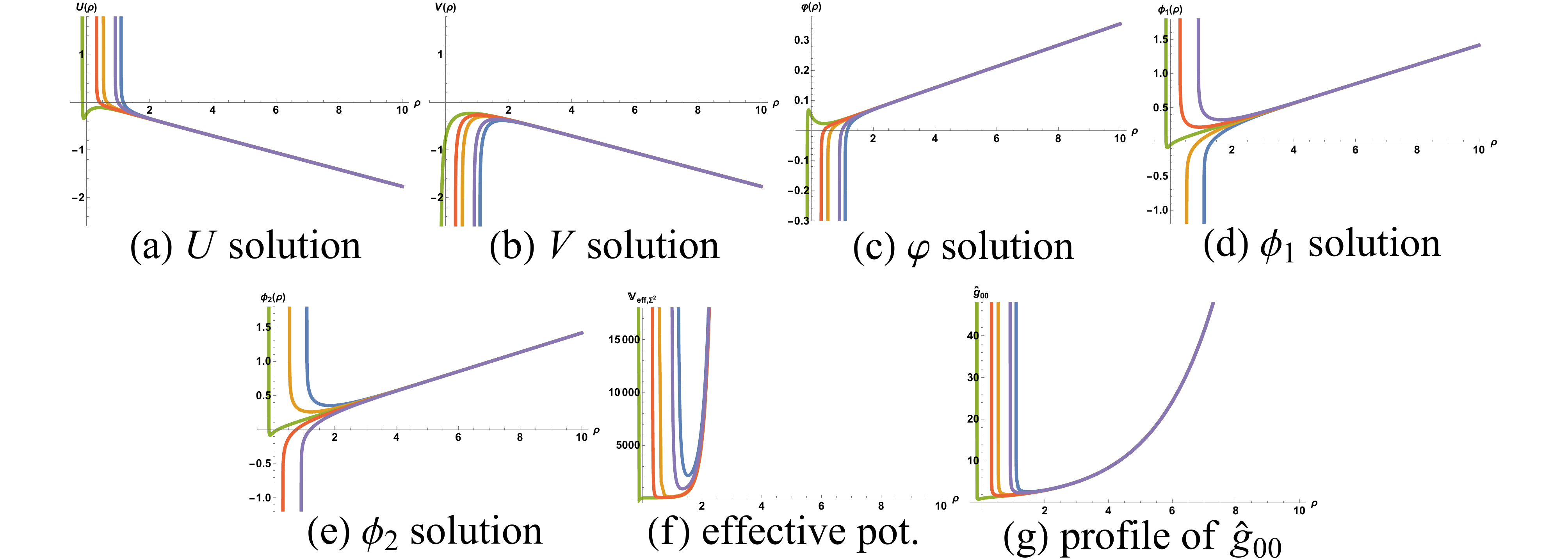}
\caption{Interpolating solutions between the locally $SO(2)\times SO(2)$ flat domain wall as $\rho\rightarrow+\infty$ and $Mkw_3\times S^2$-sliced curved domain walls for $SO(2)\times SO(2)$ twist in $SO(4,1)$ gauge group. The blue, orange, green, red, and purple curves refer to $p_1=-8.21, -2.80, 0.50, 2.50, 7$, respectively.}
\label{15_S2_SO(2)xSO(2)_SO(41)gg_flows}
\end{figure}

\begin{figure}[h!]
  \centering
    \includegraphics[width=\linewidth]{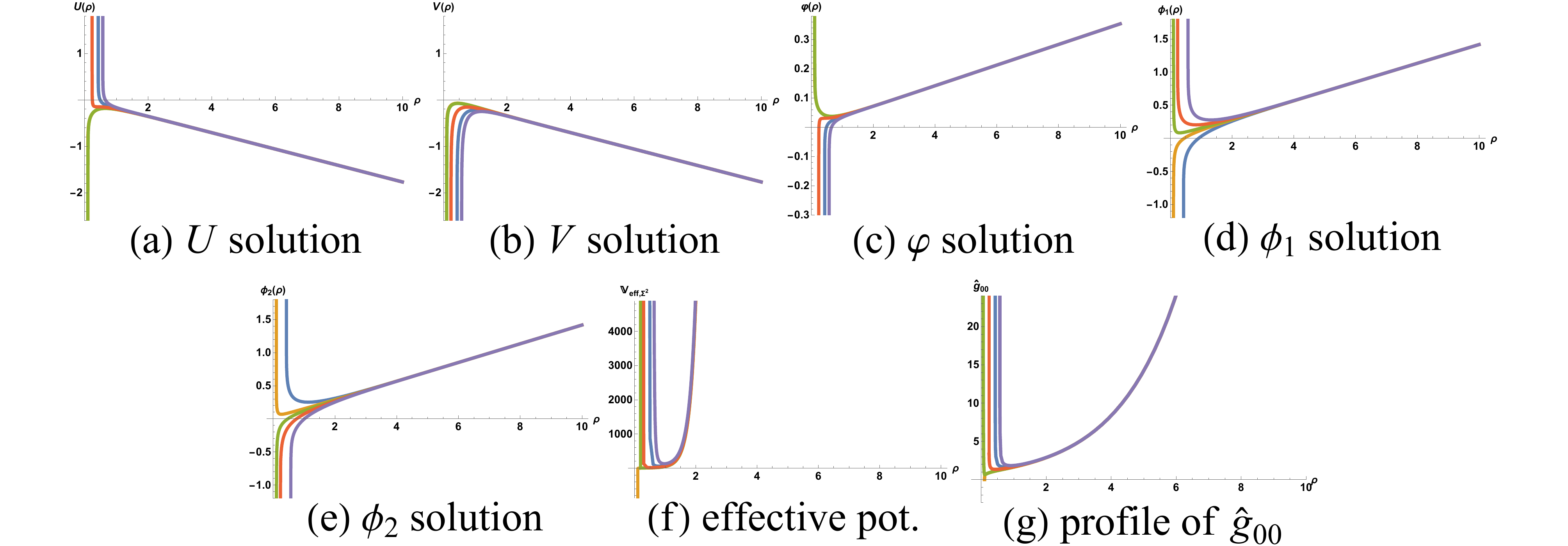}
\caption{Interpolating solutions between the locally $SO(2)\times SO(2)$ flat domain wall as $\rho\rightarrow+\infty$ and $Mkw_3\times H^2$-sliced curved domain walls for $SO(2)\times SO(2)$ twist in $SO(4,1)$ gauge group. The blue, orange, green, red, and purple curves refer to $p_1=-4, -1, 0.1, 1.7, 4$, respectively.}
\label{15_H2_SO(2)xSO(2)_SO(41)gg_flows}
\end{figure}

\begin{figure}[h!]
  \centering
    \includegraphics[width=\linewidth]{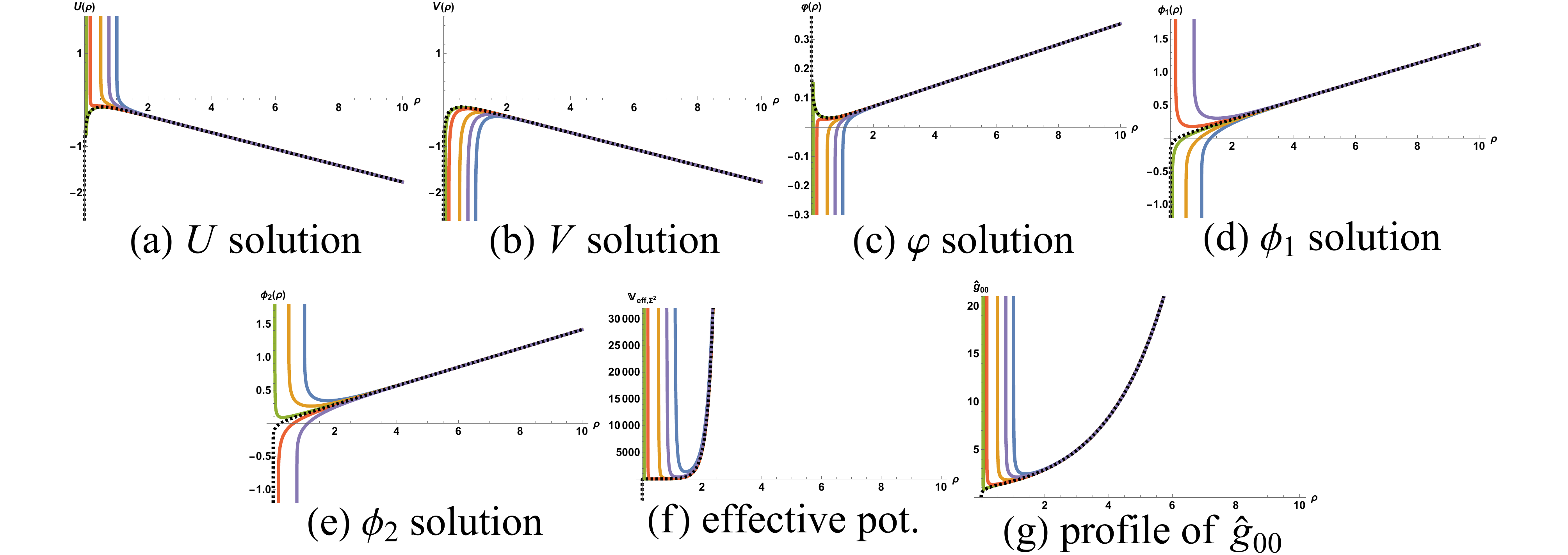}
\caption{Interpolating solutions between the locally $SO(2)\times SO(2)$ flat domain wall as $\rho\rightarrow+\infty$ and $Mkw_3\times \mathbb{R}^2$-sliced curved domain walls for $SO(2)\times SO(2)$ twist in $SO(4,1)$ gauge group. The blue, orange, green, red, and purple curves refer to $p_1=-8.2, -3.6, -0.5, 1.5, 5.6$, respectively. The dashed curve is the $SO(2)\times SO(2)$ flat domain wall obtained from $p_1=0$ which implies $p_2=k=0$.} 
\label{15_R2_SO(2)xSO(2)_SO(41)gg_flows}
\end{figure}

\begin{figure}[h!]
  \centering
    \includegraphics[width=\linewidth]{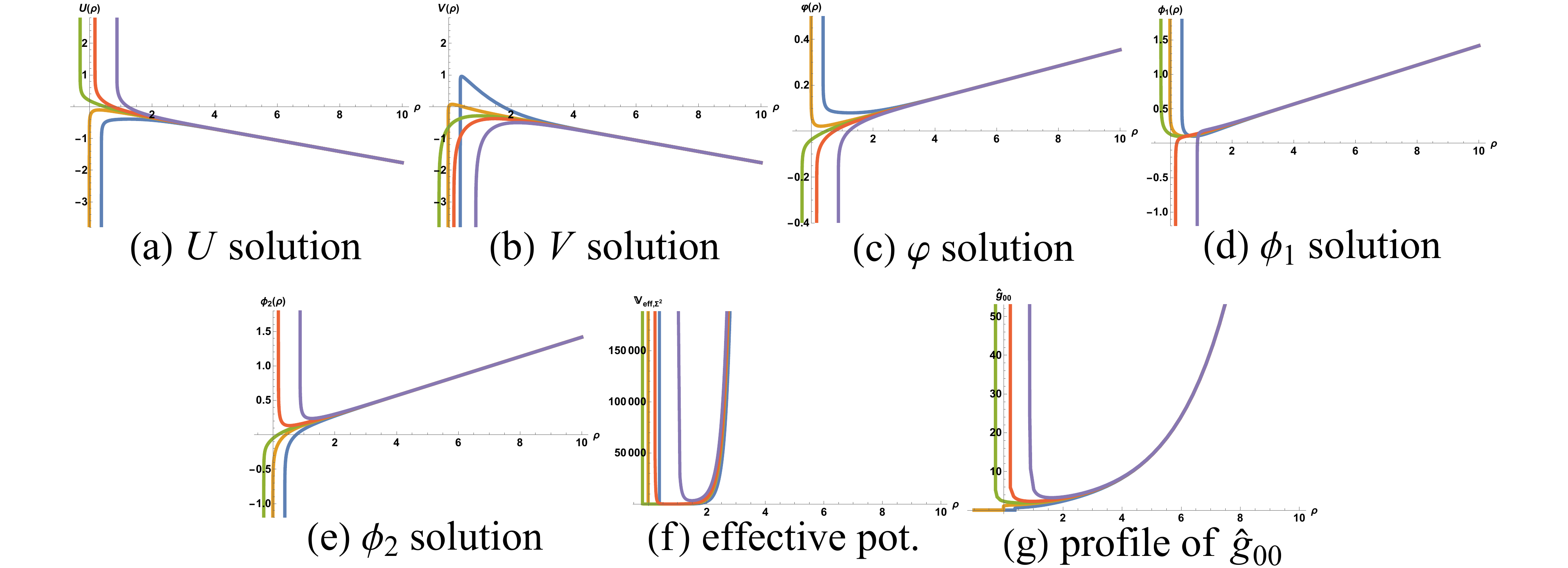}
\caption{Interpolating solutions between the locally $SO(2)\times SO(2)$ flat domain wall as $\rho\rightarrow+\infty$ and $Mkw_3\times S^2$-sliced curved domain walls for $SO(2)\times SO(2)$ twist in $SO(3,2)$ gauge group. The blue, orange, green, red, and purple curves refer to $p_1=-12.30, -0.93, 0.66, 1.86, 6$, respectively.}
\label{15_S2_SO(2)xSO(2)_SO(32)gg_flows}
\end{figure}

\begin{figure}[h!]
  \centering
    \includegraphics[width=\linewidth]{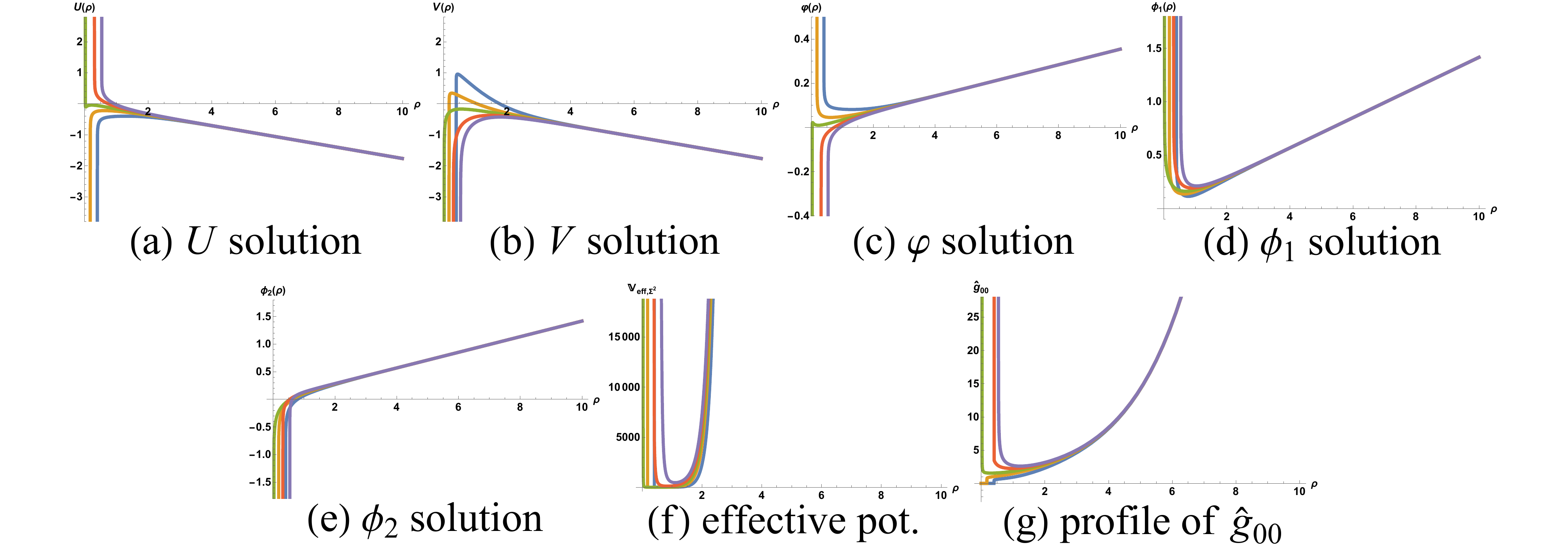}
\caption{Interpolating solutions between the locally $SO(2)\times SO(2)$ flat domain wall as $\rho\rightarrow+\infty$ and $Mkw_3\times H^2$-sliced curved domain walls for $SO(2)\times SO(2)$ twist in $SO(3,2)$ gauge group. The blue, orange, green, red, and purple curves refer to $p_1=-12, -2, 0.80, 2.81, 4.10$, respectively.}
\label{15_H2_SO(2)xSO(2)_SO(32)gg_flows}
\end{figure}

\begin{figure}[h!]
  \centering
    \includegraphics[width=\linewidth]{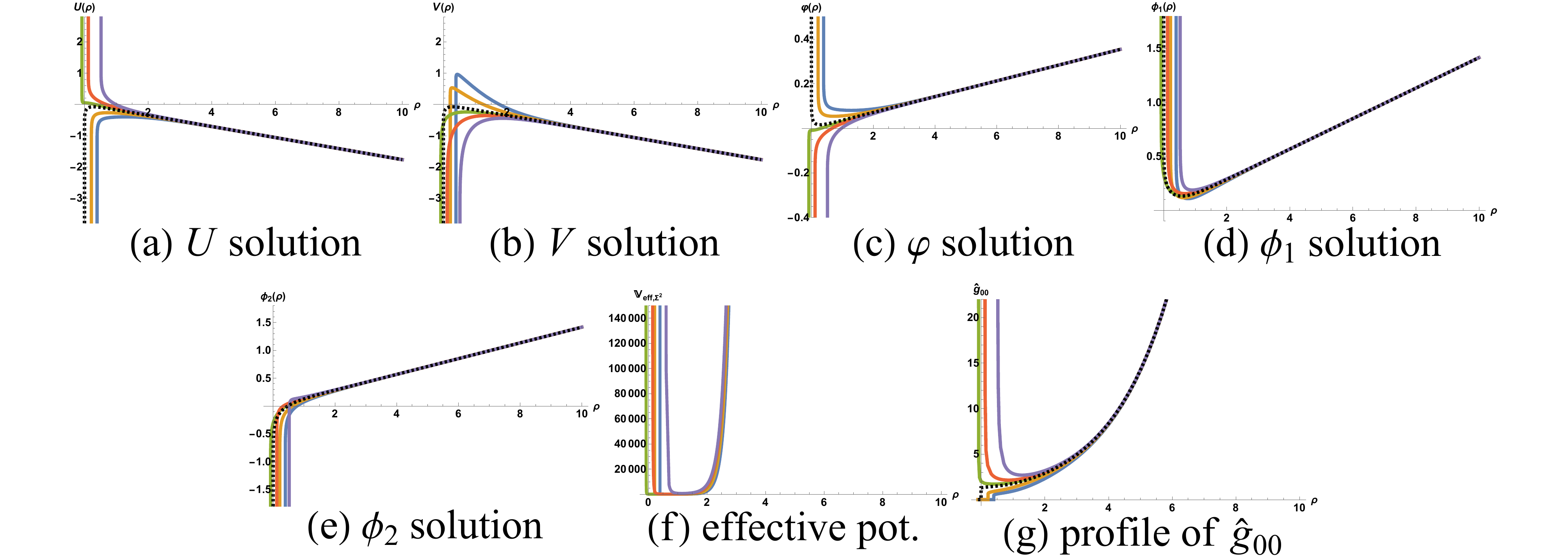}
\caption{Interpolating solutions between the locally $SO(2)\times SO(2)$ flat domain wall as $\rho\rightarrow+\infty$ and $Mkw_3\times \mathbb{R}^2$-sliced curved domain walls for $SO(2)\times SO(2)$ twist in $SO(3,2)$ gauge group. The blue, orange, green, red, and purple curves refer to $p_1=-12.30, -4, 0.78, 2, 4$, respectively. The dashed curve is the $SO(2)\times SO(2)$ flat domain wall obtained from $p_1=0$ which implies $p_2=k=0$.} 
\label{15_R2_SO(2)xSO(2)_SO(32)gg_flows}
\end{figure}

According to the DW/QFT correspondence, these physical singularities can be interpreted as holographic duals of $N=2$ three-dimensional SYM theories with $SO(2)\times SO(2)$ symmetry. The entire numerical solutions then describe holographic RG flows across dimensions between the five- and three-dimensional non-conformal field theories. 
\\
\indent We now move to the solutions with non-vanishing $\phi_1$ and $\phi_2$ in the asymptotic behavior. In this case, we require the asymptotic behavior of the solutions as $\rho\rightarrow+\infty$ to be
\begin{equation}\label{SO(2)xSO(2)_flat_DW_asym}
U\sim V\sim\frac{g \lambda  \rho }{4\sqrt{2}},\qquad \varphi\sim-\frac{g \lambda  \rho }{20 \sqrt{2}},\qquad \phi_1\sim\phi_2\sim-\frac{g \lambda  \rho }{5 \sqrt{2}}\, .
\end{equation}
We have chosen $g\lambda <0$ in order to identify the UV limit with $\rho\rightarrow +\infty$. As in the previous case, this asymptotic behavior solves the BPS equations in the limit $\rho\rightarrow+ \infty$ with the contribution from the gauge fields being small. In addition, this asymptotic solution takes the form of the $SO(2)\times SO(2)$ symmetric flat domain wall studied in \cite{6D_DW_I}. 
\\
\indent Examples of numerical solutions starting from this locally flat domain wall in the case of $SO(5)$, $SO(4,1)$, and $SO(3,2)$ gauge groups are given in figures \ref{15_S2_SO(2)xSO(2)_SO(5)gg_flows} to \ref{15_R2_SO(2)xSO(2)_SO(32)gg_flows}. In these solutions, we have set $g=-\lambda=\pm1$ with different values of $p_1$. As seen from the figures, the effective potential is bounded from above near the IR singularities in the case of $SO(5)$ gauge group for specific values of $p_1$, $-2.93\leq p_1\leq1.90$, $-0.42\leq p_1\leq1.42$, and $-1.73\leq p_1\leq1.73$, for $k=+1,-1,0$, respectively. However, upon uplifted to type IIA theory, these singularities are unphysical since the ten-dimensional metric $\hat{g}_{00}$ diverges near the singularities. For $SO(4,1)$ gauge group, only the $Mkw_3\times H^2$-sliced solution represented by the orange line in figure \ref{15_H2_SO(2)xSO(2)_SO(41)gg_flows} leads to physical singularity in both six and ten dimensions. On the other hand, although the effective potentials are not bounded from above for the curved domain walls with $p_1<0$ in $SO(3,2)$ gauge group (blue and orange curves in figures \ref{15_S2_SO(2)xSO(2)_SO(32)gg_flows}, \ref{15_H2_SO(2)xSO(2)_SO(32)gg_flows}, and \ref{15_R2_SO(2)xSO(2)_SO(32)gg_flows}), we find that  $\hat{g}_{00}\rightarrow0$ near the IR singularities. Therefore, these could lead to physically acceptable solutions in type IIA theory.
\vfil
\vfil\pagebreak
\vfil
\vfil
\vfil\pagebreak
\vfil
\vfil
\vfil\pagebreak
\vfil
\vfil
\vfil\pagebreak

For convenience, we summarize physical acceptabilities of the IR singularities according to the criteria of \cite{Gubser_Sing} and \cite{Maldacena_nogo} in table \ref{tab1}. In this table (and subsequent ones), we use the $\times$ symbol to indicate a violation of each criterion. For the cases satisfying each criterion, the conditions for the independent parameter $p_1$ are given. It should be noted that all $p_1$ conditions listed in table \ref{tab1} are applicable for $g=1$.

\begin{table}[h!]
\centering
\begin{tabular}{| c | c | c | c | c |}
\hline
UV flat & gauge & Riemann & criterion \cite{Gubser_Sing}  & criterion \cite{Maldacena_nogo} \\
domain wall& group & surface & ($\mathbf{V}_{\text{eff},\Sigma^2}\nrightarrow+\infty$) & ($\hat{g}_{00}\rightarrow0$) \\\hline
$SO(5)$ symmetric&$SO(5)$&$S^2$ & $\times$ & $p_1\leq-2$ \\
&& &  &  or $p_1\geq1.3$ \\
&&$H^2$ & $-0.5\leq p_1\leq0$ & any $p_1$ \\
&&$\mathbb{R}^2$ & $\times$ & any $p_1$ \\\hline
$SO(2)\times SO(2)$ symmetric&$SO(5)$& $S^2$ & $-2.93\leq p_1\leq1.90$ & $\times$ \\
&& $H^2$ & $-0.42\leq p_1\leq1.42$ & $\times$ \\
&& $\mathbb{R}^2$ & $-1.73\leq p_1\leq1.73$ & $\times$ \\\cline{2-5}
&$SO(4,1)$ & $S^2$ & $\times$ & $\times$ \\
& & $H^2$ & $-1\leq p_1\leq0$ & $-1\leq p_1\leq0$ \\
& & $\mathbb{R}^2$ & $p_1=0$ & $p_1=0$ \\\cline{2-5}
&$SO(3,2)$ & $S^2$ & $\times$ & $p_1\leq-0.93$ \\
& & $H^2$ & $\times$ & $p_1\leq-2$ \\
& & $\mathbb{R}^2$ & $\times$ & $p_1\leq-4$ \\\hline
\end{tabular}
\caption{Summary of satisfaction of criteria \cite{Gubser_Sing} and \cite{Maldacena_nogo} for the IR singularities of the solutions with $Mkw_3\times \Sigma^2$ slices obtained from $SO(2)\times SO(2)$ twist in $SO(5)$, $SO(4,1)$, and $SO(3,2)$ gauge groups.}\label{tab1}
\end{table}

\begin{figure}[h!]
  \centering
    \includegraphics[width=\linewidth]{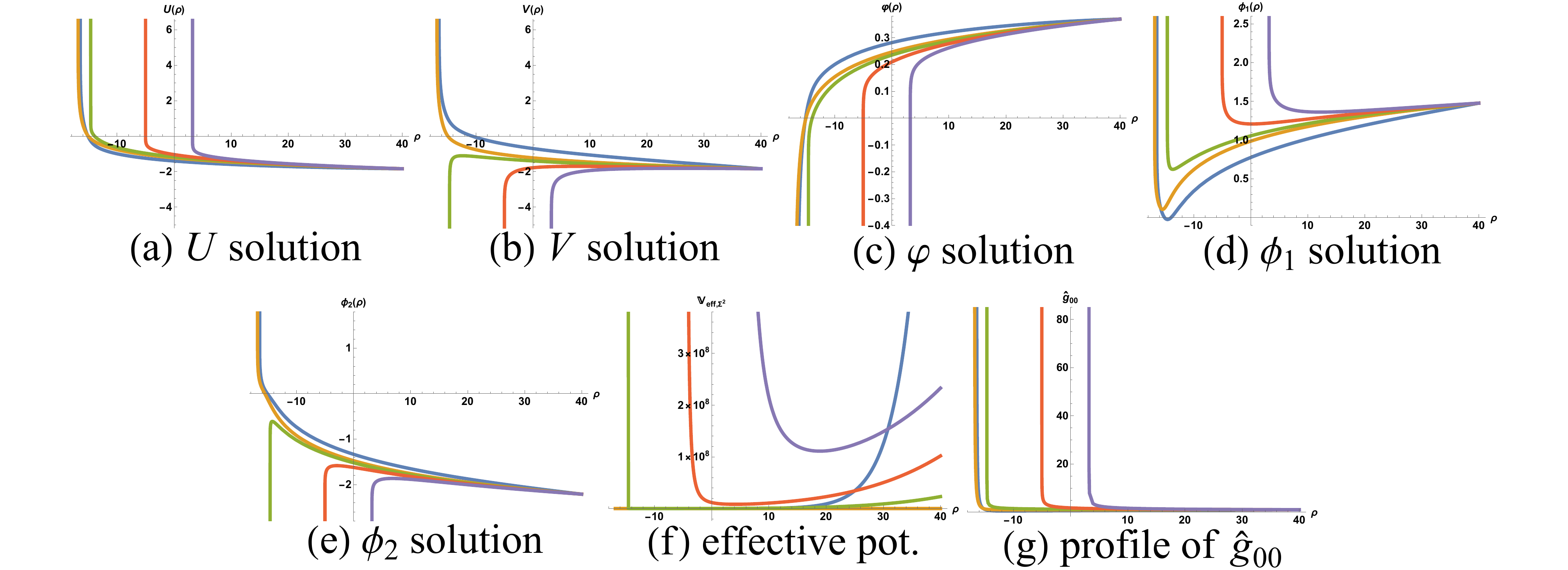}
\caption{Interpolating solutions between the locally $SO(2)\times SO(2)$ flat domain wall as $\rho\rightarrow+\infty$ and $Mkw_3\times S^2$-sliced curved domain walls for $SO(2)\times SO(2)$ twist in $CSO(4,0,1)$ gauge group. The blue, orange, green, red, and purple curves refer to $p_1=-2.50, 0, 0.30, 0.63, 0.95$, respectively.}
\label{15_S2_SO(2)xSO(2)_CSO(401)gg_flows}
\end{figure}

\begin{figure}[h!]
  \centering
    \includegraphics[width=\linewidth]{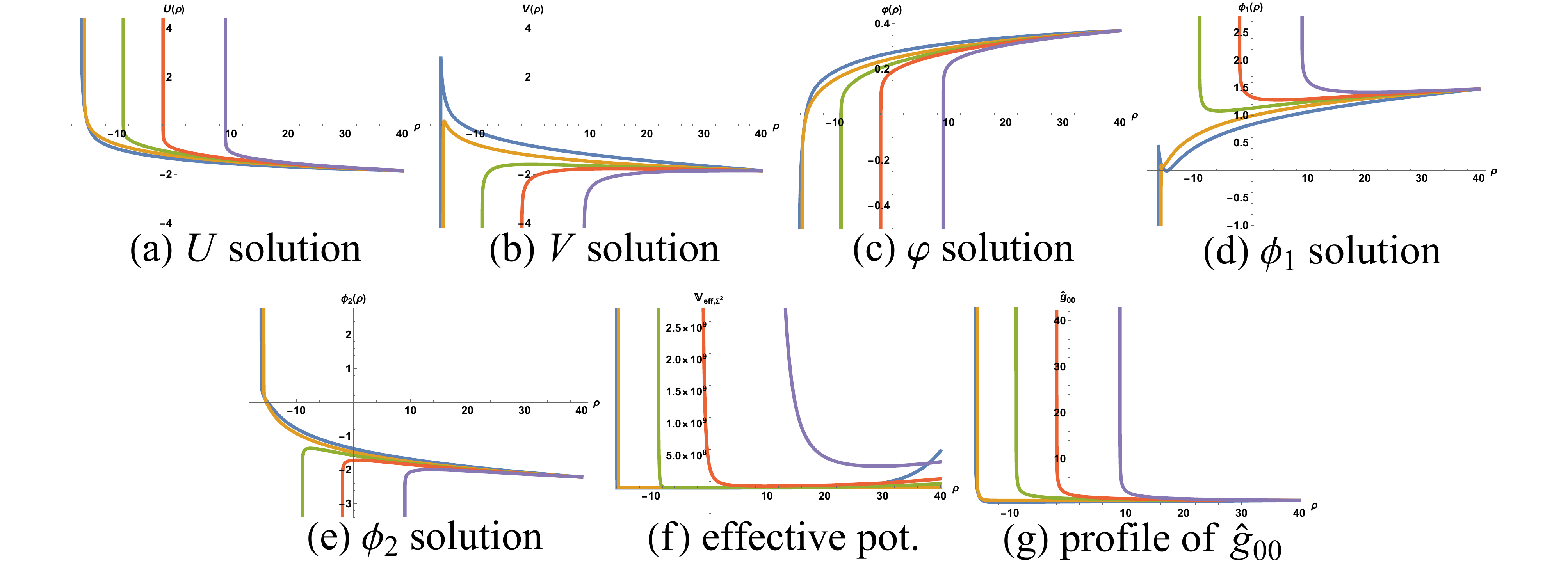}
\caption{Interpolating solutions between the locally $SO(2)\times SO(2)$ flat domain wall as $\rho\rightarrow+\infty$ and $Mkw_3\times H^2$-sliced curved domain walls for $SO(2)\times SO(2)$ twist in $CSO(4,0,1)$ gauge group. The blue, orange, green, red, and purple curves refer to $p_1=-1.50, 0, 0.50, 0.74, 1.26$, respectively.}
\label{15_H2_SO(2)xSO(2)_CSO(401)gg_flows}
\end{figure}

\begin{figure}[h!]
  \centering
    \includegraphics[width=\linewidth]{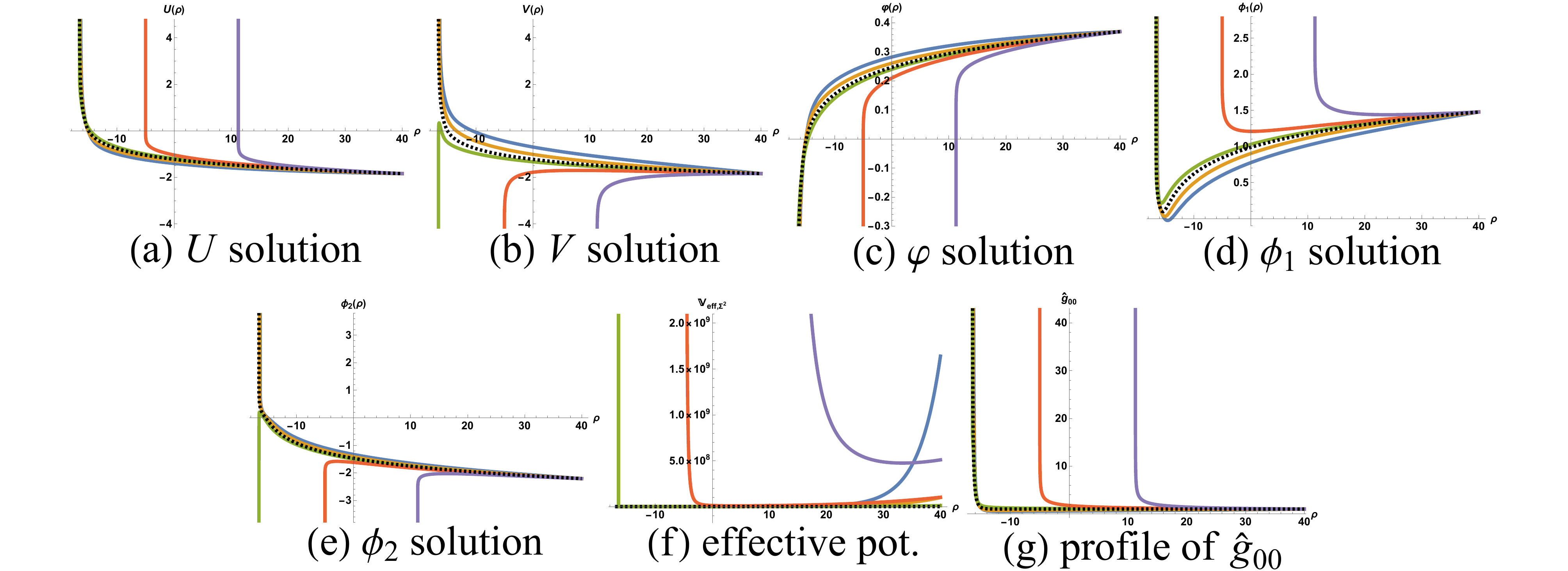}
\caption{Interpolating solutions between the locally $SO(2)\times SO(2)$ flat domain wall as $\rho\rightarrow+\infty$ and $Mkw_3\times \mathbb{R}^2$-sliced curved domain walls for $SO(2)\times SO(2)$ twist in $CSO(4,0,1)$ gauge group. The blue, orange, green, red, and purple curves refer to $p_1=-2.53, -0.62, 0.20, 0.63, 1.41$, respectively. The dashed curve is the $SO(2)\times SO(2)$ flat domain wall obtained from $p_1=0$ which implies $p_2=k=0$.} 
\label{15_R2_SO(2)xSO(2)_CSO(401)gg_flows}
\end{figure}

\begin{figure}[h!]
  \centering
    \includegraphics[width=\linewidth]{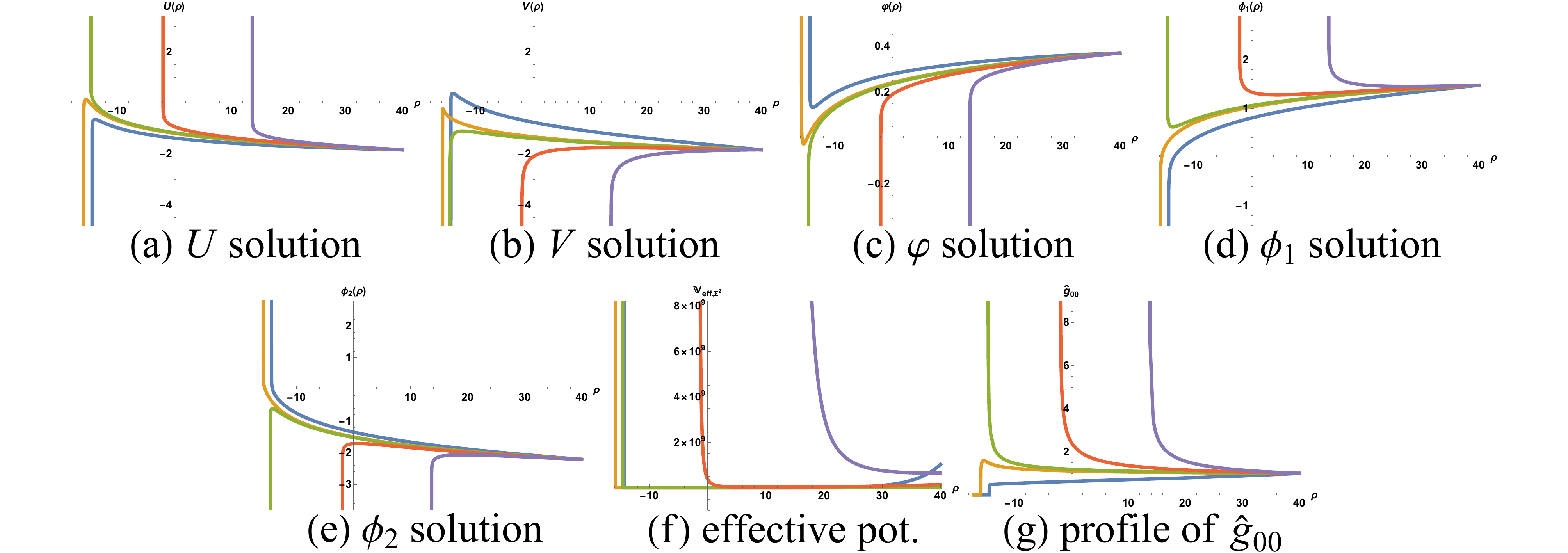}
\caption{Interpolating solutions between the locally $SO(2)\times SO(2)$ flat domain wall as $\rho\rightarrow+\infty$ and $Mkw_3\times S^2$-sliced curved domain walls for $SO(2)\times SO(2)$ twist in $CSO(2,2,1)$ gauge group. The blue, orange, green, red, and purple curves refer to $p_1=-2, 0.20, 0.30, 0.74, 1.6$, respectively.}
\label{15_S2_SO(2)xSO(2)_CSO(221)gg_flows}
\end{figure}

\begin{figure}[h!]
  \centering
    \includegraphics[width=\linewidth]{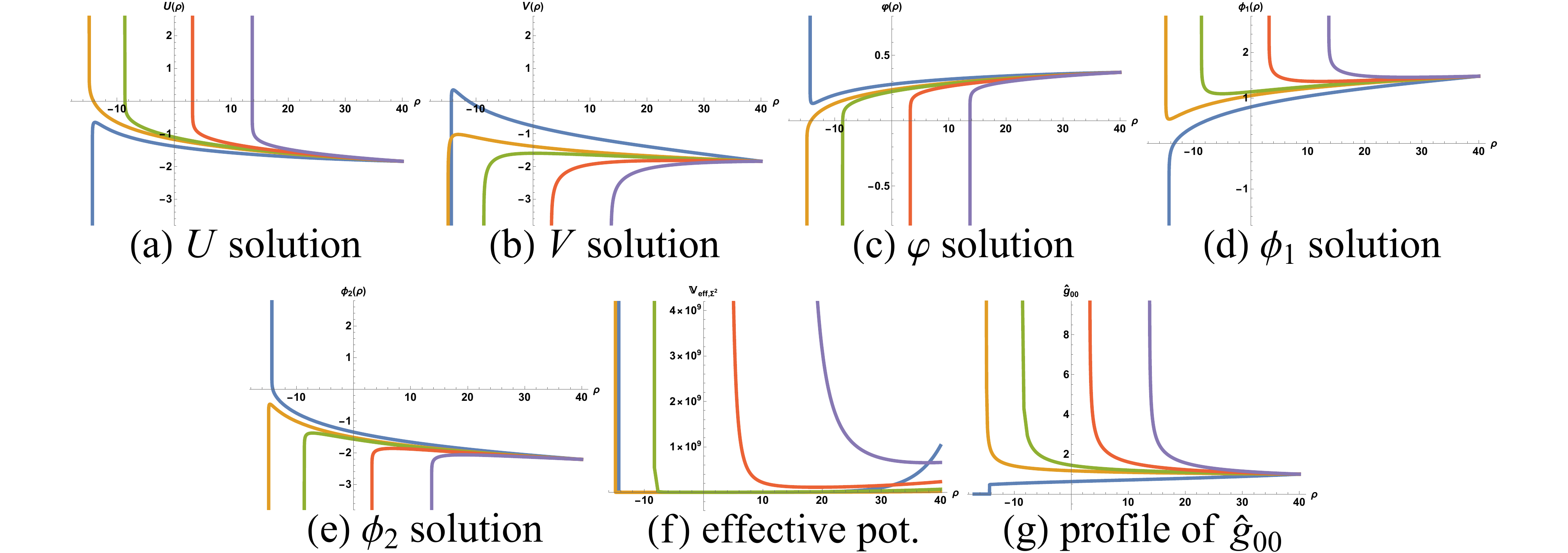}
\caption{Interpolating solutions between the locally $SO(2)\times SO(2)$ flat domain wall as $\rho\rightarrow+\infty$ and $Mkw_3\times H^2$-sliced curved domain walls for $SO(2)\times SO(2)$ twist in $CSO(2,2,1)$ gauge group. The blue, orange, green, red, and purple curves refer to $p_1=-2, 0.28, 0.51, 0.95, 1.60$, respectively.}
\label{15_H2_SO(2)xSO(2)_CSO(221)gg_flows}
\end{figure}

\begin{figure}[h!]
  \centering
    \includegraphics[width=\linewidth]{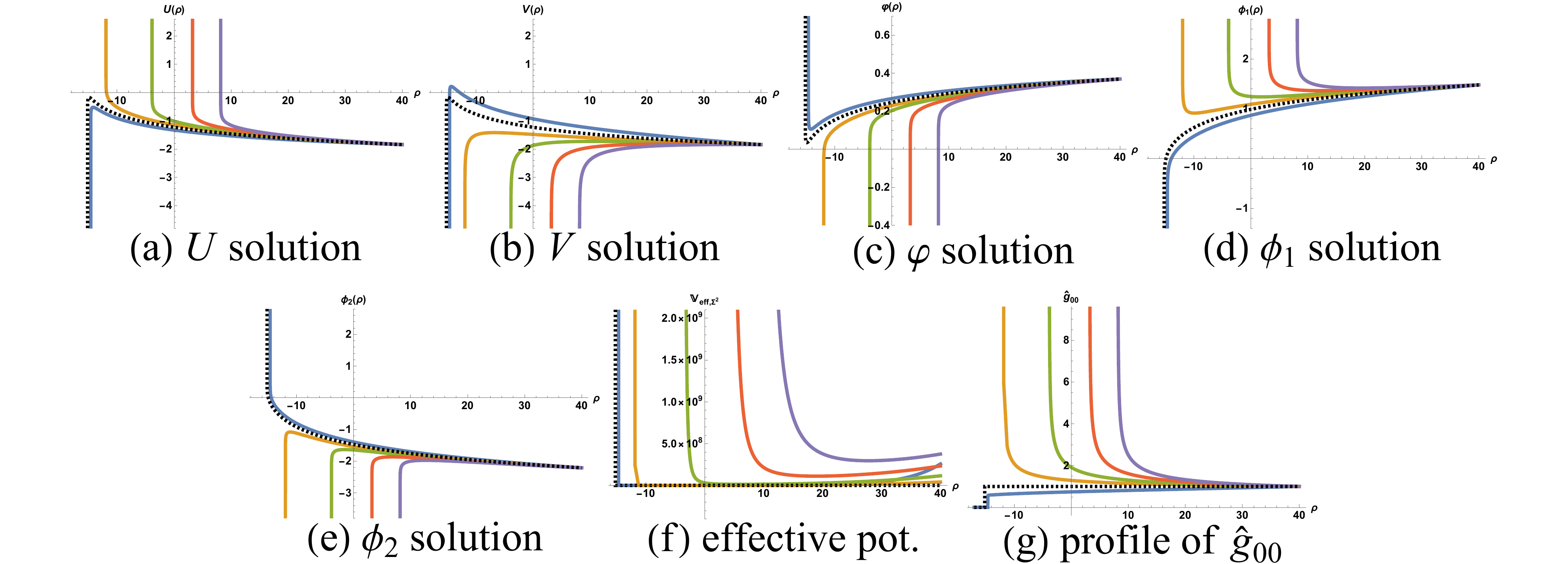}
\caption{Interpolating solutions between the locally $SO(2)\times SO(2)$ flat domain wall as $\rho\rightarrow+\infty$ and $Mkw_3\times \mathbb{R}^2$-sliced curved domain walls for $SO(2)\times SO(2)$ twist in $CSO(2,2,1)$ gauge group. The blue, orange, green, red, and purple curves refer to $p_1=-1, 0.40, 0.67, 0.95, 1.21$, respectively. The dashed curve is the $SO(2)\times SO(2)$ flat domain wall obtained from $p_1=0$ which implies $p_2=k=0$.} 
\label{15_R2_SO(2)xSO(2)_CSO(221)gg_flows}
\end{figure}

%%%%%%%%%%%%%%%%%%%%%%%%%%%%%%%%%%%%%%%%%%%%%%%%%%%%%%%%%%%%%%%%%%%%%%%%%%%%%%%%%%%%%
\paragraph{Solutions from $CSO(4,0,1)$ and $CSO(2,2,1)$ gauge groups}
We now consider the case of non-semisimple gauge groups with $\lambda=0$. These gauge groups are given by $CSO(4,0,1)$ and $CSO(2,2,1)$. As in the previous case, we defined a new radial coordinate $\rho$ via $\frac{d\rho}{dr}=e^{-\frac{5}{2}\phi_1-\frac{15}{2}\phi_2}$. By choosing $g\kappa<0$ and considering the limit in which the contributions from the gauge fields are small compared to other terms in the BPS equations, we require that as $\rho\rightarrow+\infty$, the solutions should have an asymptotic behavior of the form 
\begin{eqnarray}
U&\sim& V\sim\frac{3}{4}\ln \rho-\frac{5}{4}\ln(-g\kappa\rho) ,\qquad \varphi\sim\frac{1}{4}\ln(-g\kappa\rho)-\frac{3}{20}\ln \rho,\nonumber\\\phi_1&\sim&\frac{9}{10}\ln \rho-\frac{1}{2}\ln(-g\kappa\rho),\quad\qquad\ \phi_2\sim-\frac{3}{5}\ln\rho\label{15_SO(2)xSO(2)_fDW_lam0}
\end{eqnarray}
which is an $SO(2)\times SO(2)$ symmetric flat domain wall. We find numerical solutions starting from this $SO(2)\times SO(2)$ flat domain wall to other singular geometries in the IR as shown in figures \ref{15_S2_SO(2)xSO(2)_CSO(401)gg_flows} to \ref{15_R2_SO(2)xSO(2)_CSO(221)gg_flows}. In these solutions, we have set $g=-\lambda=\pm1$ with different values of $p_1$. From these figures, we find that only $Mkw_3\times S^2$- and $Mkw_3\times \mathbb{R}^2$-sliced curved domain walls with $p_1\leq0$ in $CSO(4,0,1)$ gauge group (blue and orange curves in figures \ref{15_S2_SO(2)xSO(2)_CSO(401)gg_flows} and \ref{15_R2_SO(2)xSO(2)_CSO(401)gg_flows}) admit physical IR singularities with the effective potential bounded from above near the singularities. However, these lead to uplifted solutions with unphysical singularities as indicated by the behaviors of $\hat{g}_{00}$ in the IR limit.  
\\
\indent For $CSO(2,2,1)$ gauge group, we find that the $Mkw_3\times S^2$-sliced solutions represented by blue and orange curves in figure \ref{15_S2_SO(2)xSO(2)_CSO(221)gg_flows} lead to uplifted type IIA solutions with physical IR singularities with $\hat{g}_{00}\rightarrow 0$ although the effective potential is not bounded from above. For $Mkw_3\times H^2$- and $Mkw_3\times \mathbb{R}^2$-sliced solutions, only the solutions represented by the blue curve in figures \ref{15_H2_SO(2)xSO(2)_CSO(221)gg_flows} and \ref{15_R2_SO(2)xSO(2)_CSO(221)gg_flows} admit physical IR singularities upon uplifted to type IIA theory. These solutions can be interpreted as $N=2$ three-dimensional non-conformal field theories obtained from twisted compactifications of five-dimensional non-conformal field theory. We summarize all solutions from these two non-semisimple gauge groups in table \ref{tab2}.
\vfil
\vfil\pagebreak
\vfil
\vfil
\vfil\pagebreak
\vfil
\vfil
\vfil

\begin{table}[h!]
\centering
\begin{tabular}{| c | c | c | c |}
\hline
 \multirow{2}{*}{gauge group} & Riemann & criterion \cite{Gubser_Sing}  & criterion \cite{Maldacena_nogo} \\
& surface & ($\mathbf{V}_{\text{eff},\Sigma^2}\nrightarrow+\infty$) & ($\hat{g}_{00}\rightarrow0$) \\\hline
$CSO(4,0,1)$& $S^2$, $\mathbb{R}^2$ & $p_1\leq 0$ & $\times$ \\
& $H^2$ & $\times$ & $\times$ \\\hline
$CSO(2,2,1)$ & $S^2$ & $\times$ & $p_1\leq0.24$ \\
 & $H^2$ & $\times$ & $p_1\leq-1$ \\
 & $\mathbb{R}^2$ & $\times$ & $p_1\leq0$ \\\hline
\end{tabular}
\caption{Summary of satisfaction of criteria \cite{Gubser_Sing} and \cite{Maldacena_nogo} for the IR singularities of the interpolating solutions between the locally $SO(2)\times SO(2)$ symmetric flat domain wall and $Mkw_3\times \Sigma^2$-sliced curved domain walls for $SO(2)\times SO(2)$ twist in $CSO(4,0,1)$ and $CSO(2,2,1)$ gauge groups.}\label{tab2}
\end{table}

%%%%%%%%%%%%%%%%%%%%%%%%%%%%%%%%%%%%%%%%%%%%%%%%%%%%%%%%%%%%%%%%%%%%%%%%%%%%%%%%%%%%%%%%%%%%%%%%%%%%%%%%%%%%%%%%%%%%%%%%%%%%%%%%%%%%%%%%%
\subsubsection{Solutions with non-vanishing $\varsigma_1$ and $\varsigma_2$}
As previously shown, in order to keep the shift scalars $\varsigma_1$ and $\varsigma_2$ non-vanishing, we need to impose the condition $p_2=p_1/\kappa$ together with an algebraic condition
\begin{equation}\label{15_Sig2_keep_shift_con}
e^{4\phi_2}\varsigma_1+\kappa e^{4\phi_1}\varsigma_2=0\, .
\end{equation}
However, in this case, the $SO(5)\times SO(5)$ composite connections along the radial direction are non-vanishing. Therefore, the Killing spinors are given by \eqref{15_sig2_full_Killing} subject to the projectors \eqref{SO(2)xSO(2)Projcon+}, \eqref{SO(2)xSO(2)Projcon-}, and \eqref{15_DW_Proj}. For convenience, we also collect all the BPS equations here
\begin{eqnarray}
U'&=&\frac{ge^{\varphi}}{4\sqrt{2}}(2e^{-4\phi_1}+2\kappa e^{-4\phi_2}+\lambda e^{8(\phi_1+\phi_2)})-\frac{p}{8\sqrt{2}}e^{-2V-\varphi}(e^{4\phi_1}+\kappa e^{4\phi_2}),\label{15_SO(2)d_twist_BPS_eq1}\\
V'&=&\frac{ge^{\varphi}}{4\sqrt{2}}(2e^{-4\phi_1}+2\kappa e^{-4\phi_2}+\lambda e^{8(\phi_1+\phi_2)})+\frac{3p}{8\sqrt{2}}e^{-2V-\varphi}(e^{4\phi_1}+\kappa e^{4\phi_2}),\label{15_SO(2)d_twist_BPS_eq2}\\
\varphi'&=&-\frac{ge^{\varphi}}{20\sqrt{2}}(2e^{-4\phi_1}+2\kappa e^{-4\phi_2}+\lambda e^{8(\phi_1+\phi_2)})+\frac{p}{40\sqrt{2}}e^{-2V-\varphi}(e^{4\phi_1}+\kappa e^{4\phi_2}),\qquad\label{15_SO(2)d_twist_BPS_eq3}\ \\
\phi'_1&=&\frac{ge^{\varphi}}{5\sqrt{2}}(3e^{-4\phi_1}-2\kappa e^{-4\phi_2}-\lambda e^{8(\phi_1+\phi_2)})-\frac{p}{20\sqrt{2}}e^{-2V-\varphi}(3e^{4\phi_1}-2\kappa e^{4\phi_2}),\label{15_SO(2)d_twist_BPS_eq4}\\
\phi'_2&=&\frac{ge^{\varphi}}{5\sqrt{2}}(3\kappa e^{-4\phi_2}-2e^{-4\phi_1}-\lambda e^{8(\phi_1+\phi_2)})+\frac{p}{20\sqrt{2}}e^{-2V-\varphi}(2e^{4\phi_1}-3\kappa e^{4\phi_2}),\label{15_SO(2)d_twist_BPS_eq5}\qquad \\
\varsigma'_1&=&\frac{ge^{\varphi}}{\sqrt{2}}(2e^{-4\phi_1}-2\kappa e^{-4\phi_2}-\lambda e^{8(\phi_1+\phi_2)})\varsigma_1-\frac{p}{2\sqrt{2}}e^{-2V-\varphi}(e^{4\phi_1}-\kappa e^{4\phi_2})\varsigma_1.\label{15_SO(2)d_twist_BPS_eq6}
\end{eqnarray}
These equations are compatible with all the bosonic field equations. In addition, these equations also imply that $W'=0$, so we can choose $W=0$ making the Killing spinors reduce to \eqref{DW_Killing_spinor}. As in the previous case, there are no $AdS_4$ fixed points from these equations. 

\begin{figure}[h!]
  \centering
    \includegraphics[width=0.7\linewidth]{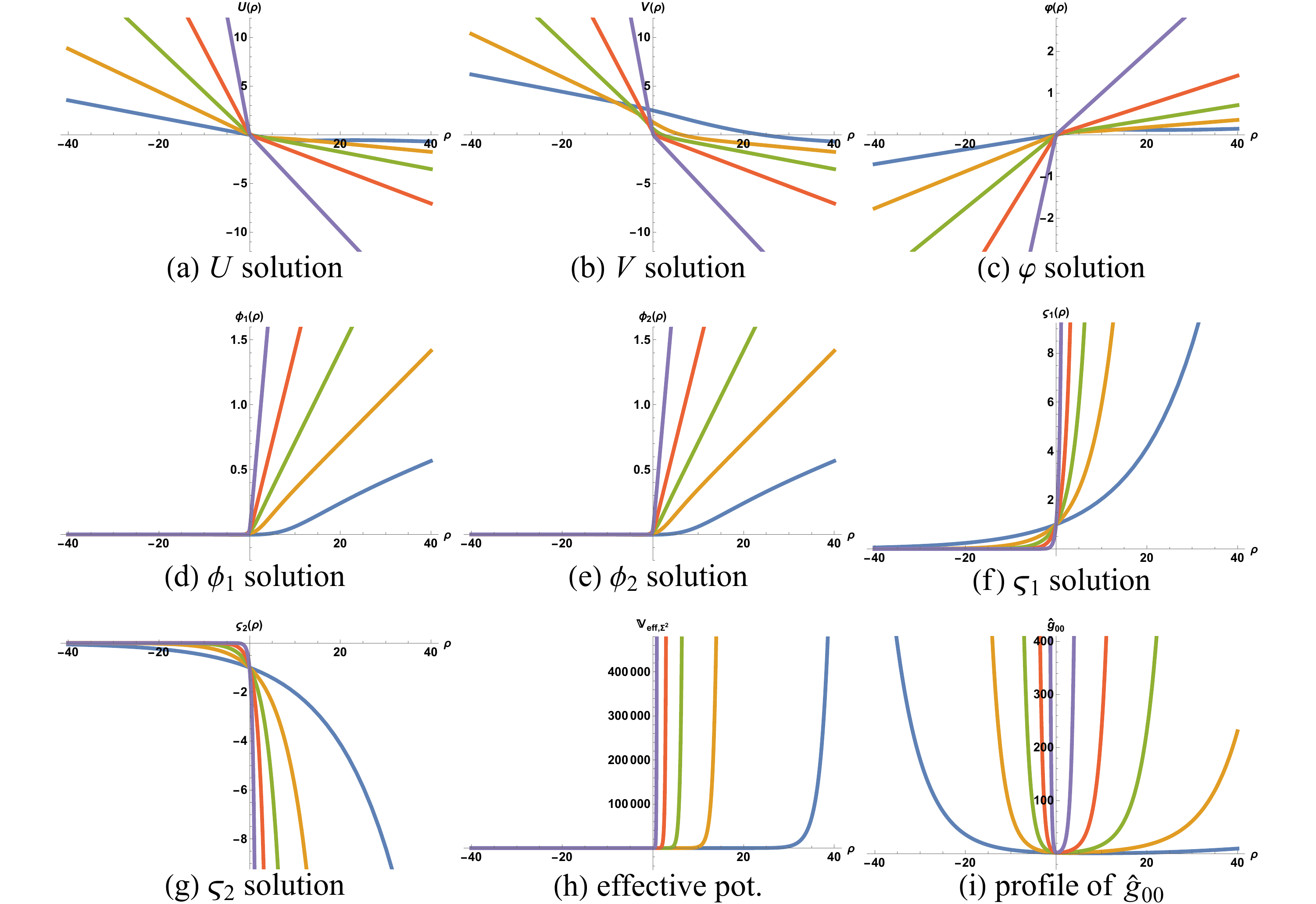}
\caption{Interpolating solutions between the locally $SO(2)\times SO(2)$ flat domain wall as $\rho\rightarrow+\infty$ and $Mkw_3\times S^2$-sliced curved domain walls with shift scalars in $SO(5)$ gauge group. The blue, orange, green, red, and purple curves refer to $g=-0.10, -0.25, -0.50, -1, -2.80$, respectively.}
\label{15_S2_SO(2)d_SO(5)gg_flows}
\end{figure}

\begin{figure}[h!]
  \centering
    \includegraphics[width=0.7\linewidth]{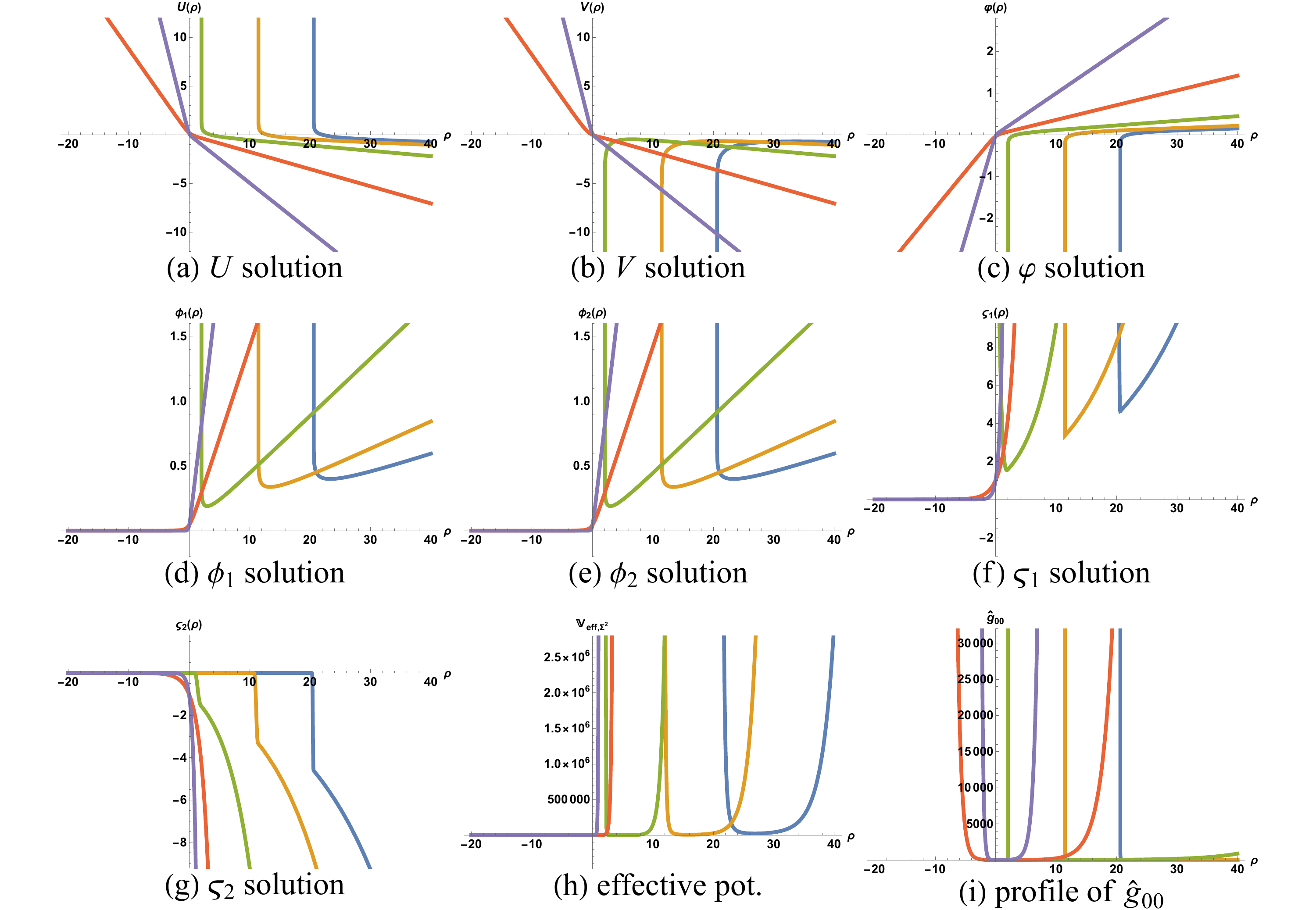}
\caption{Interpolating solutions between the locally $SO(2)\times SO(2)$ flat domain wall as $\rho\rightarrow+\infty$ and $Mkw_3\times H^2$-sliced curved domain walls with shift scalars in $SO(5)$ gauge group. The blue, orange, green, red, and purple curves refer to $g=-0.10, -0.15, -0.31, -1, -2.80$, respectively.}
\label{15_H2_SO(2)d_SO(5)gg_flows}
\end{figure}

\begin{figure}[h!]
  \centering
    \includegraphics[width=0.7\linewidth]{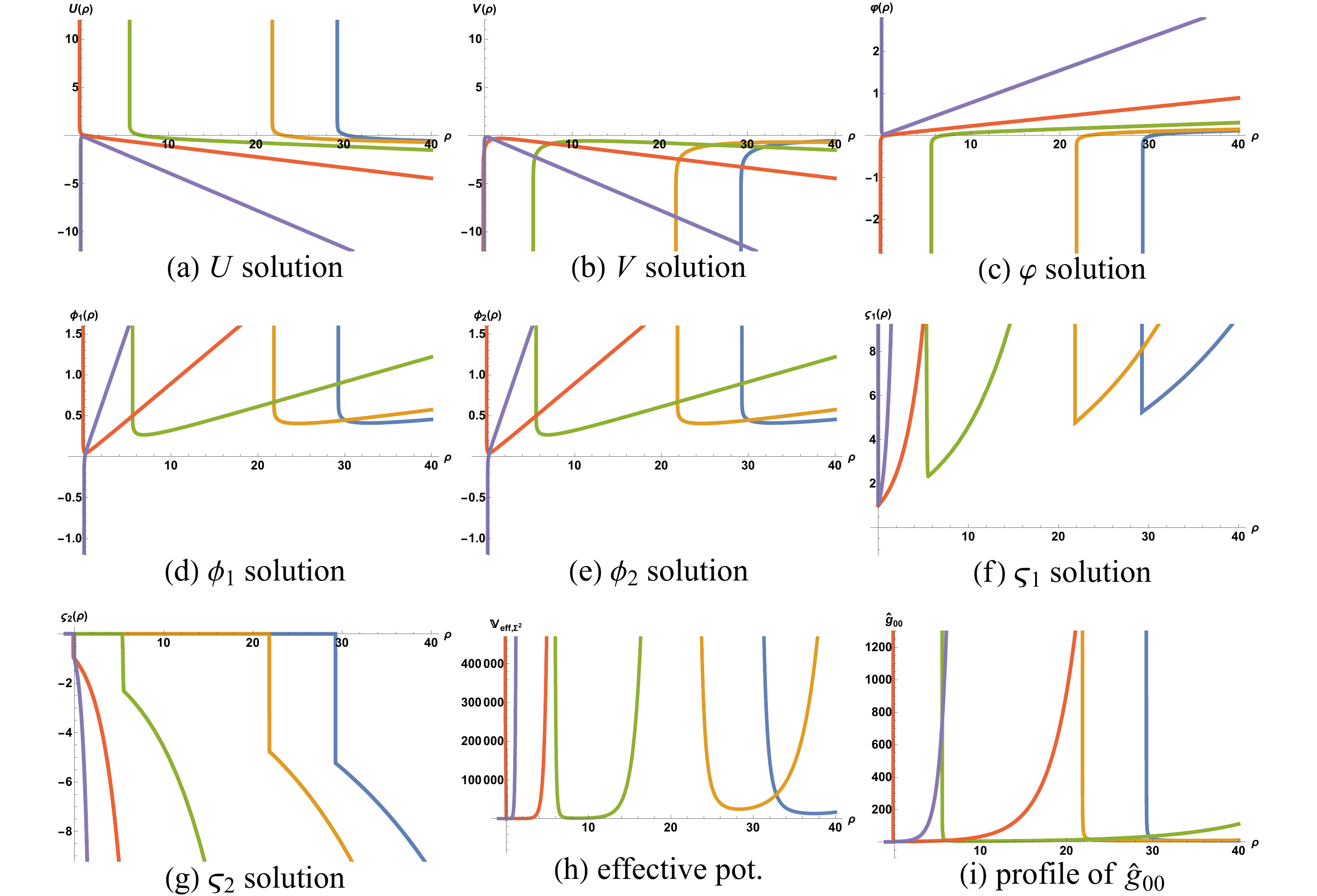}
\caption{Interpolating solutions between the locally $SO(2)\times SO(2)$ flat domain wall as $\rho\rightarrow+\infty$ and $Mkw_3\times S^2$-sliced curved domain walls with shift scalars in $SO(4,1)$ gauge group. The blue, orange, green, red, and purple curves refer to $g=0.08, 0.10, 0.20, 0.63, 2.19$, respectively.}
\label{15_S2_SO(2)d_SO(41)gg_flows}
\end{figure}

\begin{figure}[h!]
  \centering
    \includegraphics[width=0.7\linewidth]{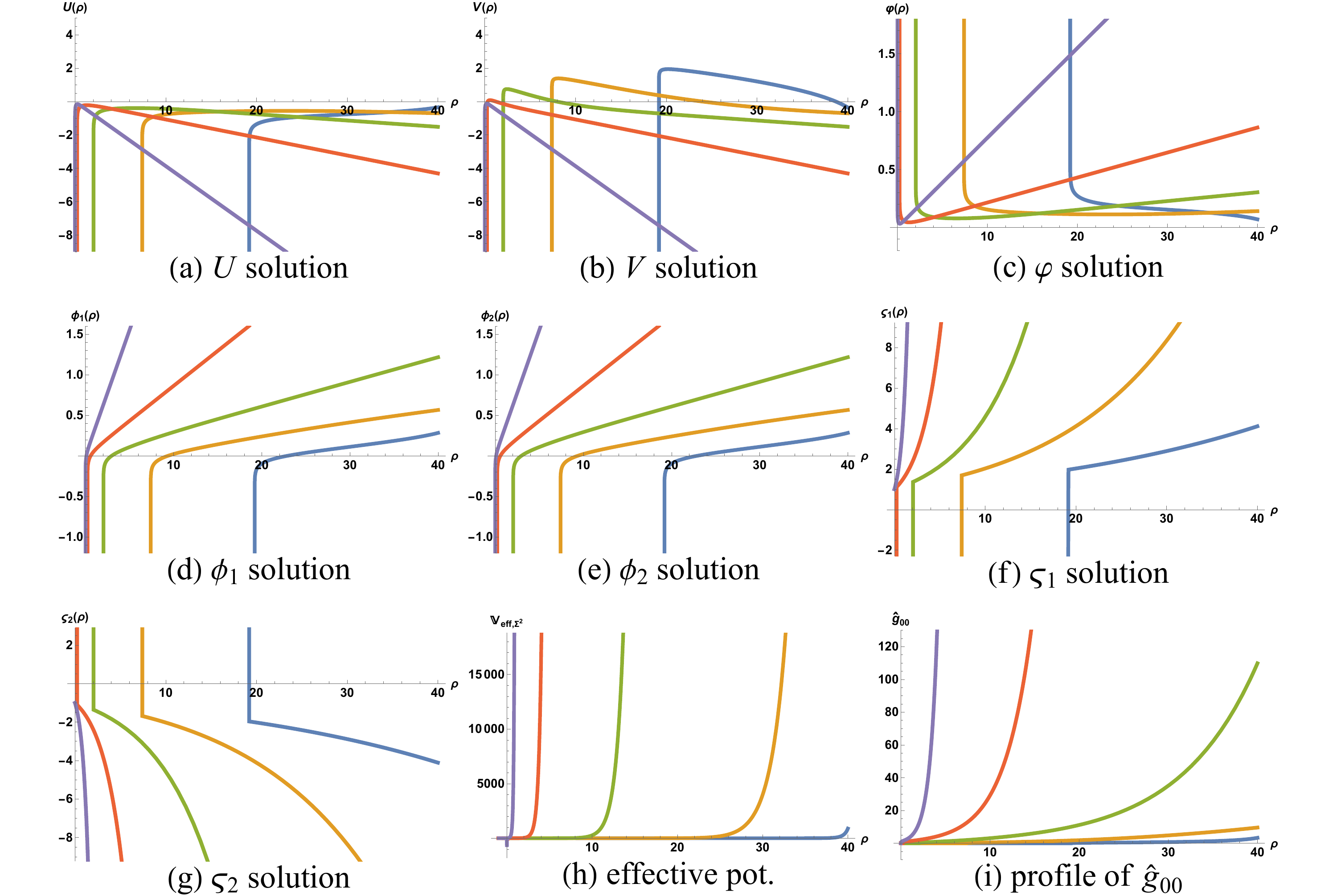}
\caption{Interpolating solutions between the locally $SO(2)\times SO(2)$ flat domain wall as $\rho\rightarrow+\infty$ and $Mkw_3\times H^2$-sliced curved domain walls with shift scalars in $SO(4,1)$ gauge group. The blue, orange, green, red, and purple curves refer to $g=0.05, 0.10, 0.20, 0.61, 2.19$, respectively.}
\label{15_H2_SO(2)d_SO(41)gg_flows}
\end{figure}

\begin{figure}[h!]
  \centering
    \includegraphics[width=0.7\linewidth]{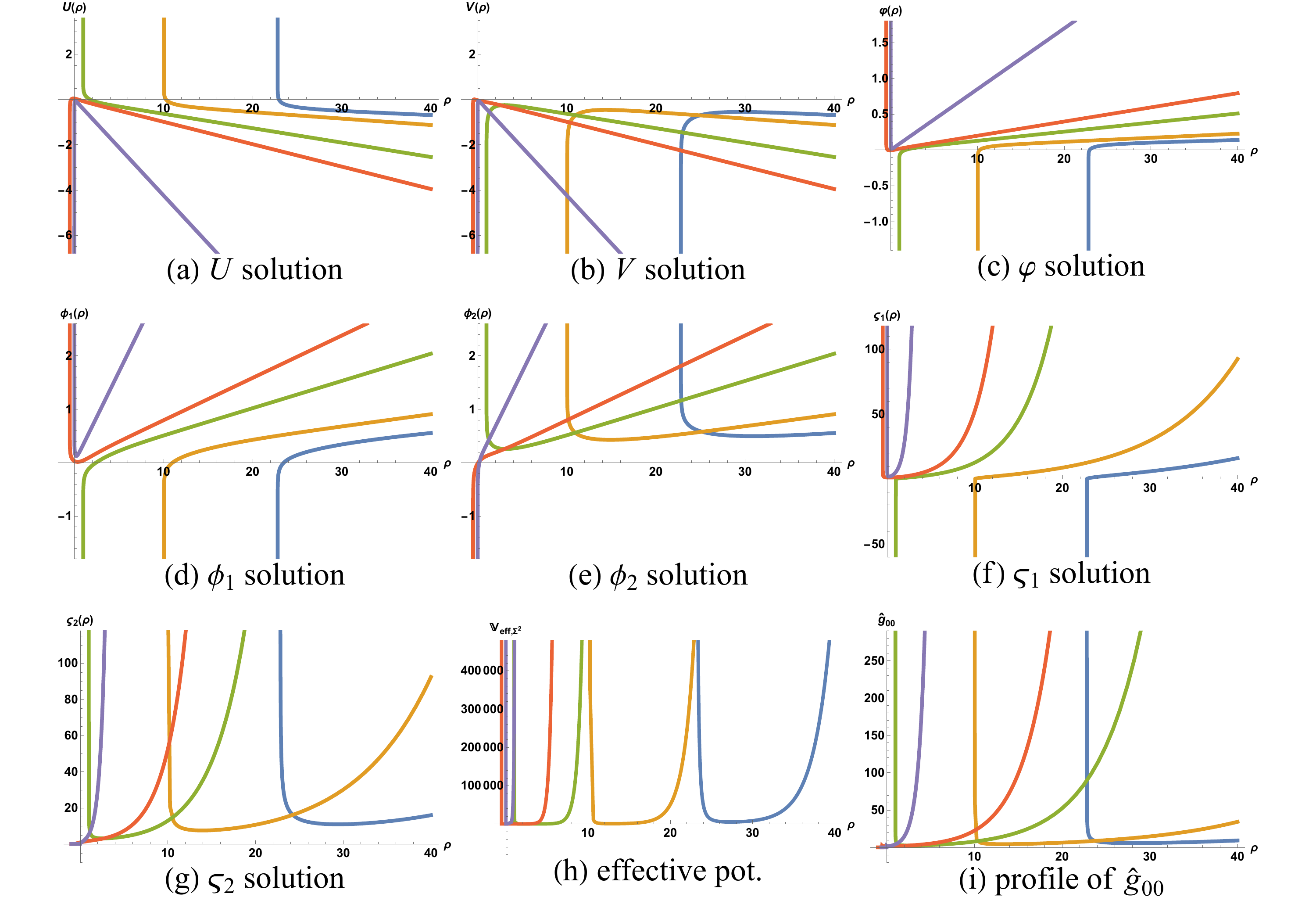}
\caption{Interpolating solutions between the locally $SO(2)\times SO(2)$ flat domain wall as $\rho\rightarrow+\infty$ and $Mkw_3\times S^2$-sliced curved domain walls with shift scalars in $SO(3,2)$ gauge group. The blue, orange, green, red, and purple curves refer to $g=-0.10, -0.16, -0.36, -0.56, -2.40$, respectively.}
\label{15_S2_SO(2)d_SO(32)gg_flows}
\end{figure}

\begin{figure}[h!]
  \centering
    \includegraphics[width=0.7\linewidth]{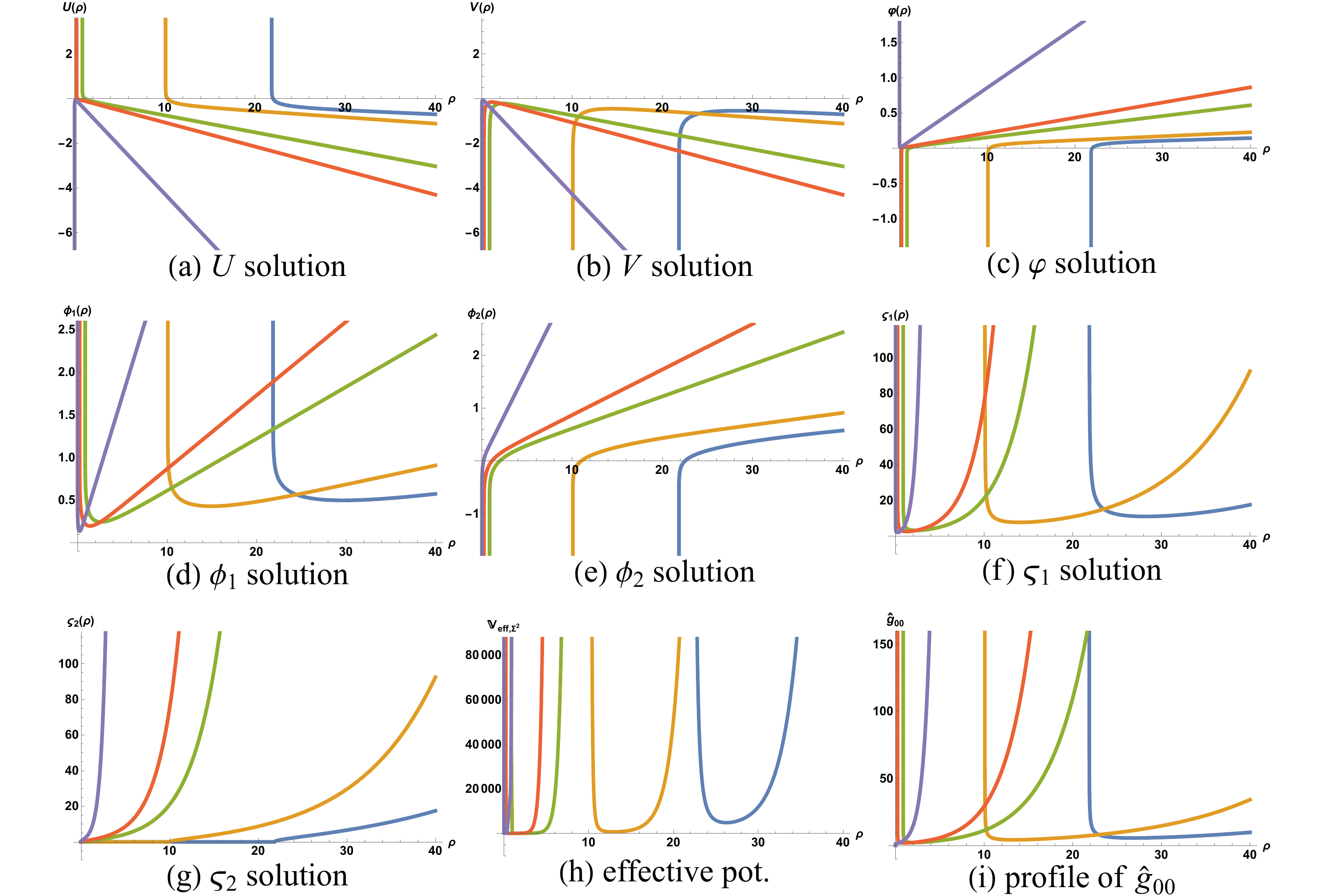}
\caption{Interpolating solutions between the locally $SO(2)\times SO(2)$ flat domain wall as $\rho\rightarrow+\infty$ and $Mkw_3\times H^2$-sliced curved domain walls with shift scalars in $SO(3,2)$ gauge group. The blue, orange, green, red, and purple curves refer to $g=-0.10, -0.16, -0.43, -0.61, -2.42$, respectively.}
\label{15_H2_SO(2)d_SO(32)gg_flows}
\end{figure}

\paragraph{Solutions from $SO(5)$, $SO(4,1)$, and $SO(3,2)$ gauge groups}
Simililar to the previous analysis, we will look for solutions that have asymptotic behaviors of flat domain walls. With the asymptotic behavior given by $SO(5)$ flat domain wall in \eqref{SO(5)_flat_DW_asym}, we find $\varsigma_1=0$ and $\varsigma_2=0$ along the entire flow solutions. Therefore, the solutions are similar to those found in the previous case, and we will not give the corresponding numerical solutions here to avoid repetition. 
\\
\indent We now consider the asymptotic behavior of the form
\begin{equation}\label{SO(2)d_flat_DW_asym}
U\sim V\sim\frac{g \lambda  \rho }{4\sqrt{2}},\qquad \varphi\sim-\frac{g \lambda  \rho }{20 \sqrt{2}},\qquad \phi_1\sim\phi_2\sim-\frac{g \lambda  \rho }{5 \sqrt{2}},\qquad\varsigma_1\sim e^{-\frac{g \lambda  \rho }{\sqrt{2}}}
\end{equation}
which is a locally flat domain wall with $SO(2)\times SO(2)$ symmetry. We will choose $g\lambda<0$ in order to identify the UV five-dimensional field theory with $\rho\rightarrow+\infty$. Examples of numerical solutions interpolating between the flat domain wall \eqref{SO(2)d_flat_DW_asym} and singular geometries in the IR are shown in figures \ref{15_S2_SO(2)d_SO(5)gg_flows} to \ref{15_H2_SO(2)d_SO(32)gg_flows} for different values of $g$. We also recall that the magnetic charge $p_2$ is related to $p_1$ via $p_2=\kappa p_1$. The magnetic charge $p_1$ is in turn fixed in terms of $g$ by the twist condition $2gp_1=k$. We have found the solutions by choosing $g$ to be the independent parameter.  In addition, in the case of $Mkw_3\times \mathbb{R}^2$-sliced solutions with $k=0$, we have $p_2=p_1=0$. Therefore, in this case, the solutions are essentially flat domain walls. Accordingly, we will not further consider these solutions. 
\\
\indent We can see from the figures that in the case of $SO(5)$ gauge group, all $Mkw_3\times S^2$-sliced solutions in figure \ref{15_S2_SO(2)d_SO(5)gg_flows} and the $Mkw_3\times H^2$-sliced solutions represented by red and purple curves in figure \ref{15_H2_SO(2)d_SO(5)gg_flows} admit physical IR singularities by the criterion of \cite{Gubser_Sing}. However, all of these solutions lead to ten-dimensional solutions with unphysical singularities when uplifted to type IIA theory. On the other hand, all $Mkw_3\times H^2$-sliced solutions in $SO(4,1)$ gauge group (figure \ref{15_H2_SO(2)d_SO(41)gg_flows}) contain physically acceptable IR singularities in both six and ten dimensions while for $Mkw_3\times S^2$-sliced solutions (figure \ref{15_S2_SO(2)d_SO(41)gg_flows}), only the solution represented by the purple curve leads to a physical IR singularity. For $SO(3,2)$ gauge group, all of the solutions have unphysical singularities according to the criterion of \cite{Gubser_Sing}, but the solutions represented by the red and purple curves in figure \ref{15_S2_SO(2)d_SO(32)gg_flows} and the purple curve in figure \ref{15_H2_SO(2)d_SO(32)gg_flows} lead to ten-dimensional solutions with physical IR singularities. The nature of all the singularities appearing in the numerical solutions is summarized in table \ref{tab3}. 
\vfil
\vfil\pagebreak

\begin{table}[h!]
\centering
\begin{tabular}{| c | c | c | c |}
\hline
 \multirow{2}{*}{gauge group} & Riemann & criterion \cite{Gubser_Sing}  & criterion \cite{Maldacena_nogo} \\
& surface & ($\mathbf{V}_{\text{eff},\Sigma^2}\nrightarrow+\infty$) & ($\hat{g}_{00}\rightarrow0$) \\\hline
$SO(5)$& $S^2$ & any $g$ & $\times$ \\
& $H^2$ & $g\leq-1$ & $\times$ \\\hline
$SO(4,1)$ & $S^2$& $g\geq2.19$ & $g\geq2.19$ \\
 & $H^2$& any $g$ & any $g$ \\\hline
$SO(3,2)$ & $S^2$ & $\times$ & $g\leq-0.56$ \\
& $H^2$ & $\times$ & $g\leq-2.42$ \\\hline
\end{tabular}
\caption{Summary of satisfaction of criteria \cite{Gubser_Sing} and \cite{Maldacena_nogo} for the IR singularities of the interpolating solutions between the locally $SO(2)\times SO(2)$ symmetric flat domain wall and $Mkw_3\times \Sigma^2$-sliced curved domain walls with shift scalars in $SO(5)$, $SO(4,1)$, and $SO(3,2)$ gauge groups.}\label{tab3}
\end{table}

\begin{figure}[h!]
  \centering
    \includegraphics[width=0.7\linewidth]{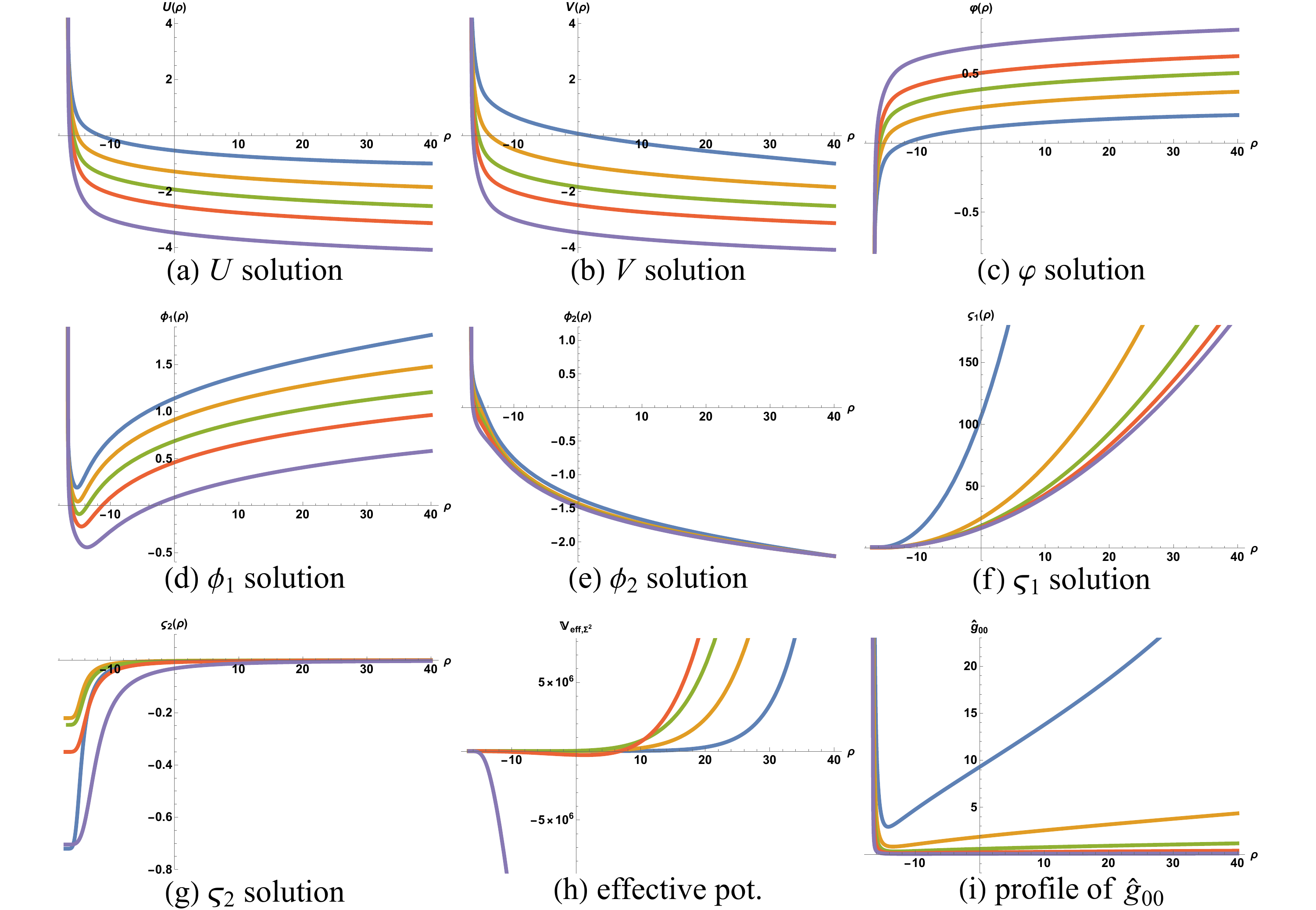}
\caption{Interpolating solutions between the locally $SO(2)\times SO(2)$ flat domain wall as $\rho\rightarrow+\infty$ and $Mkw_3\times S^2$-sliced curved domain walls with shift scalars in $CSO(4,0,1)$ gauge group. The blue, orange, green, red, and purple curves refer to $g=-0.51, -1, -1.72, -2.80, -6$, respectively.}
\label{15_S2_SO(2)d_CSO(401)gg_flows}
\end{figure}

\begin{figure}[h!]
  \centering
    \includegraphics[width=0.7\linewidth]{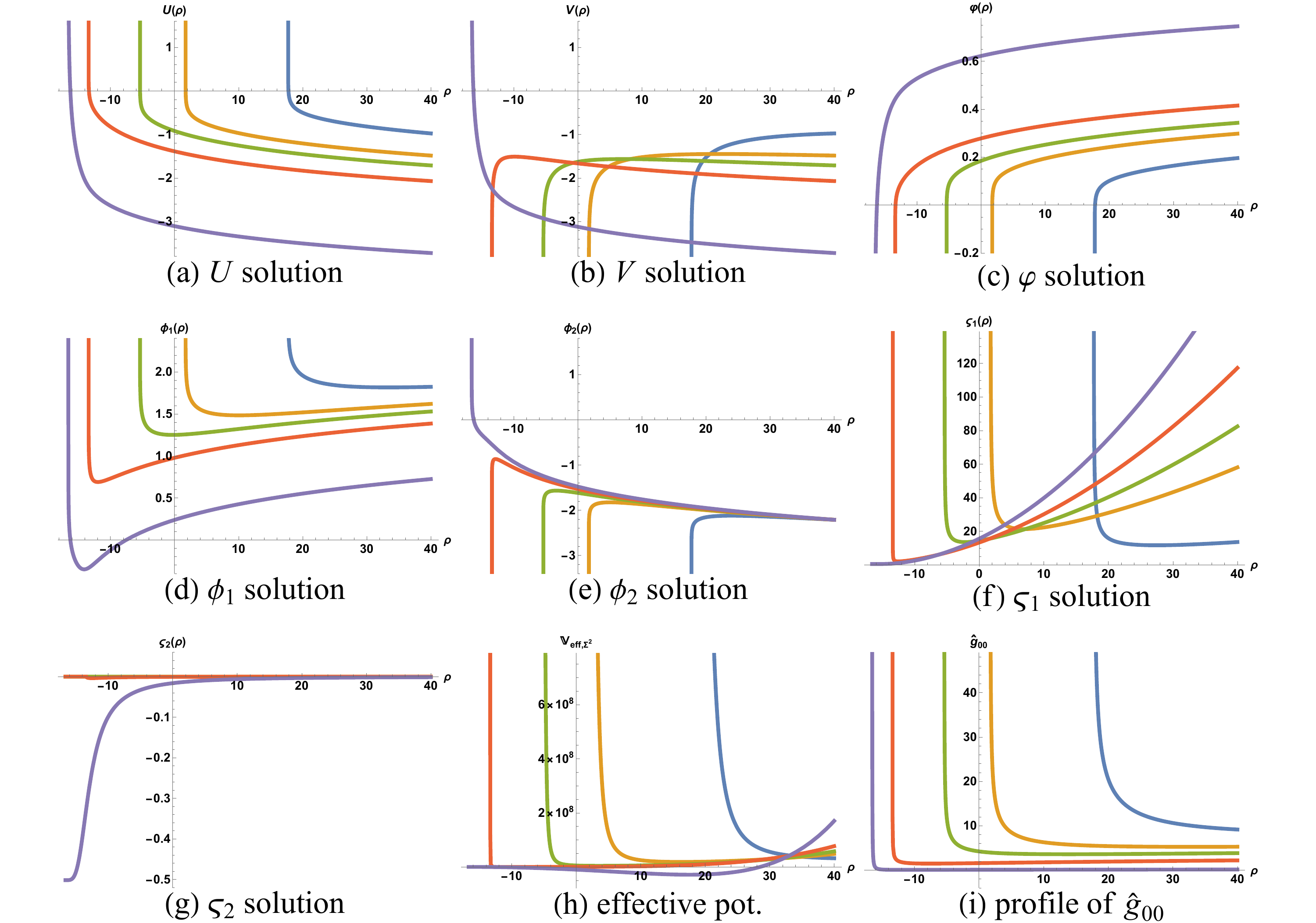}
\caption{Interpolating solutions between the locally $SO(2)\times SO(2)$ flat domain wall as $\rho\rightarrow+\infty$ and $Mkw_3\times H^2$-sliced curved domain walls with shift scalars in $CSO(4,0,1)$ gauge group. The blue, orange, green, red, and purple curves refer to $g=-0.50, -0.75, -0.90, -1.20, -4.50$, respectively.}
\label{15_H2_SO(2)d_CSO(401)gg_flows}
\end{figure}

\begin{figure}[h!]
  \centering
    \includegraphics[width=0.7\linewidth]{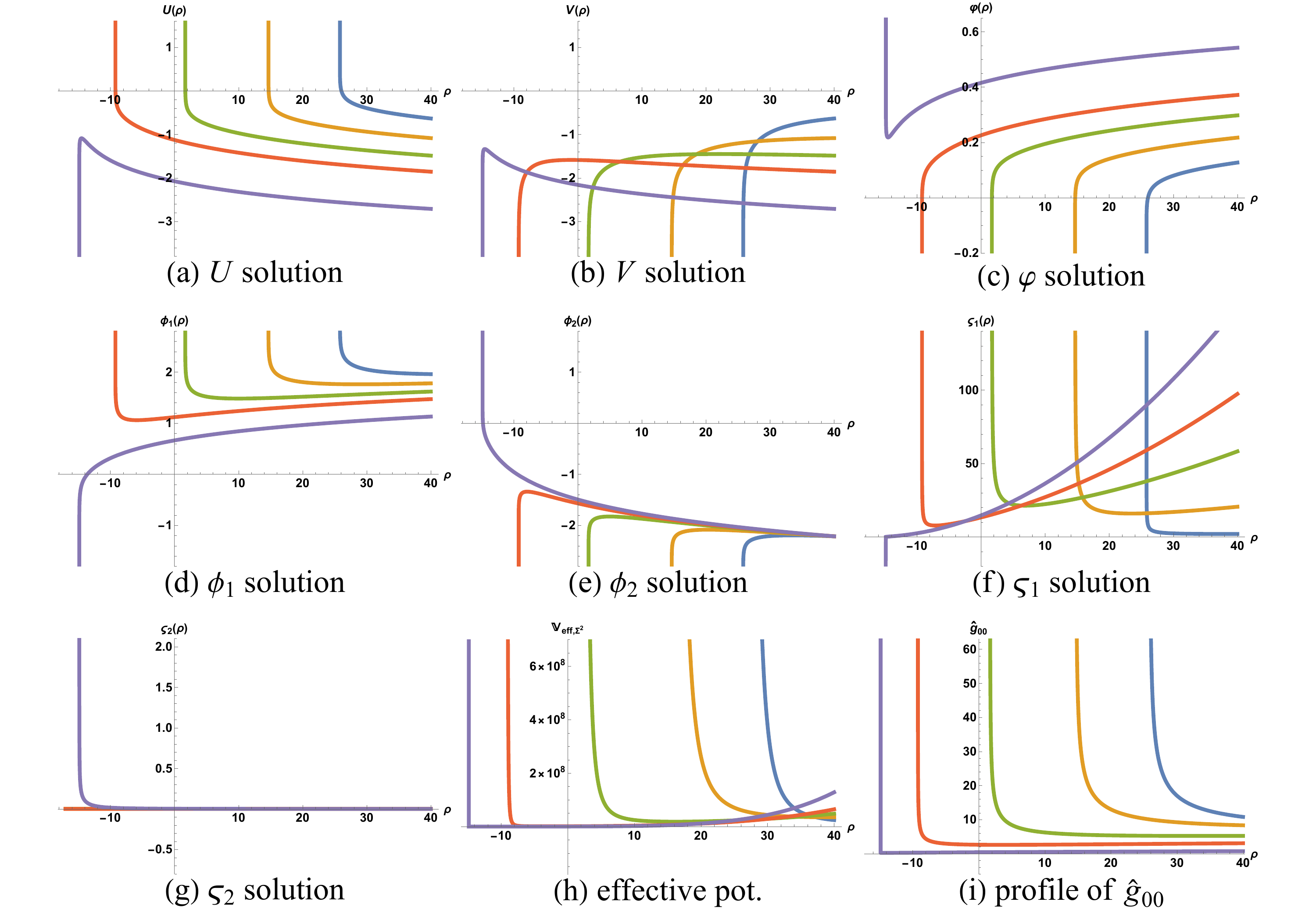}
\caption{Interpolating solutions between the locally $SO(2)\times SO(2)$ flat domain wall as $\rho\rightarrow+\infty$ and $Mkw_3\times S^2$-sliced curved domain walls with shift scalars in $CSO(2,2,1)$ gauge group. The blue, orange, green, red, and purple curves refer to $g=0.38, 0.55, 0.75, 1, 2$, respectively.}
\label{15_S2_SO(2)d_CSO(221)gg_flows}
\end{figure}

\begin{figure}[h!]
  \centering
    \includegraphics[width=0.7\linewidth]{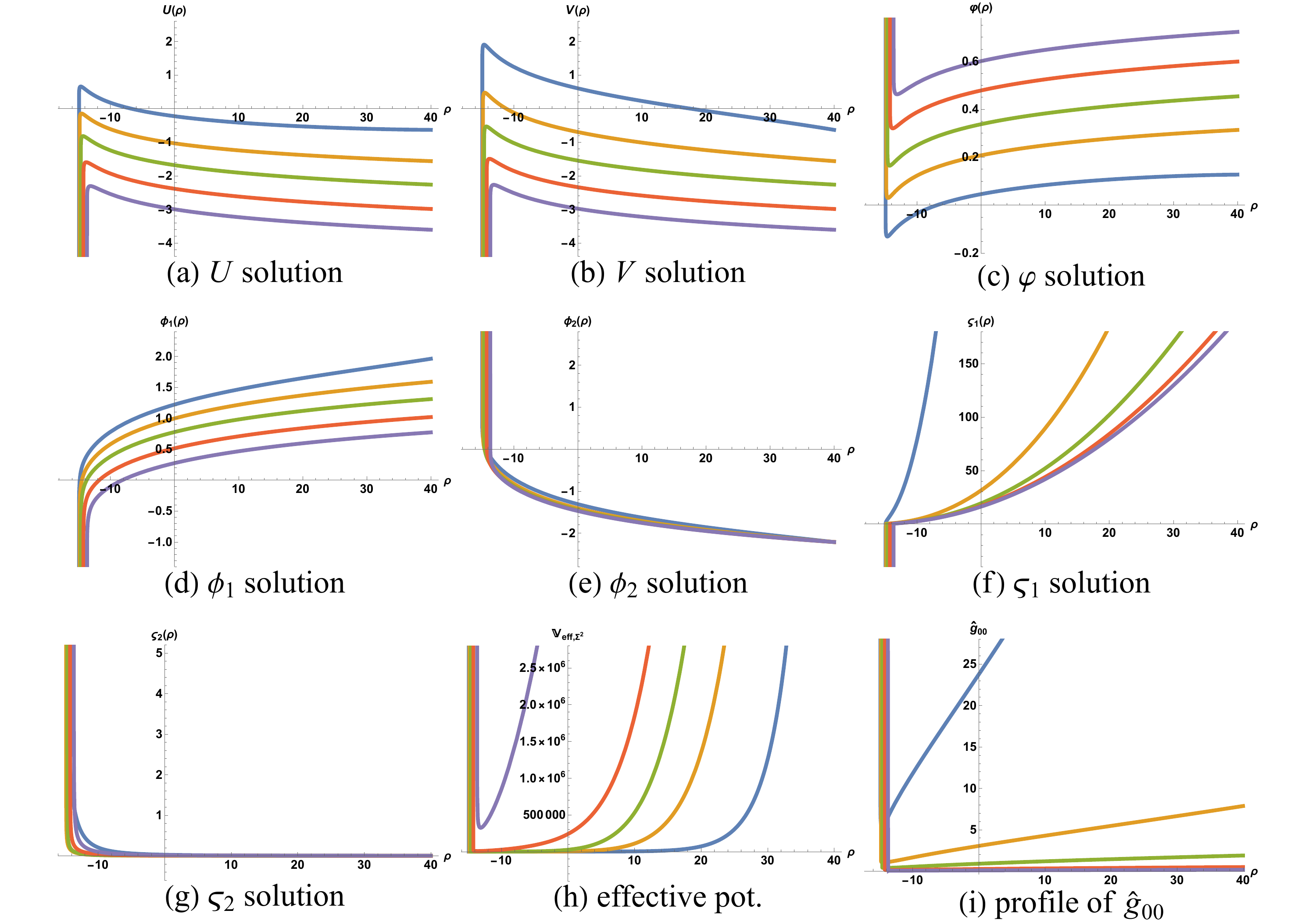}
\caption{Interpolating solutions between the locally $SO(2)\times SO(2)$ flat domain wall as $\rho\rightarrow+\infty$ and $Mkw_3\times H^2$-sliced curved domain walls with shift scalars in $CSO(2,2,1)$ gauge group. The blue, orange, green, red, and purple curves refer to $g=0.38, 0.80, 1.40, 2.50, 4.10$, respectively.}
\label{15_H2_SO(2)d_CSO(221)gg_flows}
\end{figure}

\paragraph{Solutions from $CSO(4,0,1)$ and $CSO(2,2,1)$ gauge groups}
We end this section with solutions from $CSO(4,0,1)$ and $CSO(2,2,1)$ gauge groups with non-vanishing shift scalars. In this case, we change the radial coordinate to $\rho$ defined by $\frac{d\rho}{dr}=e^{-\frac{5}{2}\phi_1-\frac{15}{2}\phi_2}$ and require the asymptotic behavior of the solutions to take the form of a flat domain wall given by
\begin{eqnarray}
U&\sim& V\sim\frac{3}{4}\ln \rho-\frac{5}{4}\ln(-g\kappa\rho) ,\qquad \varphi\sim\frac{1}{4}\ln(-g\kappa\rho)-\frac{3}{20}\ln \rho,\nonumber\\\phi_1&\sim&\frac{9}{10}\ln \rho-\frac{1}{2}\ln(-g\kappa\rho),\quad\qquad\ \phi_2\sim-\frac{3}{5}\ln\rho,\qquad\varsigma_1\sim\rho ^{\frac{\frac{k}{g^2}+8}{4 \sqrt{2}}}\, .\label{15_SO(2)d_fDW_lam0}
\end{eqnarray}
We again choose $g\kappa<0$ to identify the UV theory with $\rho\rightarrow+\infty$. 
\\
\indent As in the previous case, the $Mkw_3\times \mathbb{R}^2$-sliced solutions lead to the standard flat domain walls, so we will not consider these solutions any further. Examples of numerical solutions interpolating between the locally flat domain wall \eqref{15_SO(2)d_fDW_lam0} and singular geometries in the IR are given in figures \ref{15_S2_SO(2)d_CSO(401)gg_flows} to \ref{15_H2_SO(2)d_CSO(221)gg_flows} for different values of $g$. It turns out that all of these solutions are uplifted to type IIA solutions with unphysical IR singularities although all $Mkw_3\times S^2$-sliced solutions in $CSO(4,0,1)$ gauge group (figure \ref{15_S2_SO(2)d_CSO(401)gg_flows}) lead to effective potentials that are bounded from above, see a summary in table \ref{tab4}. 

\clearpage\newpage

\begin{table}[h!]
\centering
\begin{tabular}{| c | c | c | c |}
\hline
 \multirow{2}{*}{gauge group} & Riemann & criterion \cite{Gubser_Sing}  & criterion \cite{Maldacena_nogo} \\
& surface & ($\mathbf{V}_{\text{eff},\Sigma^2}\nrightarrow+\infty$) & ($\hat{g}_{00}\rightarrow0$) \\\hline
$CSO(4,0,1)$& $S^2$ & any $g$ & $\times$ \\
& $H^2$ & $g\leq-4.50$ & $\times$ \\\hline
$CSO(2,2,1)$ & $S^2$, $H^2$ & $\times$ & $\times$ \\\hline
\end{tabular}
\caption{Summary of satisfaction of criteria \cite{Gubser_Sing} and \cite{Maldacena_nogo} for the IR singularities of the interpolating solutions between the locally $SO(2)\times SO(2)$ symmetric flat domain wall and $Mkw_3\times \Sigma^2$-sliced curved domain walls with shift scalars in $CSO(4,0,1)$ and $CSO(2,2,1)$ gauge groups.}\label{tab4}
\end{table}

%%%%%%%%%%%%%%%%%%%%%%%%%%%%%%%%%%%%%%%%%%%%%%%%%%%%%%%%%%%%%%%%%%%%%%%%%%%%%%%%%%%%%%%%%%%%%%%%%%%%%%%%%%%%%%%%%%%%%%%%%%%%%%%%%%%%%%%%%
\subsection{D4-branes wrapped on a Riemannian three-manifold}\label{15_Sig3_Sec}
We now carry out a similar analysis for supersymmetric solutions describing D4-branes wrapped on a Riemannian three-manifold $\Sigma^3_k$. These solutions would lead to two-dimensional field theories arising from twisted compactifications of the maximal SYM theory in five dimensions. As in the previous case, we will consider the case in which $\Sigma^3_k$ has constant curvature characterized by a parameter $k=1,0,-1$ corresponding to a three-dimensional sphere $S^3$, a flat space $\mathbb{R}^3$, or a hyperbolic space $H^3$, respectively. 
\\
\indent The ansatz for the metric takes the form of
\begin{equation}\label{SO(3)6Dmetric}
ds_6^2=e^{2U(r)}\eta_{\bar{\mu}\bar{\nu}}dx^{\bar{\mu}} dx^{\bar{\nu}}+dr^2+e^{2V(r)}ds^2_{\Sigma^{3}_{k}}
\end{equation}
where $\bar{\mu},\bar{\nu}=0,1$ are space-time indices of a two-dimensional Minkowski space $Mkw_2$. The metric on $\Sigma^3_{k}$ is given by
\begin{equation}
ds^2_{\Sigma^3_{k}}=d\psi^2+f_{k}(\psi)^2(d\theta^2+\sin^2{\theta}d\zeta^2)\label{Sigma3metric}
\end{equation}
with the function $f_k(\psi)$ defined as in \eqref{fFn}. Using the vielbein
\begin{eqnarray}
e^{\hat{\bar{\mu}}}&=& e^{U}dx^{\bar{\mu}}, \qquad e^{\hat{r}}=dr,\qquad 
e^{\hat{\psi}}= e^{V}d\psi, \nonumber \\ 
e^{\hat{\theta}}&=& e^{V}f_{k}(\psi)d\theta, \qquad e^{\hat{\zeta}}=e^{V}f_{k}(\psi)\sin{\theta}d\zeta,\label{AdS4xSigma3bein}
\end{eqnarray}
we find non-vanishing components of the spin connection as follows
\begin{eqnarray}
{\omega_{\hat{\bar{\mu}}}}^{\hat{\bar{\nu}}\hat{r}}&=&U'\delta_{\hat{\bar{\mu}}}^{\hat{\bar{\nu}}}, \qquad {\omega_{\hat{i}}}^{\hat{j}\hat{r}}= V'\delta_{\hat{i}}^{\hat{j}},\qquad
{\omega_{\hat{\theta}}}^{\hat{\theta}\hat{\psi}}=e^{-V}\frac{f'_{k}(\psi)}{f_{k}(\psi)},\nonumber \\ {\omega_{\hat{\zeta}}}^{\hat{\zeta}\hat{\psi}}&=& e^{-V}\frac{f'_{k}(\psi)}{f_{k}(\psi)}, \qquad {\omega_{\hat{\zeta}}}^{\hat{\zeta}\hat{\theta}}=e^{-V}\frac{\cot\theta}{f_{k}(\psi)}\label{AdS4xSigma3SpinCon}
\end{eqnarray}
where $\hat{i}=\hat{\psi}, \hat{\theta}, \hat{\zeta}$ is a flat index on $\Sigma^3_{k}$. We will perform two types of topological twists by turning on $SO(3)$ and $SU(2)$ gauge fields in order to eliminate the following components of the spin connection ${\omega_{\hat{\theta}}}^{\hat{\theta}\hat{\psi}}$, ${\omega_{\hat{\zeta}}}^{\hat{\zeta}\hat{\psi}}$, and ${\omega_{\hat{\zeta}}}^{\hat{\zeta}\hat{\theta}}$.

\subsubsection{Solutions with $SO(3)$ twist}\label{15_Sig3_SO(3)_section}
\begin{figure}[h!]
  \centering
    \includegraphics[width=0.8\linewidth]{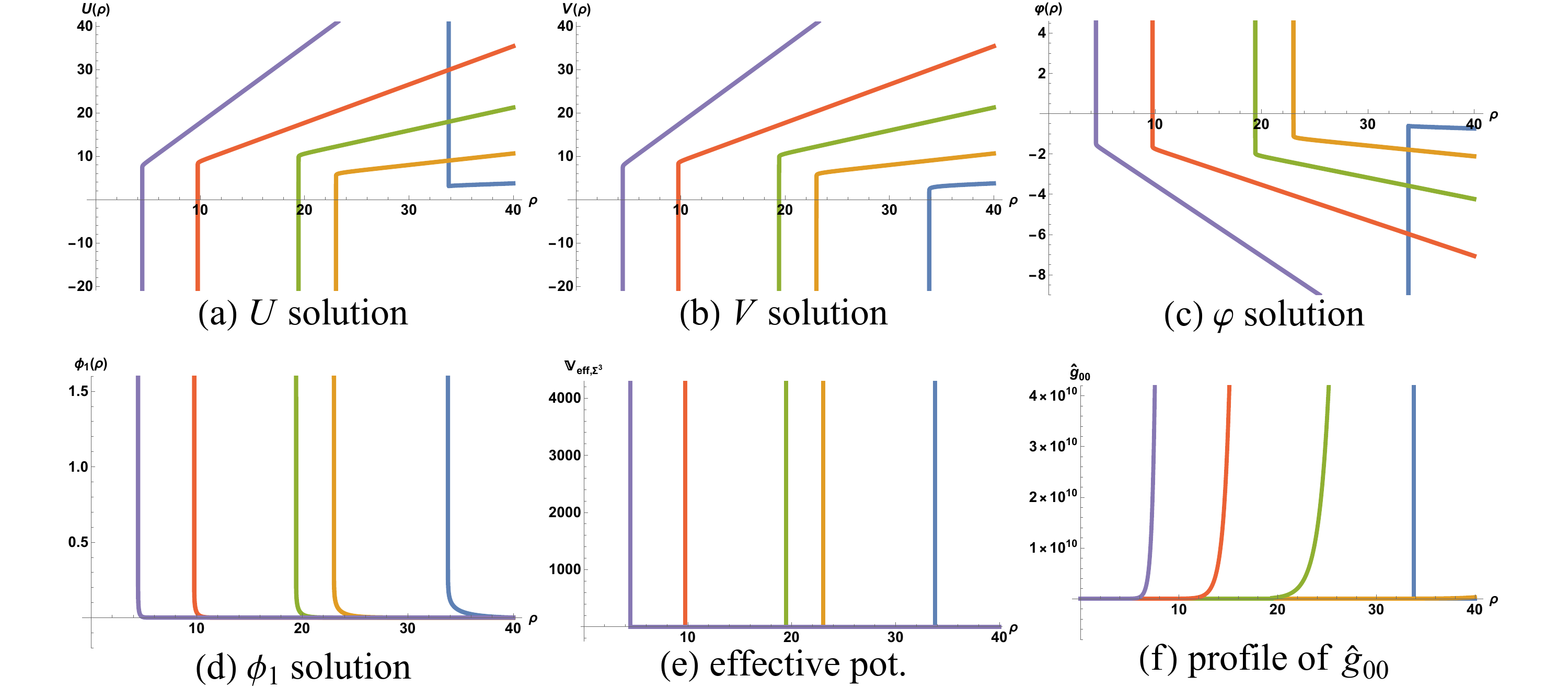}
\caption{Interpolating solutions between the locally $SO(5)$ flat domain wall as $\rho\rightarrow+\infty$ and $Mkw_2\times S^3$-sliced curved domain walls for $SO(3)$ twist in $SO(5)$ gauge group. The blue, orange, green, red, and purple curves refer to $g=0.10, 0.30, 0.63, 1, 2$, respectively.}
\label{15_S3_special_SO(3)_SO(5)gg_flows}
\end{figure}

\begin{figure}[h!]
  \centering
    \includegraphics[width=0.8\linewidth]{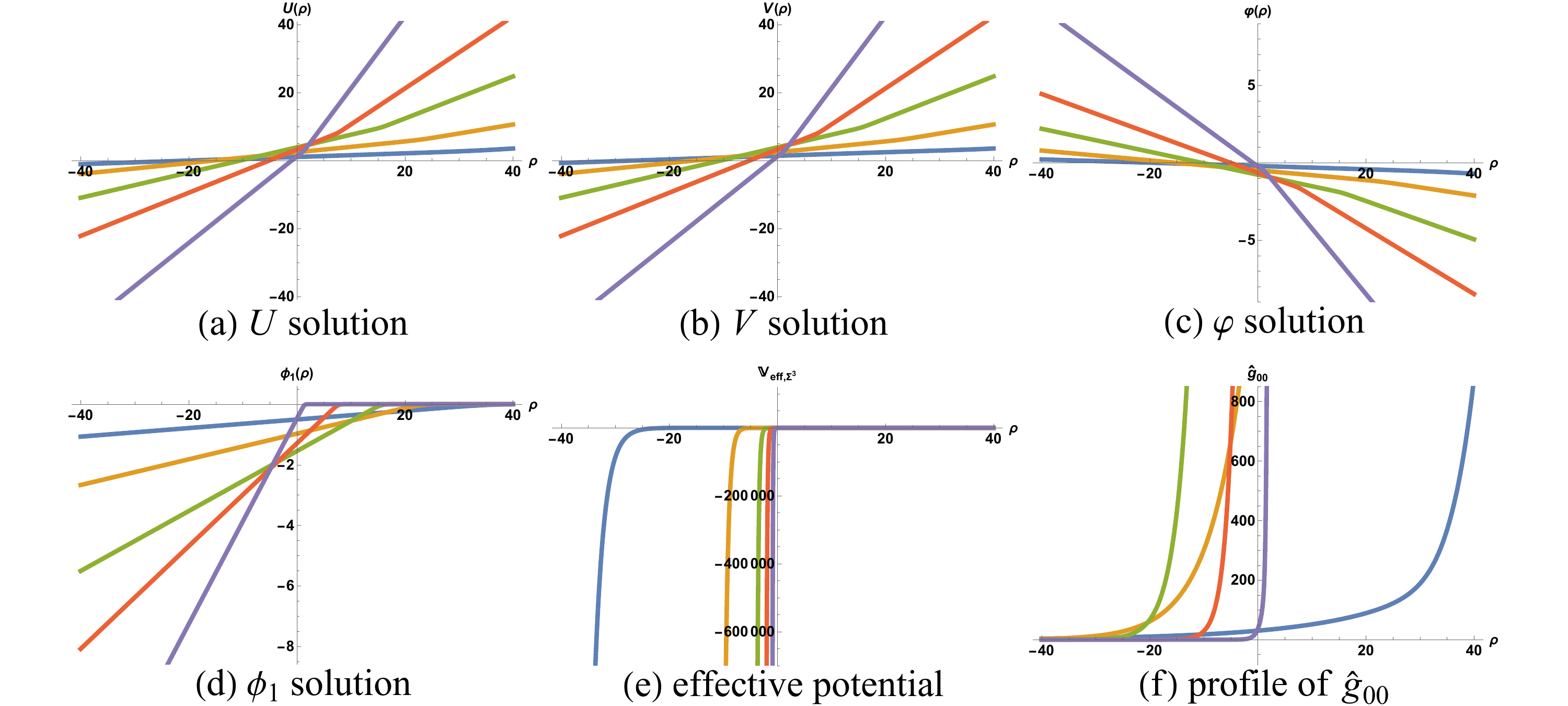}
\caption{Interpolating solutions between the locally $SO(5)$ flat domain wall as $\rho\rightarrow+\infty$ and $Mkw_2\times H^3$-sliced curved domain walls for $SO(3)$ twist in $SO(5)$ gauge group. The blue, orange, green, red, and purple curves refer to $g=0.1, 0.3, 0.7, 1.2, 2.4$, respectively.}
\label{15_H3_special_SO(3)_SO(5)gg_flows}
\end{figure}

\begin{figure}[h!]
  \centering
    \includegraphics[width=0.8\linewidth]{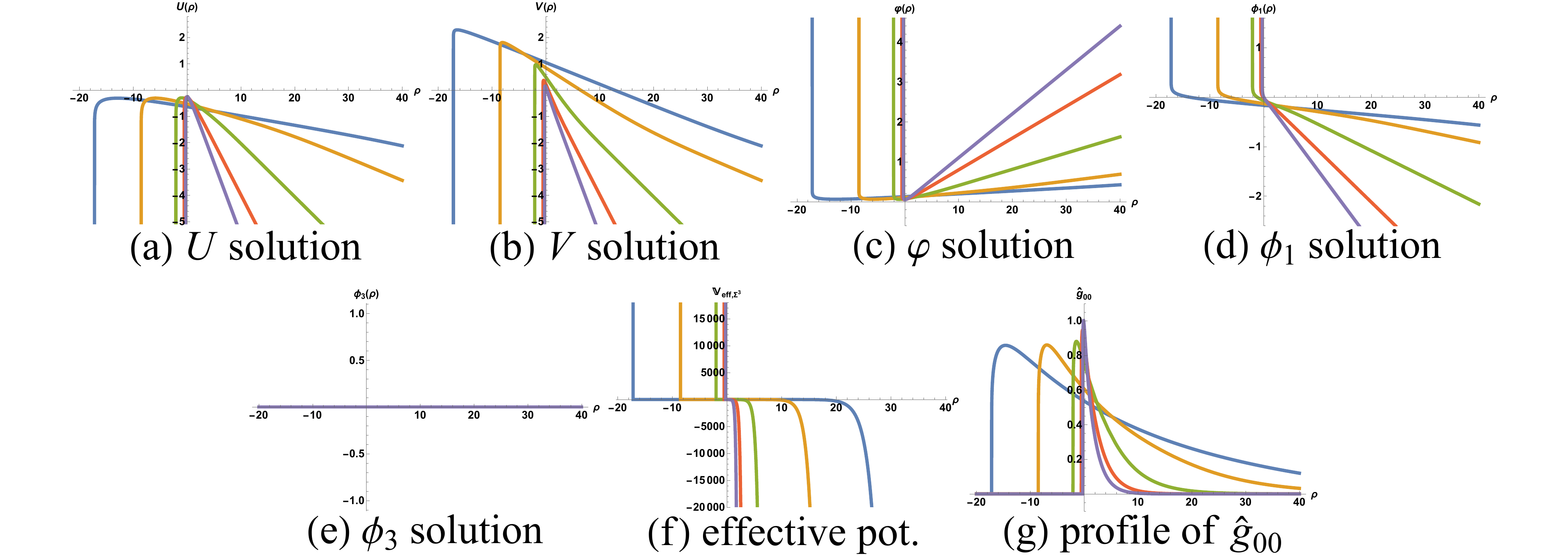}
\caption{Interpolating solutions between the locally $SO(3)$ flat domain wall as $\rho\rightarrow+\infty$ and $Mkw_2\times S^3$-sliced curved domain walls for $SO(3)$ twist in $SO(3,2)$ gauge group. The blue, orange, green, red, and purple curves refer to $g=-0.10, -0.16, -0.38, -1, -3$, respectively.}
\label{15_S3_SO(3)_SO(32)gg_flows}
\end{figure}

\begin{figure}[h!]
  \centering
    \includegraphics[width=0.8\linewidth]{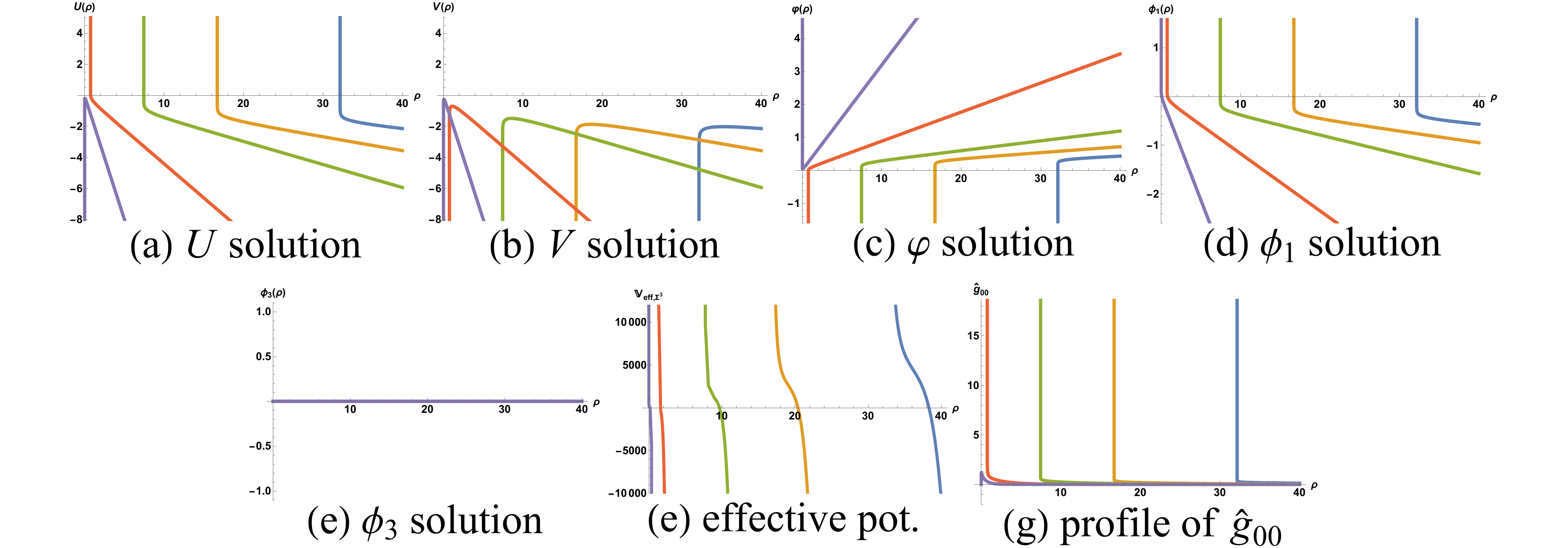}
\caption{Interpolating solutions between the locally $SO(3)$ flat domain wall as $\rho\rightarrow+\infty$ and $Mkw_2\times H^3$-sliced curved domain walls for $SO(3)$ twist in $SO(3,2)$ gauge group. The blue, orange, green, red, and purple curves refer to $g=-0.10, -0.17, -0.28, -0.82, -3$, respectively.}
\label{15_H3_SO(3)_SO(32)gg_flows}
\vfil\pagebreak
\end{figure}

We first consider the case of $SO(3)$ twist with $SO(3)$ embedded in $SO(5)\subset GL(5)$ as $\mathbf{5}\rightarrow \mathbf{3}+\mathbf{1}+\mathbf{1}$. In this case, the gauge groups under consideration must contain this $SO(3)$ subgroup. All of these gauge groups are characterized by the embedding tensor of the form
\begin{equation}\label{SO(3)_Ytensor}
Y_{mn}=\text{diag}(1,1,1,\kappa,\lambda).
\end{equation}
These gauge groups are given by $SO(5)$ ($\kappa=\lambda=1$), $SO(4,1)$ ($\kappa=-\lambda=1$), $SO(3,2)$ ($\kappa=\lambda=-1$), $CSO(4,0,1)$ ($\kappa=1$, $\lambda=0$), $CSO(3,1,1)$ ($\kappa=-1$, $\lambda=0$), and $CSO(3,0,2)$ ($\kappa=\lambda=0$). 
\\
\indent We then turn on the $SO(3)$ gauge fields corresponding to $X_{12}$, $X_{13}$, and $X_{23}$ generators by the following ansatz
\begin{equation}\label{15_Sig3_SO(3)_gaugeAnt}
{A_{\hat{\theta}}}^{12}=e^{-V}\frac{p}{4k}\frac{f'_{k}(\psi)}{f_{k}(\psi)},\ \ \ {A_{\hat{\zeta}}}^{13}=e^{-V}\frac{p}{4k}\frac{f'_{k}(\psi)}{f_{k}(\psi)}, \ \ \ {A_{\hat{\zeta}}}^{23}=e^{-V}\frac{p}{4k}\frac{\cot\theta}{f_{k}(\psi)}\, .\
\end{equation}
In addition to the dilaton, there are four $SO(3)$ singlet scalars corresponding to the following non-compact generators
\begin{eqnarray}
\overline{\mathcal{Y}}_1&=&2\,\hat{\boldsymbol{t}}^+_{1\dot{1}}+2\,\hat{\boldsymbol{t}}^+_{2\dot{2}}+2\,\hat{\boldsymbol{t}}^+_{3\dot{3}}-3\,\hat{\boldsymbol{t}}^+_{4\dot{4}}-3\,\hat{\boldsymbol{t}}^+_{5\dot{5}},\qquad\overline{\mathcal{Y}}_2\,=\,\hat{\boldsymbol{t}}^+_{4\dot{5}},\nonumber\\\overline{\mathcal{Y}}_3&=&\hat{\boldsymbol{t}}^+_{4\dot{4}}-\hat{\boldsymbol{t}}^+_{5\dot{5}},\qquad\quad \overline{\mathcal{Y}}_4\,=\,\boldsymbol{s}_{45}\, .\label{15_SO(3)_non_com}
\end{eqnarray}
With the coset representative
\begin{equation}\label{15_SO(3)_coset}
V=e^{\varphi\boldsymbol{d}+\phi_1\overline{\mathcal{Y}}_1+\phi_2\overline{\mathcal{Y}}_2+\phi_3\overline{\mathcal{Y}}_3+\varsigma\overline{\mathcal{Y}}_4},
\end{equation}
the scalar potential takes the form
\begin{eqnarray}
\mathbf{V}&=&-\frac{g^2e^{2(\varphi-8\phi_1)}}{4}\left[3+6e^{20\phi_1}\left((\kappa+\lambda)\cosh{2\phi_2}\cosh{4\phi_3}-(\kappa-\lambda)\sinh{4\phi_3}\right)\phantom{\frac{1}{2}}\right.\nonumber\\
&&\left.+\frac{e^{40\phi_1}}{4}\left(\kappa^2+10\kappa\lambda+\lambda^2-(3\kappa^2-2\kappa\lambda+3\lambda^2)\cosh{8\phi_3}\phantom{\frac{1}{2}}\right.\right.\nonumber\\
&&\left.\left.\phantom{\frac{1}{2}}-2(\kappa+\lambda)^2\cosh{4\phi_2}\cosh^2{4\phi_3}+4(\kappa^2-\lambda^2)\cosh{2\phi_2}\sinh{8\phi_3}\right)\right].\label{15genSO(3)Pot}
\end{eqnarray}
We also note that the shift scalar $\varsigma$ does not appear in the scalar potential. The $SO(5)\times SO(5)$ composite connections are given as follows
\begin{eqnarray}
{Q_{r}}^{45}&=&\varsigma'-4 \varsigma(\varphi'-3\phi'_1)-\phi'_2\sinh 4\phi_3,\label{Q1_r_Sig3}\\
{Q_{r}}^{\dot{4}\dot{5}}&=&-\varsigma'+4 \varsigma(\varphi'-3\phi'_1)-\phi'_2\sinh 4\phi_3,\label{Q2_r_Sig3}\\
{Q_{\hat{\theta}}}^{12}&=&{Q_{\hat{\theta}}}^{\dot{1}\dot{2}}\ =\ -e^{-V}\frac{gp}{k}\frac{f'_{k}(\psi)}{f_{k}(\psi)},\label{Q1_Sig3}\\
{Q_{\hat{\zeta}}}^{13}&=&{Q_{\hat{\zeta}}}^{\dot{1}\dot{3}}\ =\ -e^{-V}\frac{gp}{k}\frac{f'_{k}(\psi)}{f_{k}(\psi)},\label{Q2_Sig3}\\{Q_{\hat{\zeta}}}^{23}&=&{Q_{\hat{\zeta}}}^{\dot{2}\dot{3}}\ =\ -e^{-V}\frac{gp}{k}\frac{\cot\theta}{f_{k}(\psi)}\, .\label{Q3_Sig3}
\end{eqnarray}
The relevant topological twist can be achieved by imposing the projectors
\begin{equation}\label{SO(3)Projcon+}
\hat{\gamma}_{\hat{\theta}\hat{\psi}}\epsilon_{+\alpha}={(\gamma_{12})_\alpha}^\beta\epsilon_{+\beta} \quad\text{ and }\quad\hat{\gamma}_{\hat{\zeta}\hat{\theta}}\epsilon_{+\alpha}={(\gamma_{23})_\alpha}^\beta\epsilon_{+\beta}
\end{equation}
as well as 
\begin{equation}\label{SO(3)Projcon-}
\hat{\gamma}_{\hat{\theta}\hat{\psi}}\epsilon_{-\dot{\alpha}}={(\gamma_{\dot{1}\dot{2}})_{\dot{\alpha}}}^{\dot{\beta}}\epsilon_{-\dot{\beta}} \quad\text{ and }\quad\hat{\gamma}_{\hat{\zeta}\hat{\theta}}\epsilon_{-\dot{\alpha}}={(\gamma_{\dot{2}\dot{3}})_{\dot{\alpha}}}^{\dot{\beta}}\epsilon_{-\dot{\beta}}
\end{equation}
together with a twist condition 
\begin{equation}\label{simple_twist_con}
gp=k\, .
\end{equation}
In order to consistently truncate out all the two- and three-form fields, we need to set $\varsigma=0$ for $p\neq0$. This condition leads to a consistent set of BPS equations that are compatible with all the second-ordered field equations. 
\\
\indent With $\varsigma=0$ and the projector $\hat{\gamma}_r\epsilon_\pm=\epsilon_\mp$, we find the following set of BPS equations
\begin{eqnarray}
U'&=&\frac{ge^{\varphi-8\phi_1}}{4\sqrt{2}}\left[3+e^{20\phi_1}\left((\kappa+\lambda)\cosh{2\phi_2}\cosh{4\phi_3}-(\kappa-\lambda)\sinh{4\phi_3}\right)\right]\nonumber\\&&-\frac{3p}{8\sqrt{2}}e^{-2V-\varphi+8\phi_1},\label{15_Sig3_SO(3)_BPS_eq1}\\
V'&=&\frac{ge^{\varphi-8\phi_1}}{4\sqrt{2}}\left[3+e^{20\phi_1}\left((\kappa+\lambda)\cosh{2\phi_2}\cosh{4\phi_3}-(\kappa-\lambda)\sinh{4\phi_3}\right)\right]\nonumber\\&&+\frac{5p}{8\sqrt{2}}e^{-2V-\varphi+8\phi_1},
%\\
\end{eqnarray}
\begin{eqnarray}
\varphi'&=&-\frac{ge^{\varphi-8\phi_1}}{20\sqrt{2}}\left[3+e^{20\phi_1}\left((\kappa+\lambda)\cosh{2\phi_2}\cosh{4\phi_3}-(\kappa-\lambda)\sinh{4\phi_3}\right)\right]\nonumber\\&&+\frac{3p}{40\sqrt{2}}e^{-2V-\varphi+8\phi_1},
\\
\phi'_1&=&\frac{ge^{\varphi-8\phi_1}}{10\sqrt{2}}\left[2-e^{20\phi_1}\left((\kappa+\lambda)\cosh{2\phi_2}\cosh{4\phi_3}-(\kappa-\lambda)\sinh{4\phi_3}\right)\right]\nonumber\\&&-\frac{p}{10\sqrt{2}}e^{-2V-\varphi+8\phi_1},\\
\phi'_2&=&-\frac{ge^{\varphi+12\phi_1}}{\sqrt{2}}(\kappa+\lambda)\sinh{2\phi_2}\,\text{sech }{4\phi_3},\label{15_SO(3)_phi2_BPS}\\
\phi'_3&=&\frac{ge^{\varphi+12\phi_1}}{2\sqrt{2}}\left((\kappa-\lambda)\cosh{4\phi_3}-(\kappa+\lambda)\cosh{2\phi_2}\sinh{4\phi_3}\right)\label{15_SO(3)_phi3_BPS}
\end{eqnarray}        
with the Killing spinors given by
\begin{equation}\label{15_sig3_SO(3)_full_Killing}
\epsilon_+(r)=e^{\frac{1}{2}\left[U(r)+ W(r)\gamma_{45}\right]}\epsilon_+^{0}\qquad\text{ and }\qquad\epsilon_-(r)=e^{\frac{1}{2}\left[U(r)- W(r)\gamma_{\dot{4}\dot{5}}\right]}\epsilon_-^{0}\, .
\end{equation}
The function $W(r)$ is determined by the condition
\begin{equation}\label{15_Sig2_W_Im}
W'=-\frac{ge^{\varphi+12\phi_1}}{2\sqrt{2}}(\kappa+\lambda)\sinh{2\phi_2}\tanh{4\phi_3}\, .
\end{equation}
There do not exist any $AdS_3\times \Sigma^3$ fixed points in these equations. Before giving the solutions, we note that for the $k=0$ case, the twist condition implies $p=0$. This leads to standard flat domain walls preserving $16$ supercharges. As in the previous case, we will not further consider this type of solution. For $k=\pm 1$, the solutions preserve only four supercharges. Therefore, the dual field theories in the IR are expected to have $N=(2,2)$ supersymmetry in two dimensions.    
\\
\indent We begin with the simplest case of $CSO(3,0,2)$ gauge group with $\kappa=\lambda=0$. In this case, the BPS equations give $\phi'_2=\phi'_3=W'=0$, so that we can choose $\phi_2=\phi_3=W=0$ without loss of generality. The other BPS equations give $U'=-5\varphi'$ and $\phi'_1=-\frac{4}{3}\varphi'$ giving rise to the solutions 
\begin{equation}
U=-5\varphi \qquad  \textrm{and}\qquad  \phi_1=C_1-\frac{4}{3}\varphi
\end{equation} 
with $C_1$ being an integration constant. We have neglected an additive integration constant for $U$ which can be absorbed by rescaling the coordinates on $Mkw_2$.
\\
\indent Taking a linear combination $V'+5\varphi'$ and changing to a new radial coordinate $\rho$ given by $\frac{d\rho}{dr}=e^{4 \varphi+8\phi_1-V}$, we find
\begin{equation}
V= \ln \left[\frac{k \rho }{\sqrt{2}g}+C\right]-5 \varphi.
\end{equation}
The integration constant $C$ can also be set to zero by shifting the coordinate $\rho$. With all these results and $C=0$, the equation for $\varphi'$ gives
\begin{equation}
\varphi=\frac{6}{5}C_1-\frac{3}{40} \ln \left[\frac{\rho ^3+C_0}{3 \rho }\right]
\end{equation}
in which $C_0$ is another integration constant. 
\\
\indent For $\rho\rightarrow +\infty$, the contribution of the gauge fields to the BPS equations becomes small, and we find
\begin{eqnarray}
U\sim V,\qquad \varphi \sim -\frac{3}{20} \ln \rho,\qquad \phi_1\sim\frac{1}{5}\ln \rho
\end{eqnarray}
which takes the form of a flat domain wall.
\\
\indent There is another singularity as $\rho\rightarrow 0$. For $C_0\neq 0$, we find
\begin{eqnarray}
U&\sim&-\frac{3}{8} \ln\rho,\qquad\qquad V\sim\frac{5}{8}\ln \rho,\nonumber\\ 
\varphi&\sim& \frac{3}{40}\ln \rho,\qquad\phi_1\sim-\frac{1}{10} \ln \rho\, .
\end{eqnarray}
This leads to $\mathbf{V}_{\textrm{eff},\Sigma^3}\rightarrow +\infty$, so the singularity is unphysical by the criterion of \cite{Gubser_Sing}.
\\
\indent For $C_0=0$, we have the following asymptotic behavior 
\begin{eqnarray}
U&\sim&\frac{3}{4} \ln\rho,\qquad\qquad V\sim\frac{7}{4}\ln \rho,\nonumber\\ 
\varphi&\sim& -\frac{3}{20} \ln \rho,\qquad\phi_1\sim\frac{1}{5} \ln \rho\, .
\end{eqnarray}
In this case, we find $\mathbf{V}_{\textrm{eff},\Sigma^3}\rightarrow 0$ which is physically acceptable.
\\
\indent By uplifting the solution to type IIA theory, we can find the behavior of the ten-dimensional metric component $\hat{g}_{00}$ near the singularity. For $C_0\neq 0$ and $\rho\rightarrow 0$, we have
\begin{equation}
\hat{g}_{00}\sim e^{-3\phi_1+\varphi+2U}\sim \rho^{-\frac{3}{8}}\rightarrow +\infty
\end{equation}  
which implies that the singularity is unphysical. On the other hand with $C_0=0$, we find
\begin{equation}
\hat{g}_{00}\sim e^{-3\phi_1+\varphi+2U}\sim \rho^{\frac{3}{5}}\rightarrow 0
\end{equation}
indicating that the singularity is physical. In this case, the solution should describe $N=(2,2)$ two-dimensional field theory arising from D4-branes wrapped on $\Sigma^3$.
\\
\indent In other gauge groups, we can solve the BPS equations numerically. We will consider two types of solutions interpolating between asymptotically flat domain walls with $SO(5)$ and $SO(3)$ symmetries. Examples of numerical solutions with the $SO(5)$ asymptotic behavior given by \eqref{SO(5)_flat_DW_asym} are shown in figures \ref{15_S3_special_SO(3)_SO(5)gg_flows} and \ref{15_H3_special_SO(3)_SO(5)gg_flows}. We also note that all of these solutions are obtained in the case of $SO(5)$ gauge group which admits the $SO(5)$ symmetric flat domain wall solution. From these figures, we see that all $Mkw_2\times H^3$-sliced solutions are uplifted to type IIA solutions with physically acceptable IR singularities. In figure \ref{15_S3_special_SO(3)_SO(5)gg_flows}, although all $Mkw_2\times S^3$-sliced solutions lead to IR singularities with the effective potential unbounded above, these solutions, except for the one represented by the blue curve in figure \ref{15_S3_special_SO(3)_SO(5)gg_flows}, can be uplifted to physical type IIA solutions. 
\vfil
\vfil\pagebreak
\vfil
\vfil
Another class of solutions interpolates between a locally $SO(3)$ symmetric flat domain wall with the asymptotic behavior given by 
\begin{equation}\label{15_SO(3)_flat_DW_asym}
U\sim V\sim\frac{3 g \rho }{4 \sqrt{2}},\qquad \varphi\sim-\frac{3 g \rho }{20 \sqrt{2}},\qquad \phi_1\sim\frac{g \rho }{5 \sqrt{2}},\qquad \phi_2\sim\phi_3\sim0
\end{equation}
and another singular geometry in the IR. The new radial coordinate $\rho$ is defined by $\frac{d\rho}{dr}=e^{\varphi-8\phi_1}$. By choosing $g<0$ in order to identify the UV field theory in five dimensions with $\rho\rightarrow +\infty$, we find a number of numerical solutions starting from this locally $SO(3)$ symmetric flat domain wall. It turns out that all solutions in $SO(5)$, $SO(4,1)$, $CSO(4,0,1)$, and $CSO(3,1,1)$ gauge groups are uplifted to type IIA solutions with unphysical IR singularities. In the case of $SO(3,2)$ gauge group examples of numerical solutions are shown in figures \ref{15_S3_SO(3)_SO(32)gg_flows} and \ref{15_H3_SO(3)_SO(32)gg_flows}. In these figures, all $Mkw_2\times S^3$-sliced solutions lead to ten-dimensional solutions with physical singularities for any values of $g$. On the other hand, the $Mkw_2\times H^3$-sliced solutions admit physical IR singularities only for large values of $g$, see table \ref{tab5}.
\vfil
\vfil

\begin{table}[h!]
\centering
\begin{tabular}{| c | c | c | c | c |}
\hline
UV flat&  \multirow{2}{*}{gauge group} & Riemannian  & criterion \cite{Gubser_Sing}  & criterion \cite{Maldacena_nogo} \\
domain wall& & three-manifold  & ($\mathbf{V}_{\text{eff},\Sigma^3}\nrightarrow+\infty$) & ($\hat{g}_{00}\rightarrow0$) \\\hline
$SO(5)$ symmetric & $SO(5)$& $S^3$ & $\times$ & $g\geq0.2$ \\
&& $H^3$ & any $g$ & any $g$ \\\hline
$SO(3)$ symmetric &$SO(5)$& $S^3$ & any $g$ & $\times$ \\
&& $H^3$ & $g\leq-1.8$ & $\times$ \\\cline{2-5}
&$SO(4,1)$ & $S^3$, $H^3$& $\times$ & $\times$ \\\cline{2-5}
&$SO(3,2)$ & $S^3$ & $\times$ & any $g$ \\
&& $H^3$ & $\times$ & $g\leq-3$ \\\cline{2-5}
&$CSO(4,0,1)$& $S^3$ & any $g$ & $\times$ \\
&& $H^3$ & $\times$ & $\times$ \\\cline{2-5}
&$CSO(3,1,1)$ & $S^3$, $H^3$& $\times$ & $\times$ \\\cline{2-5}
&$CSO(3,0,2)$ & \multirow{2}{*}{$S^3$, $H^3$}& \multirow{2}{*}{$C_0=0$} & \multirow{2}{*}{$C_0=0$} \\
&(analytic) & &  &  \\\hline
\end{tabular}
\caption{Summary of satisfaction of criteria \cite{Gubser_Sing} and \cite{Maldacena_nogo} for the IR singularities of the $Mkw_2\times \Sigma^3$-sliced curved domain walls obtained from $SO(3)$ twist in $SO(5)$, $SO(4,1)$, $SO(3,2)$, $CSO(4,0,1)$, $CSO(3,1,1)$, and $CSO(3,0,2)$ gauge groups.}\label{tab5}
\end{table}
%%%%%%%%%%%%%%%%%%%%%%%%%%%%%%%%%%%%%%%%%%%%%%%%%%%%%%%%%%%%%%%%%%%%%%%%%%%%%%%%%%%%%%%%%%%%%%%%%%%%%%%%%%%%%%%%%%%%%%%%%%%%%%%%%%%%%%%%%
\subsubsection{Solutions with $SU(2)$ twist}\label{15_Sig3_SO(3)sd_section}
\begin{figure}[h!]
  \centering
    \includegraphics[width=0.8\linewidth]{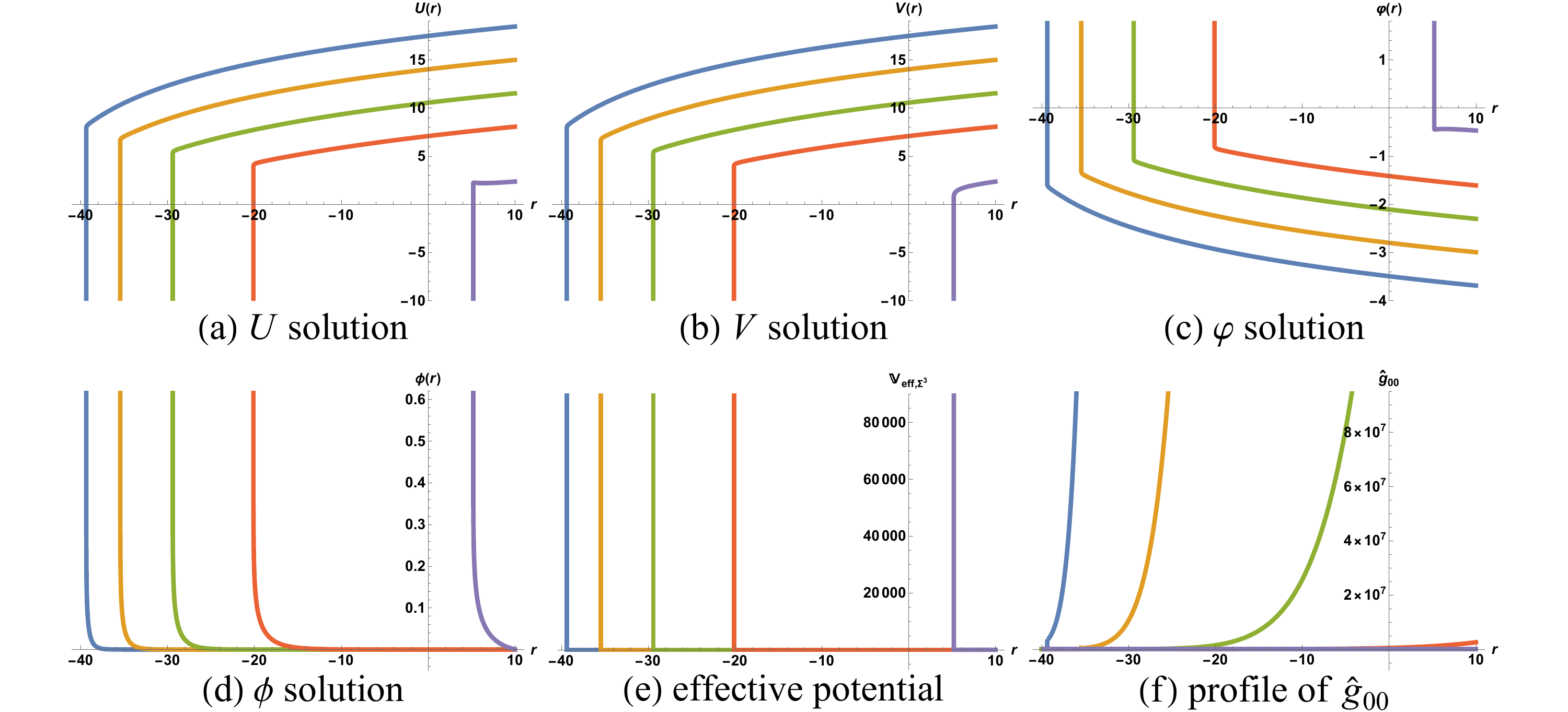}
\caption{Interpolating solutions between the locally $SO(5)$ flat domain wall as $r\rightarrow+\infty$ and $Mkw_2\times S^3$-sliced curved domain walls for $SU(2)$ twist in $SO(5)$ gauge group. The blue, orange, green, red, and purple curves refer to $g=4, 2, 1, 0.50, 0.16$, respectively.}
\label{15_S3_special_SO(3)d_SO(5)gg_flows}
\end{figure}

\begin{figure}[h!]
  \centering
    \includegraphics[width=0.8\linewidth]{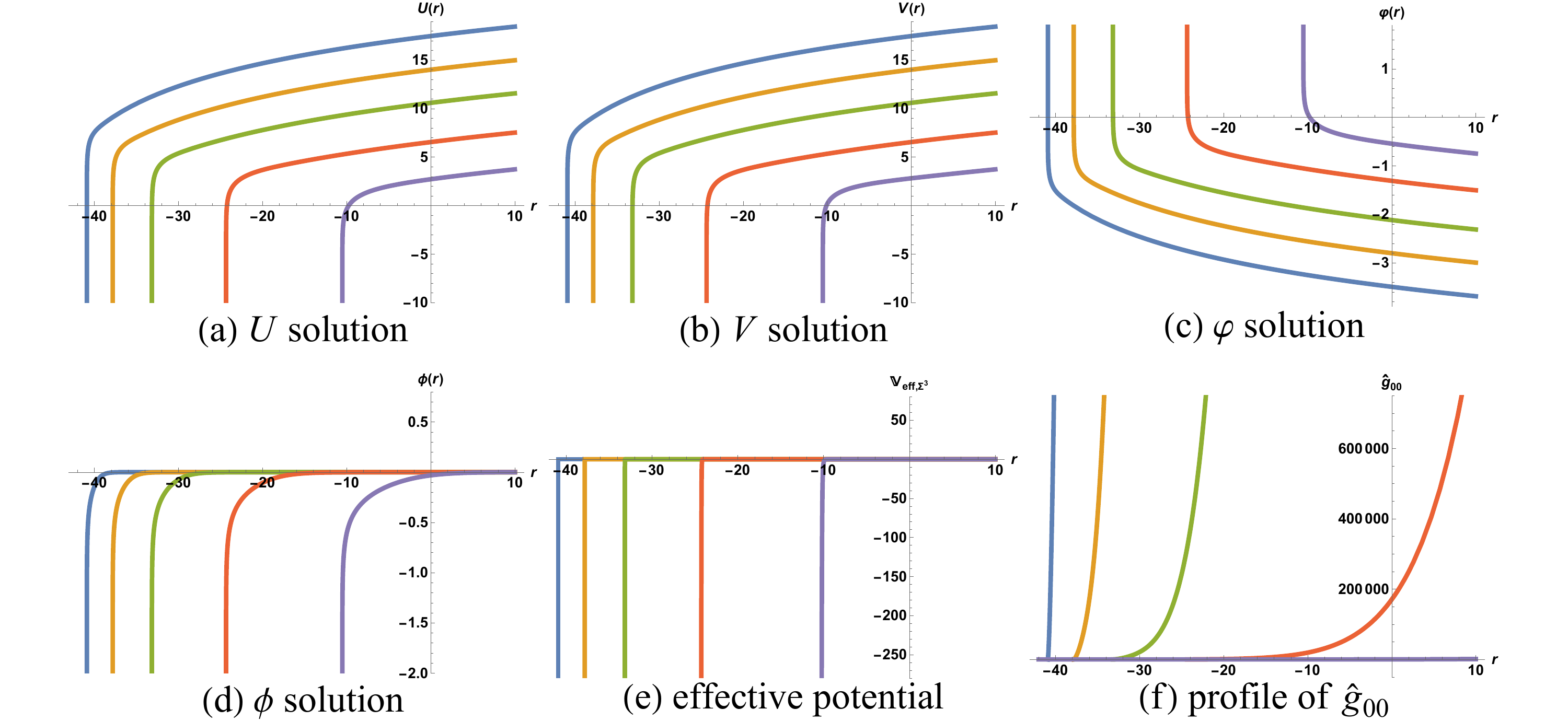}
\caption{Interpolating solutions between the locally $SO(5)$ flat domain wall as $r\rightarrow+\infty$ and $Mkw_2\times H^3$-sliced curved domain walls for $SU(2)$ twist in $SO(5)$ gauge group. The blue, orange, green, red, and purple curves refer to $g=4, 2, 1, 0.45, 0.21$, respectively.}
\label{15_H3_special_SO(3)d_SO(5)gg_flows}
\end{figure}

We end this section by considering a topological twist given by turning on $SU(2)$ gauge fields. We identify this $SU(2)$ with the self-dual $SU(2)_+$ subgroup of $SU(2)_+\times SU(2)_-\sim SO(4)\subset SO(5)$ generated by $X_{12}+X_{34}$, $X_{13}+X_{24}$, and $X_{23}+X_{14}$. We will accordingly turn on the following gauge fields
\begin{eqnarray}
{A_{\hat{\theta}}}^{12}={A_{\hat{\theta}}}^{34}={A_{\hat{\zeta}}}^{13}={A_{\hat{\zeta}}}^{24}&=& e^{-V}\frac{p}{8k}\frac{f'_{k}(\psi)}{f_{k}(\psi)},\nonumber \\ {A_{\hat{\zeta}}}^{23}={A_{\hat{\zeta}}}^{14}&=& e^{-V}\frac{p}{8k}\frac{\cot\theta}{f_{k}(\psi)}\, . \label{15_SO(3)diag_gauge_fields}
\end{eqnarray}
There are three relevant gauge groups to consider; $SO(5)$, $SO(4,1)$, and $CSO(4,0,1)$. These groups can be characterized by the embedding tensor of the form
\begin{equation}\label{SO(4)Ytensor}
Y_{mn}=\text{diag}(1,1,1,1,\lambda)
\end{equation}
with $\lambda=1,-1,0$, respectively. 
\\
\indent There are two singlet scalars parametrized by the coset representative
\begin{equation}\label{15_SO(4)_coset}
V=e^{\varphi\boldsymbol{d}+\phi\mathcal{Y}}
\end{equation}
with
\begin{equation}\label{15_SO(4)_non_com}
\mathcal{Y}=\hat{\boldsymbol{t}}^+_{1\dot{1}}+\hat{\boldsymbol{t}}^+_{2\dot{2}}+\hat{\boldsymbol{t}}^+_{3\dot{3}}+\hat{\boldsymbol{t}}^+_{4\dot{4}}-4\,\hat{\boldsymbol{t}}^+_{5\dot{5}}\, .
\end{equation}
The scalar potential reads
\begin{eqnarray}
\mathbf{V}=-\frac{g^2}{4}e^{2\varphi-8\phi}\left(8+8\lambda e^{20\phi}-\lambda^2e^{40\phi}\right).\label{YSO(4)Pot}
\end{eqnarray}
To implement the twist, we impose the projection conditions given in \eqref{SO(3)Projcon+} and \eqref{SO(3)Projcon-} together with
\begin{equation}\label{GammaSDProj}
{(\gamma_{12})_\alpha}^\beta\epsilon_{+\beta}={(\gamma_{34})_\alpha}^\beta\epsilon_{+\beta}\qquad \textrm{and}\qquad 
{(\gamma_{\dot{1}\dot{2}})_{\dot{\alpha}}}^{\dot{\beta}}\epsilon_{-\dot{\beta}}=
{(\gamma_{\dot{3}\dot{4}})_{\dot{\alpha}}}^{\dot{\beta}}\epsilon_{-\dot{\beta}}
\end{equation}
on the Killing spinor \eqref{DW_Killing_spinor} together with the twist condition \eqref{simple_twist_con}. With the $\hat{\gamma}_r$ projection given in \eqref{15_DW_Proj}, we find the following BPS equations 
\begin{eqnarray}
U'&=&\frac{g}{4\sqrt{2}}e^{\varphi-4\phi}(4+\lambda e^{20\phi})-\frac{3p}{8\sqrt{2}}e^{-2V-\varphi+4\phi},\label{15_M3_SO3d_BPSeq1}\\
V'&=&\frac{g}{4\sqrt{2}}e^{\varphi-4\phi}(4+\lambda e^{20\phi})+\frac{5p}{8\sqrt{2}}e^{-2V-\varphi+4\phi},\label{15_M3_SO3d_BPSeq2}\\
\varphi'&=&-\frac{g}{20\sqrt{2}}e^{\varphi-4\phi}(4+\lambda e^{20\phi})+\frac{3p}{40\sqrt{2}}e^{-2V-\varphi+4\phi},\label{15_M3_SO3d_BPSeq3}\\
\phi'&=&\frac{g}{5\sqrt{2}}e^{\varphi-4\phi}(1-\lambda e^{20\phi})-\frac{3p}{40\sqrt{2}}e^{-2V-\varphi+4\phi}\, .\label{15_M3_SO3d_BPSeq4}
\end{eqnarray}
With the additional projectors \eqref{GammaSDProj} imposed on both $\epsilon_{+\alpha}$ and $\epsilon_{-\dot{\alpha}}$, in this case, the solutions preserve only two supercharges corresponding to $N=(1,1)$ supersymmetry in two dimensions.
\\
\indent For $CSO(4,0,1)$ gauge group with $\lambda=0$, we can analytically solve these equations. The resulting solution is given by
\begin{eqnarray}\label{15_M3_SO3d_CSO401_BPS_soln}
& &U=-5\varphi,\qquad V= \ln \left[\frac{k \rho }{\sqrt{2}g}\right]-5 \varphi,\nonumber \\
& &\phi=-\varphi,\qquad \varphi=C-\frac{k \rho ^2}{20}+\frac{3}{40}\ln \rho
\end{eqnarray}
with $C$ being an integration constant. The new radial coordinate $\rho$ is defined by $\frac{d\rho}{dr}=e^{-V}$. At large $\rho$, we find that the solution becomes a locally flat domain wall with the following asymptotic behavior
\begin{equation}
U\sim V\sim\frac{k \rho ^2}{4},\qquad \varphi\sim -\frac{k \rho ^2}{20},\qquad \phi\sim\frac{k \rho ^2}{20}\, .
\end{equation} 
On the other hand, as $\rho\rightarrow 0$, the above solution reduces to
\begin{equation}
U\sim-\frac{3}{8}\ln\rho,\qquad V\sim\ln \left[\frac{k \rho }{\sqrt{2}g}\right]-\frac{3}{8}\ln\rho,\qquad\varphi\sim-\phi\sim \frac{3}{40}\ln\rho.
\end{equation} 
However, near this singularity, both the effective potential and the type IIA metric component $\hat{g}_{00}$ diverge to $+\infty$ leading to solutions with unphysical IR singularities. 
\\
\indent In $SO(5)$ and $SO(4,1)$ gauge groups with $\lambda\neq0$, the BPS equations can only be solved numerically. We begin with solutions that, as $r\rightarrow+\infty$, are asymptotic to the locally flat domain wall with $SO(5)$ symmetry of the form
\begin{equation}\label{SO(4)_SO(5)_flat_DW_asym}
U\sim V\sim 5\ln gr,\qquad\varphi\sim-\ln gr,\qquad\phi\sim \frac{1}{g^{16} r^{16}}\, .
\end{equation}
Examples of these solutions are given in figures \ref{15_S3_special_SO(3)d_SO(5)gg_flows} and \ref{15_H3_special_SO(3)d_SO(5)gg_flows} for different values of $g$. From these solutions, we find that all solutions can be uplifted to solutions of type IIA theory with physical IR singularities for any values of $g$. 
\\
\indent Another class of solutions has an asymptotic geometry in the form of a locally flat domain wall with $SO(4)$ symmetry. The corresponding asymptotic behavior is given by
\begin{equation}\label{15_SO(4)_flat_DW_asym}
U\sim V\sim -\frac{5}{4} \ln \left[-\frac{2\sqrt{2} g \rho}{5} \right],\qquad\varphi\sim-\phi\sim\frac{1}{4} \ln \left[-\frac{2\sqrt{2} g \rho}{5} \right].
\end{equation}
We have found a number of numerical solutions in both $SO(5)$ and $SO(4,1)$ gauge groups. However, from the behaviors of $\hat{g}_{00}$, it turns out that all solutions lead to unphysical ten-dimensional solutions although all of the $Mkw_2\times S^3$-sliced solutions in $SO(5)$ gauge group admit the IR singularities with the effective potential bounded from above. We will not present these solutions here but simply include them in table \ref{tab6}.
\vfil
\vfil

\begin{table}[h!]
\centering
\begin{tabular}{| c | c | c | c | c |}
\hline
UV flat&  \multirow{2}{*}{gauge group} & Riemannian  & criterion \cite{Gubser_Sing}  & criterion \cite{Maldacena_nogo} \\
domain wall& & three-manifold  & ($\mathbf{V}_{\text{eff},\Sigma^3}\nrightarrow+\infty$) & ($\hat{g}_{00}\rightarrow0$) \\\hline
$SO(5)$ symmetric & $SO(5)$& $S^3$ & $\times$ & any $g$ \\
&& $H^3$ & any $g$ & any $g$ \\\hline
$SO(4)$ symmetric &$SO(5)$& $S^3$ & any $g$ & $\times$ \\
&& $H^3$ & $\times$ & $\times$ \\\cline{2-5}
&$SO(4,1)$ & $S^3$, $H^3$& $\times$ & $\times$ \\\cline{2-5}
&$CSO(4,0,1)$ &  \multirow{2}{*}{$S^3$, $H^3$} &  \multirow{2}{*}{$\times$} &  \multirow{2}{*}{$\times$} \\
&(analytic) & &  &  \\\hline
\end{tabular}
\caption{Summary of satisfaction of criteria \cite{Gubser_Sing} and \cite{Maldacena_nogo} for the IR singularities of the $Mkw_2\times \Sigma^3$-sliced curved domain walls obtained from $SU(2)$ twist in $SO(5)$, $SO(4,1)$, and $CSO(4,0,1)$ gauge groups.}\label{tab6}
\end{table}

%%%%%%%%%%%%%%%%%%%%%%%%%%%%%%%%%%%%%%%%%%%%%%%%%%%%%%%%%%%%%%%%%%%%%%%%%%%%%%%%%%%%%%%%%%%%%%%%%%%%%%%%%%%%%%%%%%%%%%%%%%%%%%%%%%%%%%%%%
\section{Wrapped D4-branes from $CSO(p,q,4-p-q)\ltimes \mathbb{R}^4$ gauged supergravity}\label{40_Sec}
We repeat the same analysis for gaugings from $\overline{\mathbf{40}}^{-1}$ representation associated with the embedding tensor
\begin{equation}\label{40_rep_theta}
\theta^{Am}=\mathbb{T}^A_{np}U^{np,m}\quad\text{and}\quad{\theta^{A}}_m\ = \ 0\, .
\end{equation}
The tensor $U^{mn,p}$ satisfies the conditions $U^{mn,p}=U^{[mn],p}$ with $U^{[mn,p]}=0$, and the quadratic constraint determines $U^{mn,p}$ to take the form 
\begin{equation}
U^{mn,p}\ =\ v^{[m}w^{n]p}
\end{equation}
in which $v^m$ is a $GL(5)$ vector, and $w^{mn}$ is a symmetric tensor, $w^{mn}=w^{(mn)}$. 
\\
\indent Following \cite{7D_Max_Gauging}, we can use the $SL(5)$ symmetry to fix the vector $v^m=\delta^m_5$ and split the $SL(5)$ index $m=(i,5)$, $i=1,..,4$. For simplicity, we will also set $w^{i5}=w^{55}=0$. The remaining $SL(4)\subset SL(5)$ residual symmetry can be used to diagonalize the $4\times4$ block corresponding to $w^{ij}$ as
\begin{equation}
w^{ij}\ =\ \text{diag}(\underbrace{1,..,1}_p,\underbrace{-1,..,-1}_q,\underbrace{0,..,0}_r)
\end{equation}
with $p+q+r=4$. From the decomposition in \eqref{GaugeGenSplit}, the only non-vanishing gauge generators are given by  
\begin{equation}\label{GenCSO(p,q,4-p-q)X}
X_{ij}=\frac{1}{\sqrt{2}}\varepsilon_{ijkk'}w^{kl}{\boldsymbol{t}^{k'}}_l\qquad\textrm{and}\qquad X^i=w^{ij}\boldsymbol{s}_{5j}\, .
\end{equation}
These gauge generators satisfy the following commutation relations
\begin{equation}
[X^i,X^j]=0,\qquad [X_{ij},X^k]={(X_{ij})_l}^kX^l,\qquad [X_{ij},X_{kl}]={(X_{ij})_{kl}}^{k'l'}X_{k'l'}
\end{equation}
where ${(X_{ij})_{kl}}^{k'l'}=2{(X_{ij})_{[k}}^{[k'}\delta_{l]}^{l']}$. The corresponding gauge group takes the form
\begin{equation}\label{40_gauge_group}
G_0=CSO(p,q,4-p-q)\ltimes \mathbb{R}^{4}=SO(p,q) \ltimes \left(\mathbb{R}^{(p+q)(4-p-q)}\times\mathbb{R}^{4}\right).
\end{equation}
The $CSO(p,q,4-p-q)\subset SL(4)$ factor and the four-dimensional translation group from the shift symmetries $\mathbb{R}^{4}$ are respectively generated by $X_{ij}$ and $X^i$. We recall that gaugings in $\overline{\mathbf{40}}^{-1}$ representation are purely electric, so the three-form $C_{\mu\nu\rho, A}$ and magnetic two-form ${B_{\mu\nu}}^m$ fields do not appear. Furthermore, we will set the four gauge fields $A_{\mu,i}$ corresponding to the shift symmetry generators $X^i$ to zero. It should also be pointed out that this six-dimensional gauged supergravity can be obtained from a dimensional reduction of the maximal gauged supergravity in seven dimensions with $CSO(p,q,4-p-q)$ gauge group on a circle $S^1$.   
\\
\indent For scalar fields, we split the $SO(5)\times SO(5)$ vector indices as $a=(i,5)$ and $\dot{a}=(\dot{i},\dot{5})$ resulting in the decomposition of $SL(5)$ non-compact generators as
\begin{equation}
\tilde{\boldsymbol{t}}_{a\dot{b}}\rightarrow\left(\tilde{\boldsymbol{t}}_{i\dot{j}},\,\tilde{\boldsymbol{t}}_{i\dot{5}},\,\tilde{\boldsymbol{t}}_{5\dot{5}}\right).
\end{equation}
Since the $SL(5)$ generators $\tilde{\boldsymbol{t}}_{a\dot{b}}$ are traceless, the generator $\tilde{\boldsymbol{t}}_{5\dot{5}}$ is related to the trace part of $\tilde{\boldsymbol{t}}_{i\dot{j}}$ according to the relation $\tilde{\boldsymbol{t}}_{1\dot{1}}+\tilde{\boldsymbol{t}}_{2\dot{2}}+\tilde{\boldsymbol{t}}_{3\dot{3}}+\tilde{\boldsymbol{t}}_{4\dot{4}}=-\tilde{\boldsymbol{t}}_{5\dot{5}}$. It is then convenient to define new non-compact generators $\overline{\boldsymbol{t}}_{i\dot{j}}$ as
\begin{equation}
\overline{\boldsymbol{t}}_{i\dot{j}}=\tilde{\boldsymbol{t}}_{i\dot{j}}+\frac{1}{4}\tilde{\boldsymbol{t}}_{5\dot{5}}\delta_{i\dot{j}}
\end{equation}
which are symmetric and traceless. The nine scalar fields corresponding to these generators then parametrize an $SL(4)/SO(4)$ coset. The other four scalars associated with $\tilde{\boldsymbol{t}}_{i\dot{5}}=\hat{\boldsymbol{t}}^+_{i\dot{5}}$ are nilpotent scalars and will be denoted by $b_i$. The remaining scalar in $SL(5)/SO(5)$ coset is another dilatonic scalar corresponding to the $SO(1,1)$ factor in $SL(4)\times SO(1,1)\subset SL(5)$. In addition, there are additional ten axionic scalars corresponding to the antisymmetric shift generators as in the previous section. 
\\
\indent As previously mentioned, the six-dimensional gauged supergravity with $CSO(p,q,4-p-q)\ltimes \mathbb{R}^4$ gauge group can be embedded in type IIA theory via a consistent truncation on $H^{p,q}\times \mathbb{R}^{4-p-q}\times S^1$. The truncation of type IIA theory on $H^{p,q}\times \mathbb{R}^{4-p-q}$ leading to the maximal gauged supergravity in seven dimensions with $CSO(p,q,4-p-q)$ gauge group has been studied in \cite{Malek_IIA_IIB}. In particular, the $(00)$-component of the ten-dimensional metric is given, in the present notation, by 
\begin{equation}
\hat{g}_{00}=e^{-3\phi_0}\Delta^{\frac{1}{4}} g^{(7)}_{00}\, .
\end{equation}
$\phi_0$ is the dilatonic scalar corresponding to $SO(1,1)$ factor in the decomposition $SL(5)\rightarrow SL(4)\times SO(1,1)$ as mentioned above. The warp factor $\Delta$ is defined by 
\begin{equation}
\Delta=m^{ij}\eta_{ik}\eta_{jl}\mu^k\mu^l,\qquad i,j,\ldots=1,2,3,4, 
\end{equation}
in which $m^{ij}$ is the inverse of the $SL(4)$ matrix $m_{ij}=(\widetilde{\mc{V}}\widetilde{\mc{V}}^T)_{ij}$ with $\widetilde{\mc{V}}$ being $SL(4)/SO(4)$ coset representative. The coordinates $\mu^i$ satisfy $\mu^i\mu^j\eta_{ij}=1$ with $\eta_{ij}$ being the $CSO(p,q,4-p-q)$ invariant tensor.
\\
\indent By another dimensional reduction on $S^1$, we can uplift the six-dimensional metric of the domain wall solutions to type IIA theory. This results in
\begin{equation}
\hat{g}_{00}=e^{-3\phi_0}\Delta^{\frac{1}{4}}e^{2U+2\varphi}
\end{equation} 
with $\varphi$ being the dilaton corresponding to the $SO(1,1)$ factor in $GL(5)\sim SL(5)\times SO(1,1)$ as in the previous section.
%%%%%%%%%%%%%%%%%%%%%%%%%%%%%%%%%%%%%%%%%
\subsection{D4-branes wrapped on a Riemann surface with $SO(2)\times SO(2)$ twist}\label{40_Sig2_Sec}
We now look for supersymmetric solutions describing D4-branes wrapped on a Riemann surface $\Sigma^2$ with $SO(2)\times SO(2)$ twists. The ansatz for the six-dimensional metric is still given by \eqref{Sig_2_6D_metric}. The $SO(2)\times SO(2)$ corresponds to $X_{12}$ and $X_{34}$ gauge generators under which there are five singlet scalars described by the coset representative
\begin{equation}
V=e^{\varphi\boldsymbol{d}+\phi_0\widetilde{\mathcal{Y}}_0+\phi\widehat{\mathcal{Y}}_1
+\varsigma_1\widehat{\mathcal{Y}}_2+\varsigma_2\widehat{\mathcal{Y}}_3}\, .\label{40_fullSO(2)xSO(2)singlet_coset}
\end{equation}
The first scalar $\varphi(r)$ is the dilaton corresponding to the $SO(1,1)$ generator $\boldsymbol{d}$ while the second one $\phi_0$ is the dilaton corresponding to the $SO(1,1)$ factor in $SL(4)\times SO(1,1)$. The explicit form of this non-compact generator is given by
\begin{equation}\label{40repDil}
\widetilde{\mathcal{Y}}_0=\hat{\boldsymbol{t}}^+_{1\dot{1}}+\hat{\boldsymbol{t}}^+_{2\dot{2}}+\hat{\boldsymbol{t}}^+_{3\dot{3}}+\hat{\boldsymbol{t}}^+_{4\dot{4}}-4\,\hat{\boldsymbol{t}}^+_{5\dot{5}}\, .
\end{equation}
The remaining three scalars $\phi(r)$, $\varsigma_1(r)$, and $\varsigma_2(r)$ correspond to the following $SO(5,5)$ non-compact generators
\begin{equation}
\widehat{\mathcal{Y}}_1=\hat{\boldsymbol{t}}^+_{1\dot{1}}+\hat{\boldsymbol{t}}^+_{2\dot{2}}-\hat{\boldsymbol{t}}^+_{3\dot{3}}-\hat{\boldsymbol{t}}^+_{4\dot{4}},\qquad
\widehat{\mathcal{Y}}_2=\boldsymbol{s}_{12},\qquad
\widehat{\mathcal{Y}}_3=\boldsymbol{s}_{34}\, .
\end{equation}
There are two gauge groups containing $SO(2)\times SO(2)$ as a subgroup. These can be characterized by the embedding tensor of the form
\begin{equation}\label{SO(2)xSO(2)wij}
w^{ij}=\text{diag}(1,1,\kappa,\kappa)
\end{equation}
where $\kappa=1,-1$ respectively correspond to $SO(4)\ltimes \mathbb{R}^{4}$ and $SO(2,2)\ltimes \mathbb{R}^{4}$ gauge groups.
\\
\indent With the coset representative \eqref{40_fullSO(2)xSO(2)singlet_coset}, the scalar potential is given by
\begin{equation}\label{ZSO(4)Pot}
\mathbf{V}=-g^2\kappa e^{2\varphi-8\phi_0}.
\end{equation}
We will perform a topological twist by turning on the following gauge fields
\begin{equation}\label{40_SO(2)_SO(2)_SIg2_A}
{A_{\hat{\zeta}}}^{12}=-e^{-V}\frac{p_1}{2\sqrt{2}k}\frac{f'_{k}(\theta)}{f_{k}(\theta)}\qquad \textrm{and}\qquad
{A_{\hat{\zeta}}}^{34}=-e^{-V}\frac{p_2}{2\sqrt{2}k}\frac{f'_{k}(\theta)}{f_{k}(\theta)}\, .
\end{equation} 
The twist can be achieved by imposing the projection conditions given in \eqref{SO(2)xSO(2)Projcon+} and \eqref{SO(2)xSO(2)Projcon-} as well as
\begin{equation}
\hat{\gamma}_r\epsilon_\pm=\gamma^5\epsilon_\mp\label{pureZProj}
\end{equation}
on the Killing spinors \eqref{DW_Killing_spinor} together with the twist condition \eqref{15_SO(2)xSO(2)_twist_con}. 
\\
\indent As in the case of gaugings in $\mathbf{15}^{-1}$ representation, the solutions preserve four supercharges corresponding to $N=2$ supersymmetry in three dimensions. In addition, consistency of the BPS equations requires either all the shift scalars $\varsigma_1$ and $\varsigma_2$ vanish or the condition $p_2=p_1/\kappa$ must be imposed. We again consider these two possibilities separately. 
%%%%%%%%%%%%%%%%%%%%%%%%%%%%%%%%%%%%%%%%%%%%
\subsubsection{Solutions with $\varsigma_1=\varsigma_2=0$} 
\begin{figure}[h!]
  \centering
    \includegraphics[width=\linewidth]{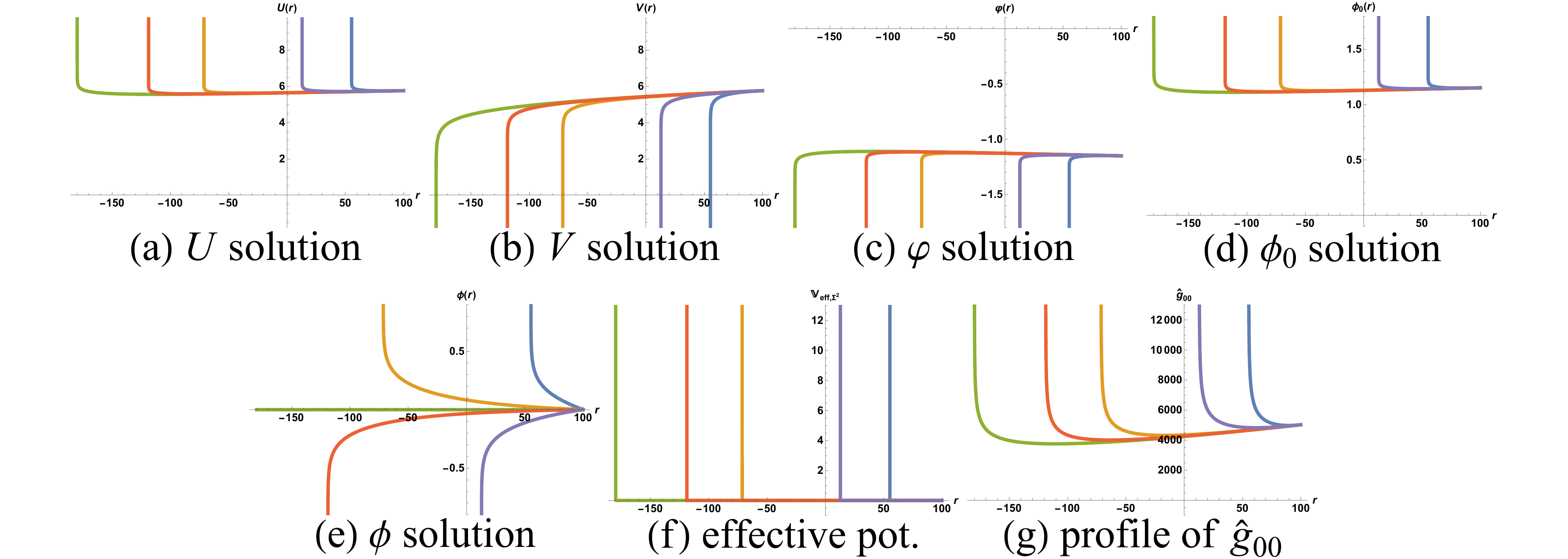}
\caption{Interpolating solutions between the locally $SO(4)$ flat domain wall as $\rho\rightarrow+\infty$ and $Mkw_3\times S^2$-sliced curved domain walls for $SO(2)\times SO(2)$ twist in $SO(4)\ltimes \mathbb{R}^{4}$ gauge group. The blue, orange, green, red, and purple curves refer to $p_1=-4.25, 0, 0.50, 0.71$, and $2.42$, respectively. }
\label{40_S2_special_SO(2)xSO(2)_SO(4)gg_flows}
\end{figure}

\begin{figure}[h!]
  \centering
    \includegraphics[width=\linewidth]{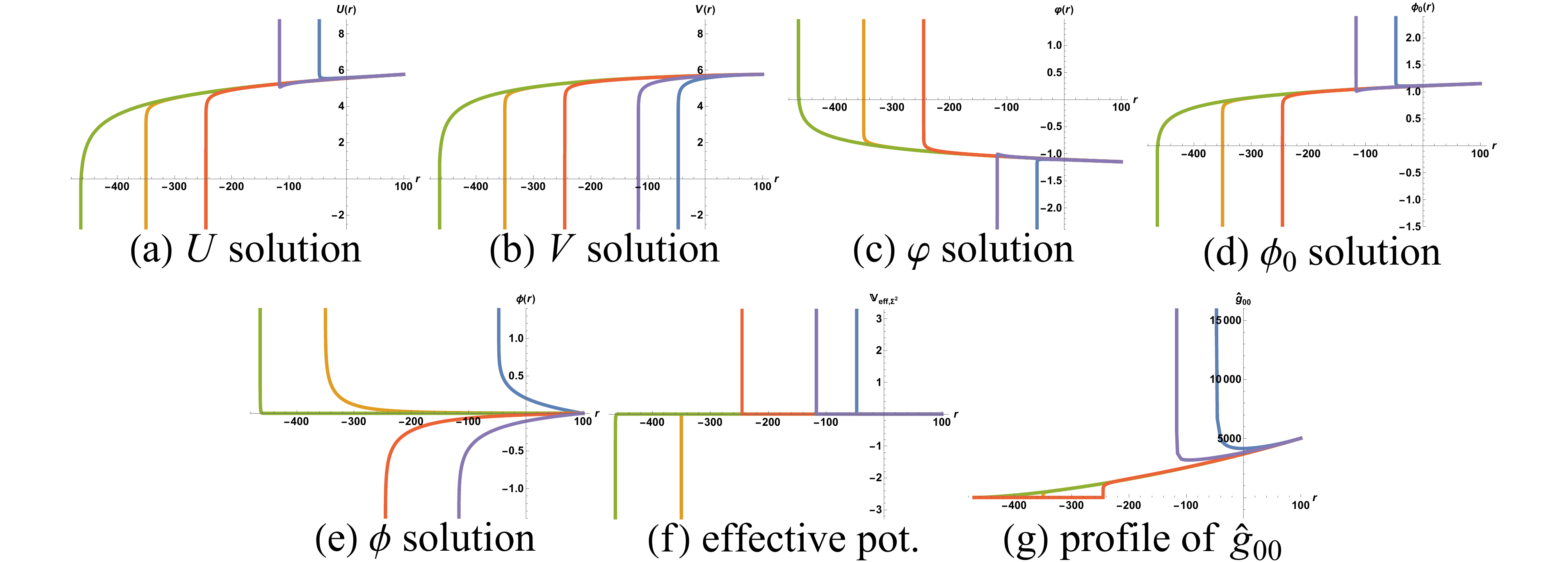}
\caption{Interpolating solutions between the locally $SO(4)$ flat domain wall as $\rho\rightarrow+\infty$ and $Mkw_3\times H^2$-sliced curved domain walls for $SO(2)\times SO(2)$ twist in $SO(4)\ltimes \mathbb{R}^{4}$ gauge group. The blue, orange, green, red, and purple curves refer to $p_1=-2.2, -0.52, -0.50, -0.30$, and $0.49$, respectively. }
\label{40_H2_special_SO(2)xSO(2)_SO(4)gg_flows}
\end{figure}

\begin{figure}[h!]
  \centering
    \includegraphics[width=\linewidth]{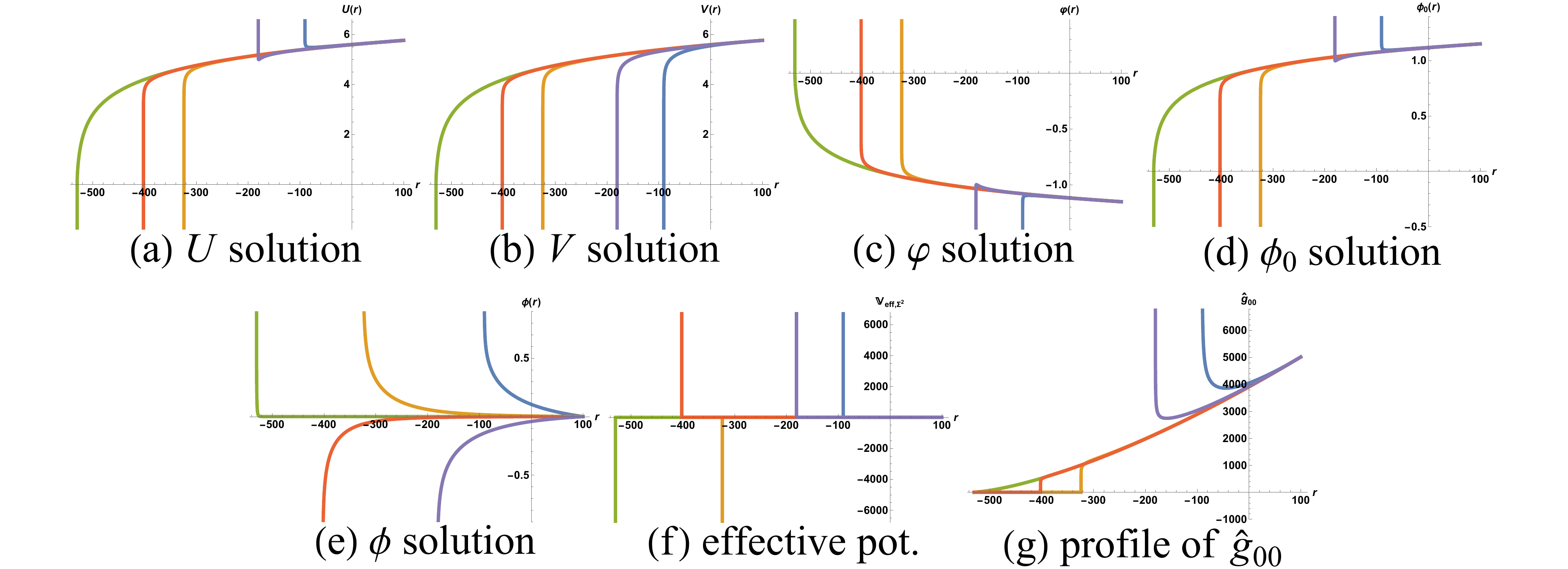}
\caption{Interpolating solutions between the locally $SO(4)$ flat domain wall as $\rho\rightarrow+\infty$ and $Mkw_3\times \mathbb{R}^2$-sliced curved domain walls for $SO(2)\times SO(2)$ twist in $SO(4)\ltimes \mathbb{R}^{4}$ gauge group. The blue, orange, green, red, and purple curves refer to $p_1=-0.810, -0.040, 0, 0.006$, and $0.34$, respectively. The solution with $p_1=0$ is the $SO(2)\times SO(2)$ flat domain wall since $p_2= 0$ due to the twist condition.}
\label{40_R2_special_SO(2)xSO(2)_SO(4)gg_flows}
\end{figure}

Setting $\varsigma_1=\varsigma_2=0$, we find the following BPS equations
\begin{eqnarray}
U'&=&\frac{g}{4}e^{\varphi-4(\phi_0+\phi)}(e^{8\phi}+\kappa)-\frac{1}{8}e^{-2V-\varphi+4\phi_0}(p_1e^{-4\phi}+p_2e^{4\phi}),\label{40_SO(2)xSO(2)_Sig2_BPS1}\\
V'&=&\frac{g}{4}e^{\varphi-4(\phi_0+\phi)}(e^{8\phi}+\kappa)+\frac{3}{8}e^{-2V-\varphi+4\phi_0}(p_1e^{-4\phi}+p_2e^{4\phi}),\\
\varphi'&=&-\frac{g}{20}e^{\varphi-4(\phi_0+\phi)}(e^{8\phi}+\kappa)+\frac{1}{40}e^{-2V-\varphi+4\phi_0}(p_1e^{-4\phi}+p_2e^{4\phi}),\\
\phi'_0&=&\frac{g}{20}e^{\varphi-4(\phi_0+\phi)}(e^{8\phi}+\kappa)-\frac{1}{40}e^{-2V-\varphi+4\phi_0}(p_1e^{-4\phi}+p_2e^{4\phi}),\\
\phi'&=&-\frac{g}{4}e^{\varphi-4(\phi_0+\phi)}(e^{8\phi}-\kappa)+\frac{1}{8}e^{-2V-\varphi+4\phi_0}(p_1e^{-4\phi}-p_2e^{4\phi}).\label{40_SO(2)xSO(2)_Sig2_BPS5}
\end{eqnarray}
We will solve these equations numerically with suitable asymptotic behaviors. We first consider the following asymptotic behavior as $r\rightarrow+\infty$,
\begin{equation}\label{40_SO(2)xSO(2)_full_IR_fDW}
U\sim V\sim \frac{5}{4} \ln (g r),\qquad \varphi\sim-\phi_0\sim-\frac{1}{4}\ln (g r),\qquad\phi\sim \frac{1}{g^5 r^5}\, .
\end{equation}
This takes the form of a locally $SO(4)$ symmetric flat domain wall. This asymptotic geometry can arise only in the case of $SO(4)\ltimes \mathbb{R}^{4}$ gauge group. Examples of numerical solutions starting from this $SO(4)$ symmetric locally flat domain wall are given in figures \ref{40_S2_special_SO(2)xSO(2)_SO(4)gg_flows}, \ref{40_H2_special_SO(2)xSO(2)_SO(4)gg_flows}, and \ref{40_R2_special_SO(2)xSO(2)_SO(4)gg_flows}, in which we have set $g=1$ for different values of $p_1$. 
\\
\indent From these figures, we find that all $Mkw_3\times S^2$-sliced domain walls contain unphysical IR singularities. For $Mkw_3\times H^2$-sliced domain walls, the solutions with $p_1=-0.52,-0.50,-0.30$ (the orange, green, and red curves in figure \ref{40_H2_special_SO(2)xSO(2)_SO(4)gg_flows}) lead to physical solutions in type IIA theory. For $Mkw_3\times \mathbb{R}^2$-sliced domain walls in figure \ref{40_R2_special_SO(2)xSO(2)_SO(4)gg_flows}, the solutions represented by the orange and red curves with $p_1=-0.040,0.006$ can also be uplifted to solutions of type IIA theory with physically acceptable IR singularities. We also note that the solution shown by the green line in figure \ref{40_R2_special_SO(2)xSO(2)_SO(4)gg_flows} corresponds to a flat domain wall since in this case, $p_1=p_2=k=0$ due to the twist condition.
\vfil
\vfil\pagebreak

\begin{table}[h!]
\centering
\begin{tabular}{| c | c | c |}
\hline
Riemann  & criterion \cite{Gubser_Sing}  & criterion \cite{Maldacena_nogo} \\
surface  & ($\mathbf{V}_{\text{eff},\Sigma^2}\nrightarrow+\infty$) & ($\hat{g}_{00}\rightarrow0$) \\\hline
 $S^2$ & $\times$ & $\times$ \\
$H^2$ & $-0.52\leq p_1\leq-0.5$ & $-0.52\leq p_1\leq-0.3$ \\
 $\mathbb{R}^2$ & $-0.04\leq p_1\leq0$ & $-0.04\leq p_1\leq0.006$ \\\hline
\end{tabular}
\caption{Summary of satisfaction of criteria \cite{Gubser_Sing} and \cite{Maldacena_nogo} for the IR singularities of the interpolating solutions between the locally $SO(4)$ symmetric flat domain wall and $Mkw_3\times S^2$-sliced curved domain walls for $SO(2)\times SO(2)$ twist in $SO(4)\ltimes \mathbb{R}^{4}$ gauge group.}\label{tab7}
\end{table}

\begin{figure}[h!]
  \centering
    \includegraphics[width=\linewidth]{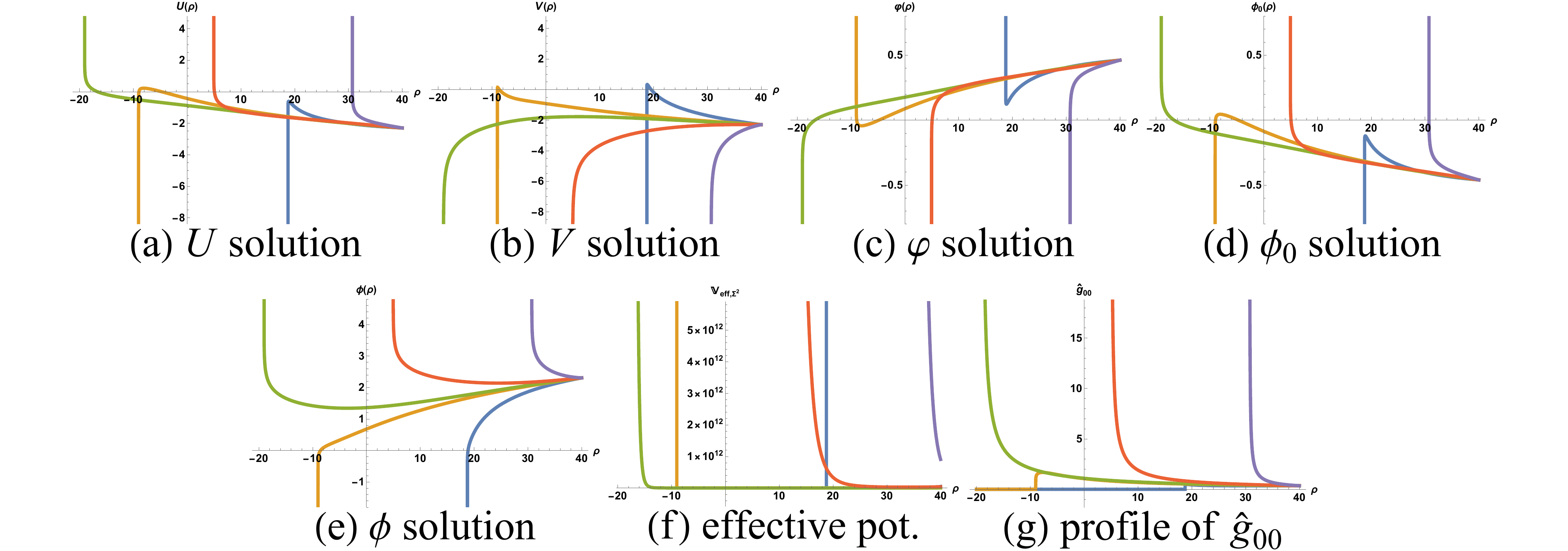}
\caption{Interpolating solutions between the locally $SO(2)\times SO(2)$ flat domain wall as $\rho\rightarrow+\infty$ and $Mkw_3\times S^2$-sliced curved domain walls for $SO(2)\times SO(2)$ twist in $SO(2,2)\ltimes \mathbb{R}^{4}$ gauge group. The blue, orange, green, red, and purple curves refer to $p_1=-1.60, -0.91, -0.82, -0.45, 1.69$, respectively. }
\label{40_S2_SO(2)xSO(2)_SO(22)gg_flows}
\end{figure}

\begin{figure}[h!]
  \centering
    \includegraphics[width=\linewidth]{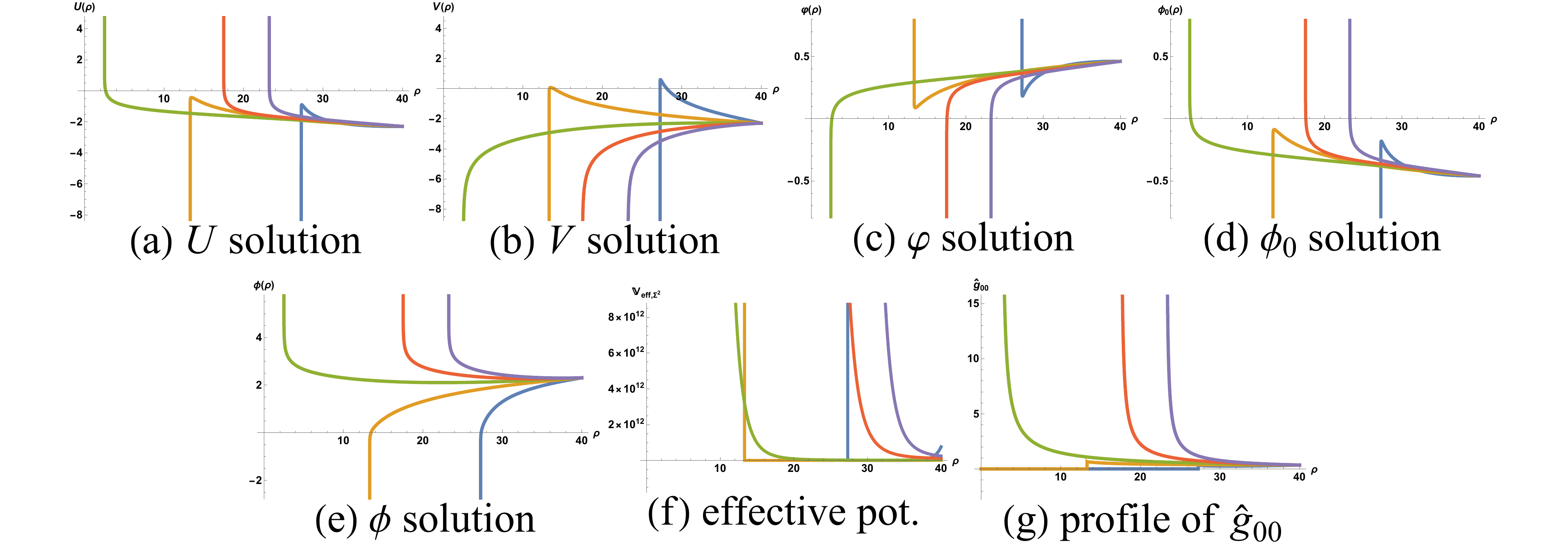}
\caption{Interpolating solutions between the locally $SO(2)\times SO(2)$ flat domain wall as $\rho\rightarrow+\infty$ and $Mkw_3\times H^2$-sliced curved domain walls for $SO(2)\times SO(2)$ twist in $SO(2,2)\ltimes \mathbb{R}^{4}$ gauge group. The blue, orange, green, red, and purple curves refer to $p_1=-1.46, 0.76, 1.50, 1.99, 2.40$, respectively. }
\label{40_H2_SO(2)xSO(2)_SO(22)gg_flows}
\end{figure}

\begin{figure}[h!]
  \centering
    \includegraphics[width=\linewidth]{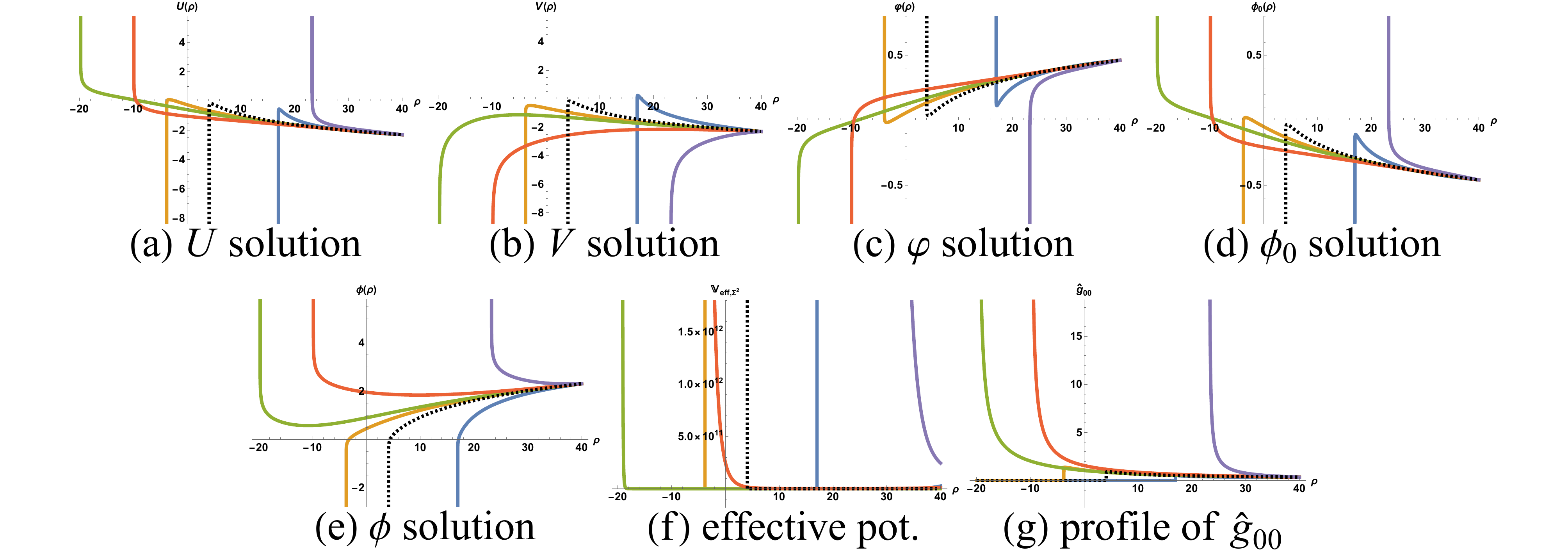}
\caption{Interpolating solutions between the locally $SO(2)\times SO(2)$ flat domain wall as $\rho\rightarrow+\infty$ and $Mkw_3\times \mathbb{R}^2$-sliced curved domain walls for $SO(2)\times SO(2)$ twist in $SO(2,2)\ltimes \mathbb{R}^{4}$ gauge group. The blue, orange, green, red, and purple curves refer to $p_1=-0.450, 0.071, 0.110, 0.300, 1.400$, respectively. The dashed curve is the $SO(2)\times SO(2)$ flat domain wall obtained from $p_1=0$ which implies $p_2=k=0$.}
\label{40_R2_SO(2)xSO(2)_SO(22)gg_flows}
\end{figure}

Another class of solutions are described by the following asymptotic behavior
\begin{equation}\label{40_SO(2)xSO(2)_flat_DW_asym}
U\sim V\sim-\phi\sim-\ln \left[-\frac{g \rho}{4}\right]\qquad\text{and}\qquad \varphi\sim-\phi_0\sim\frac{1}{5}\ln \left[-\frac{g \rho}{4}\right]
\end{equation}
with the new radial coordinate $\rho$ defined by $\frac{d\rho}{dr}=e^{-V+5\phi}$. This asymptotic geometry takes the form of an $SO(2)\times SO(2)$ symmetric flat domain wall. We find a number of numerical solutions interpolating between this flat domain wall and singular geometries in the IR for both $SO(4)\ltimes \mathbb{R}^{4}$ and $SO(2,2)\ltimes \mathbb{R}^{4}$ gauge groups. For $SO(4)\ltimes \mathbb{R}^4$ gauge group, all solutions turn out to be unphysical upon uplifted to type IIA theory, see table \ref{tab7}. For $SO(2,2)\ltimes \mathbb{R}^4$ gauge group, examples of numerical solutions are shown in figures \ref{40_S2_SO(2)xSO(2)_SO(22)gg_flows}, \ref{40_H2_SO(2)xSO(2)_SO(22)gg_flows}, and \ref{40_R2_SO(2)xSO(2)_SO(22)gg_flows}. From these figures, we see that only the solutions represented by the orange and blue curves can be uplifted to ten-dimensional solutions with physical IR singularities, see also table \ref{tab8}. 
\vfil
\vfil
\vfil\pagebreak

\begin{table}[h!]
\centering
\begin{tabular}{| c | c | c | c |}
\hline
\multirow{2}{*}{gauge group} & Riemann  & criterion \cite{Gubser_Sing}  & criterion \cite{Maldacena_nogo} \\
 & surface  & ($\mathbf{V}_{\text{eff},\Sigma^2}\nrightarrow+\infty$) & ($\hat{g}_{00}\rightarrow0$) \\\hline
$SO(4)\ltimes \mathbb{R}^{4}$& $S^2$ & $p_1\geq-1$ & $\times$ \\
& $H^2$, $\mathbb{R}^2$ & $\times$ & $\times$ \\\hline
 $SO(2,2)\ltimes \mathbb{R}^{4}$ & $S^2$& $\times$ & $p_1\leq-0.91$ \\
 & $H^2$& $\times$ & $p_1\leq0.76$ \\
 & $\mathbb{R}^2$& $\times$ & $p_1\leq0.071$ \\\hline
\end{tabular}
\caption{Summary of satisfaction of criteria \cite{Gubser_Sing} and \cite{Maldacena_nogo} for the interpolating solutions between the locally $SO(2)\times SO(2)$ symmetric flat domain wall and $Mkw_3\times H^2$-sliced curved domain walls for $SO(2)\times SO(2)$ twist in $SO(4)\ltimes \mathbb{R}^{4}$ and $SO(2,2)\ltimes \mathbb{R}^{4}$ gauge groups.}\label{tab8}
\end{table}

%%%%%%%%%%%%%%%%%%%%%%%%%%%%%%%%%%%%%%%%%%%%%%%%%%%%%%%%%%
\subsubsection{Solutions with non-vanishing $\varsigma_1$ and $\varsigma_2$}
\begin{figure}[h!]
  \centering
    \includegraphics[width=0.8\linewidth]{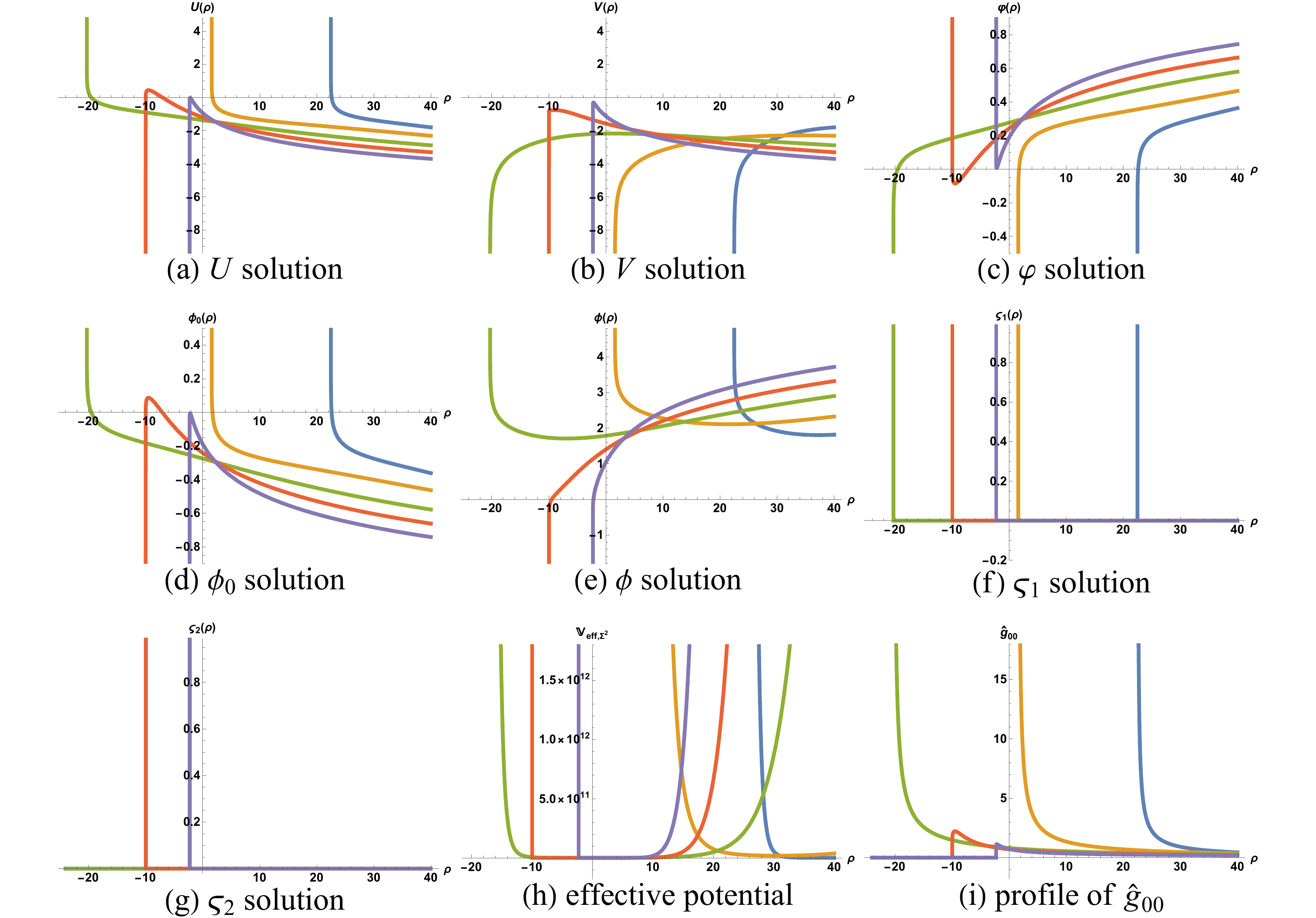}
\caption{Interpolating solutions between the locally $SO(2)\times SO(2)$ flat domain wall as $\rho\rightarrow+\infty$ and $Mkw_3\times S^2$-sliced curved domain walls with shift scalars in $SO(2,2)\ltimes \mathbb{R}^{4}$ gauge group. The blue, orange, green, red, and purple curves refer to $g=-0.61, -1.02, -1.81, -2.75, -4.10$, respectively. }
\label{40_S2_SO(2)d_SO(22)gg_flows}
\end{figure}

\begin{figure}[h!]
  \centering
    \includegraphics[width=0.8\linewidth]{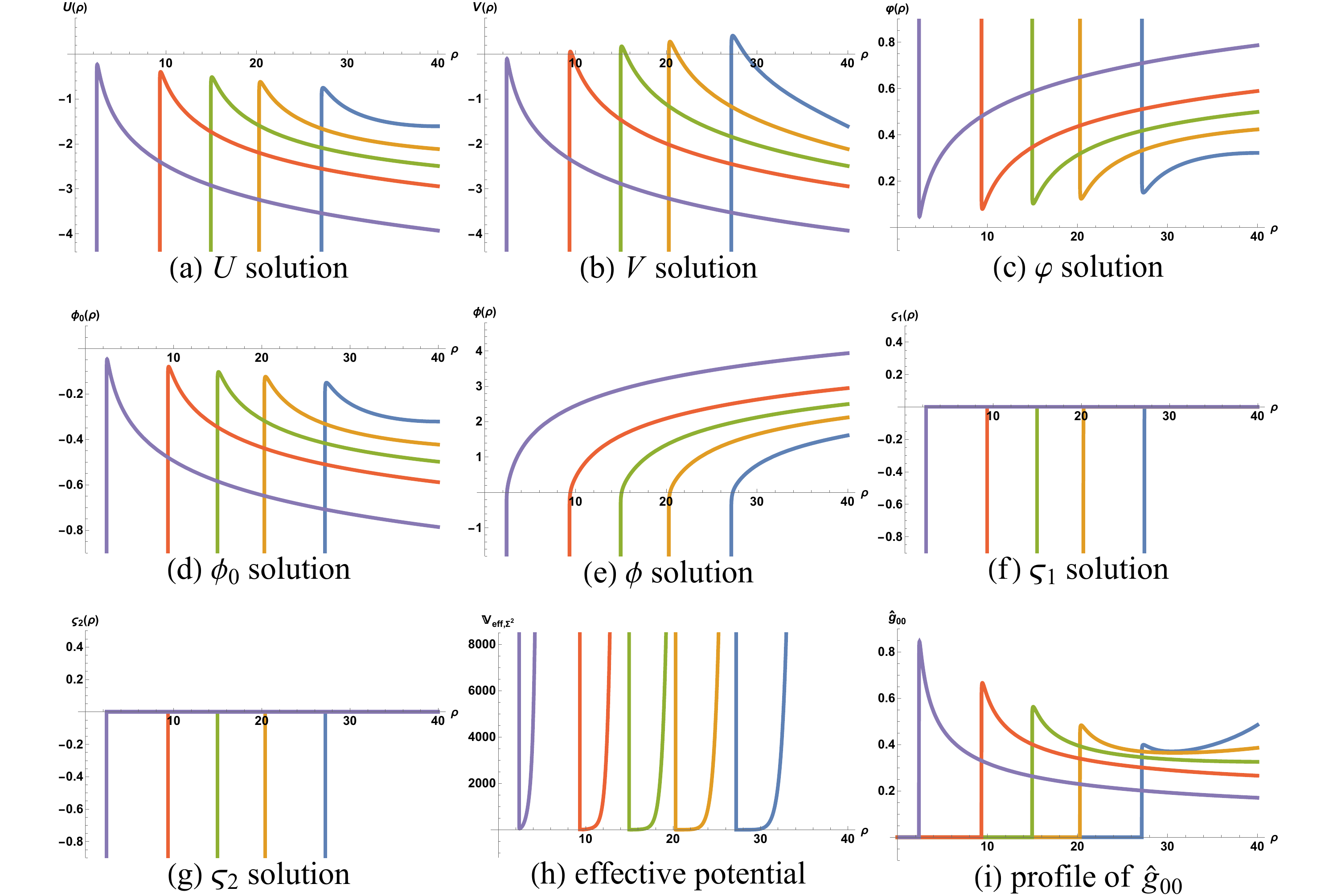}
\caption{Interpolating solutions between the locally $SO(2)\times SO(2)$ flat domain wall as $\rho\rightarrow+\infty$ and $Mkw_3\times H^2$-sliced curved domain walls with shift scalars in $SO(2,2)\ltimes \mathbb{R}^{4}$ gauge group. The blue, orange, green, red, and purple curves refer to $g=-0.50, -0.83, -1.21, -1.90, -5.10$, respectively. }
\label{40_H2_SO(2)d_SO(22)gg_flows}
\end{figure}

In this case, we set $p_2=p_1/\kappa$ as well as 
\begin{equation}
\varsigma_2=-\kappa e^{-8\phi} \varsigma_1\, .
\end{equation}
The resulting BPS equations are given by \eqref{40_SO(2)xSO(2)_Sig2_BPS1} to \eqref{40_SO(2)xSO(2)_Sig2_BPS5} with $p_2=p_1/\kappa$ together with an additional equation 
\begin{equation}\label{40_Sig2_SO(2)d_varsig1_BPSeq}
\varsigma'_1=-\frac{\varsigma}{2 \kappa }(e^{8\phi}-\kappa) e^{-\varphi-2V-4\phi-4\phi_0} (2 g \kappa  e^{2 (\varphi+V)}+p_1 e^{8 \phi_0}).
\end{equation}
For the asymptotic behavior given by $SO(4)$ symmetric flat domain wall \eqref{40_SO(2)xSO(2)_full_IR_fDW}, we find that all solutions from $SO(4)\ltimes \mathbb{R}^4$ gauge group have $\varsigma_1=\varsigma_2=0$ along the entire solutions. Since these solutions are qualitatively the same as those considered in the previous case, we will not give the resulting numerical solutions here to avoid repetition.   
\\
\indent Changing to a new radial coordinate $\rho$ via $\frac{d\rho}{dr}=e^{-V+5\phi}$, we now consider an asymptotic behavior as $\rho\rightarrow+\infty$ of the form
\begin{equation}\label{40_SO(2)d_flat_DW_asym}
U\sim V\sim-\phi\sim-\ln \left[-\frac{g \rho}{4}\right],\qquad \varphi\sim-\phi_0\sim\frac{1}{5}\ln \left[-\frac{g \rho}{4}\right],\qquad\varsigma_1\sim \frac{1}{\rho ^4}
\end{equation}
with $g<0$. This takes the form of an $SO(2)\times SO(2)$ symmetric flat domain wall. From a large number of numerical solutions, we find that all solutions from $SO(4)\ltimes \mathbb{R}^4$ gauge group are uplifted to ten-dimensional solutions with unphysical IR singularities. Therefore, we will omit these solutions and only include them in table \ref{tab9}. Examples of numerical solutions starting from the asymptotic geometry \eqref{40_SO(2)d_flat_DW_asym} from $SO(2,2)\ltimes \mathbb{R}^4$ gauge group are given in figures \ref{40_S2_SO(2)d_SO(22)gg_flows} and \ref{40_H2_SO(2)d_SO(22)gg_flows}. From these solutions, all $Mkw_3\times H^2$-sliced domain walls in figure \ref{40_H2_SO(2)d_SO(22)gg_flows} can be uplifted to solutions of type IIA theory with physical IR singularities. The $Mkw_3\times S^2$-sliced domain walls represented by the red and purple curves in figure \ref{40_S2_SO(2)d_SO(22)gg_flows} also lead to physical solutions of type IIA theory. 
\vfil
\vfil\pagebreak

\begin{table}[h!]
\centering
\begin{tabular}{| c | c | c | c |}
\hline
\multirow{2}{*}{gauge group} & Riemann  & criterion \cite{Gubser_Sing}  & criterion \cite{Maldacena_nogo} \\
 & surface  & ($\mathbf{V}_{\text{eff},\Sigma^2}\nrightarrow+\infty$) & ($\hat{g}_{00}\rightarrow0$) \\\hline
$SO(4)\ltimes \mathbb{R}^{4}$& $S^2$ & any $g$ & $\times$ \\
& $H^2$ & $g\leq-5.34$ & $\times$ \\\hline
 $SO(2,2)\ltimes \mathbb{R}^{4}$ & $S^2$& $\times$ & $g\leq-2.75$ \\
 & $H^2$& $\times$ & any $g$ \\\hline
\end{tabular}
\caption{Summary of satisfaction of criteria \cite{Gubser_Sing} and \cite{Maldacena_nogo} for the IR singularities of the interpolating solutions between the locally $SO(2)\times SO(2)$ symmetric flat domain wall and $Mkw_3\times \Sigma^2$-sliced curved domain walls with shift scalars in $SO(4)\ltimes \mathbb{R}^{4}$ and $SO(2,2)\ltimes \mathbb{R}^{4}$ gauge groups.}\label{tab9}
\end{table}

%%%%%%%%%%%%%%%%%%%%%
\subsection{D4-branes wrapped on a Riemann surface with $SO(2)$ twist}\label{40_Sig2_SO(2)_section}
\begin{figure}[h!]
  \centering
    \includegraphics[width=\linewidth]{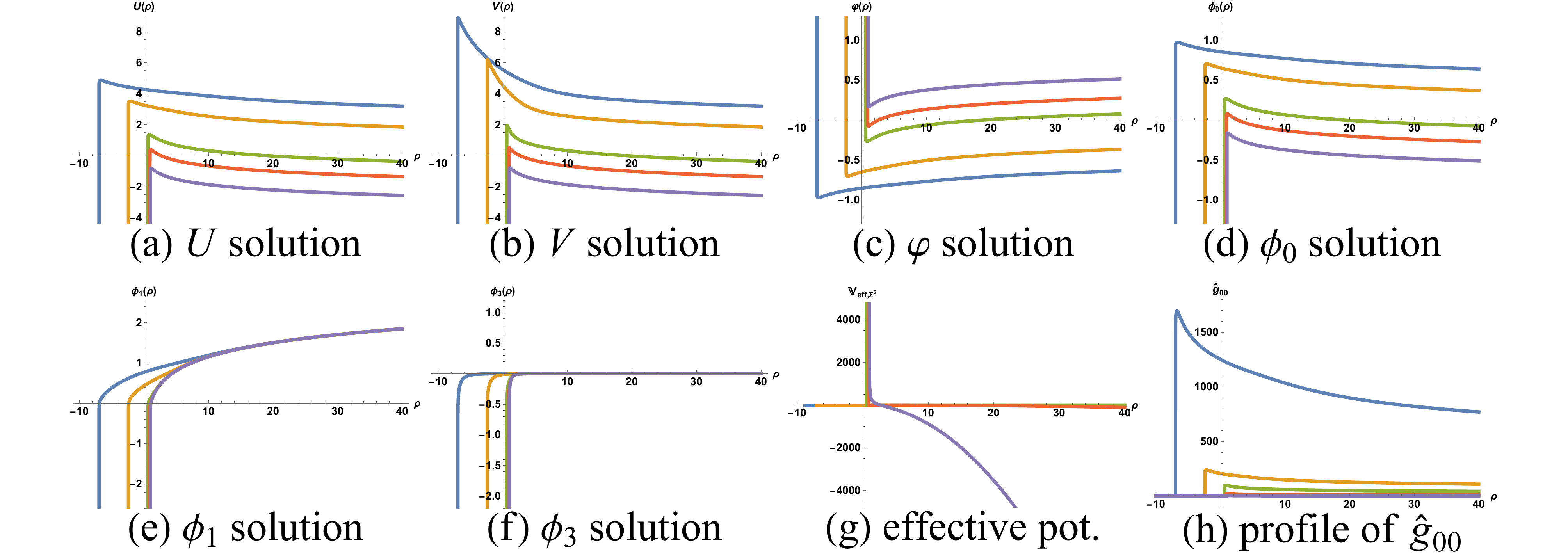}
\caption{Interpolating solutions between the locally $SO(2)$ flat domain wall as $\rho\rightarrow+\infty$ and $Mkw_3\times S^2$-sliced curved domain walls for $SO(2)$ twist in $SO(3,1)\ltimes \mathbb{R}^{4}$ gauge group. The blue, orange, green, red, and purple curves refer to $g=-0.01, -0.05, -0.46, -1.23, -4.10$, respectively.}
\label{40_S2_SO(2)_SO(31)gg_flows}
\end{figure}

\begin{figure}[h!]
  \centering
    \includegraphics[width=\linewidth]{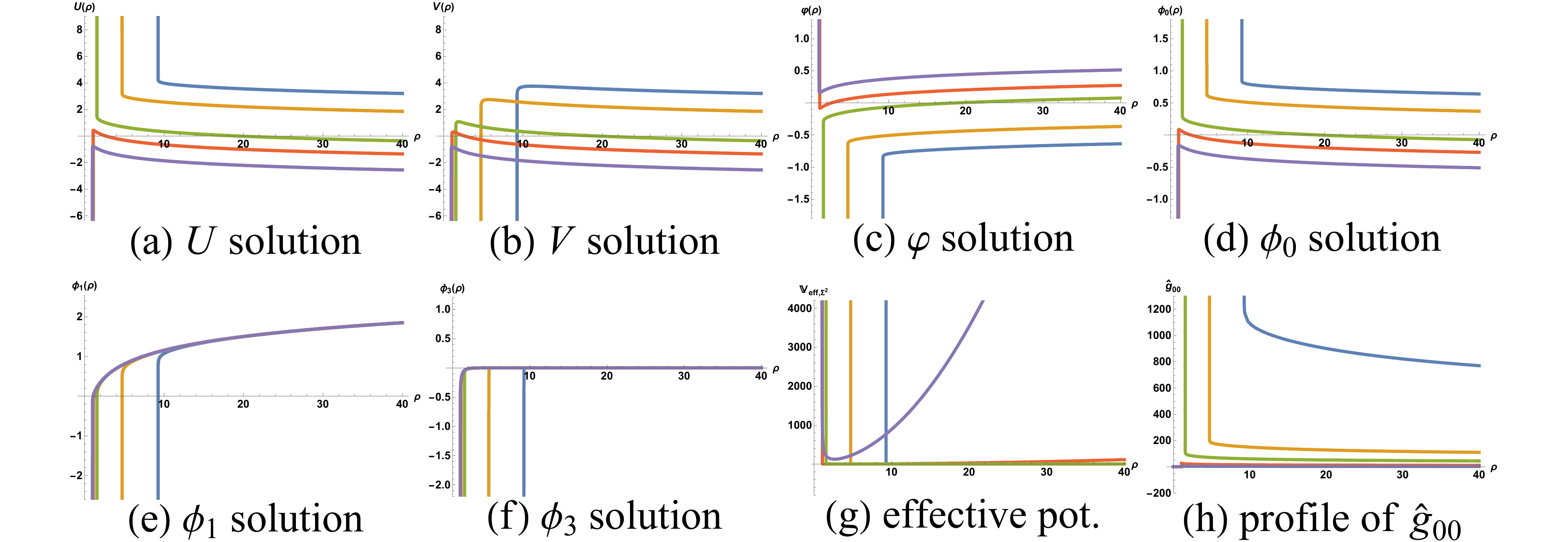}
\caption{Interpolating solutions between the locally $SO(2)$ flat domain wall as $\rho\rightarrow+\infty$ and $Mkw_3\times H^2$-sliced curved domain walls for $SO(2)$ twist in $SO(3,1)\ltimes \mathbb{R}^{4}$ gauge group. The blue, orange, green, red, and purple curves refer to $g=-0.01, -0.05, -0.46, -1.23, -4.10$, respectively.}
\label{40_H2_SO(2)_SO(31)gg_flows}
\end{figure}

\begin{figure}[h!]
  \centering
    \includegraphics[width=\linewidth]{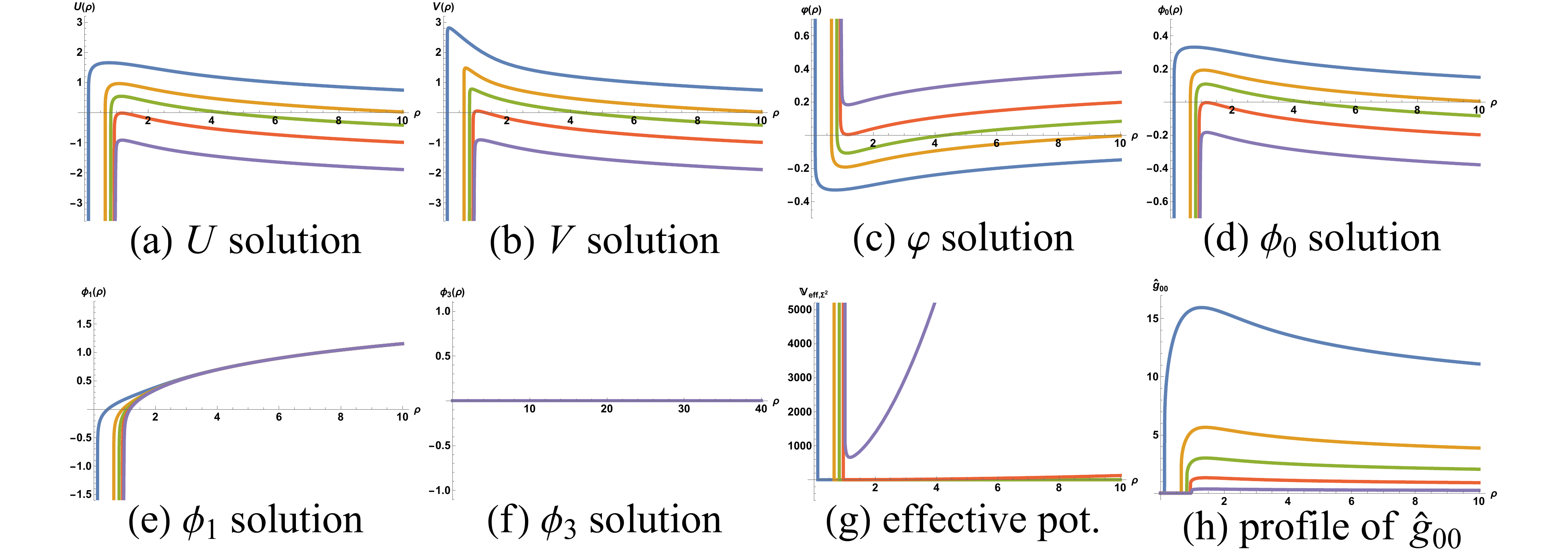}
\caption{Interpolating solutions between the locally $SO(2)$ flat domain wall as $\rho\rightarrow+\infty$ and $Mkw_3\times S^2$-sliced curved domain walls for $SO(2)$ twist in $SO(2,2)\ltimes \mathbb{R}^{4}$ gauge group. The blue, orange, green, red, and purple curves refer to $g=-0.30, -0.62, -0.96, -1.70, -4.20$, respectively.}
\label{40_S2_SO(2)_SO(22)gg_flows}
\end{figure}

\begin{figure}[h!]
  \centering
    \includegraphics[width=\linewidth]{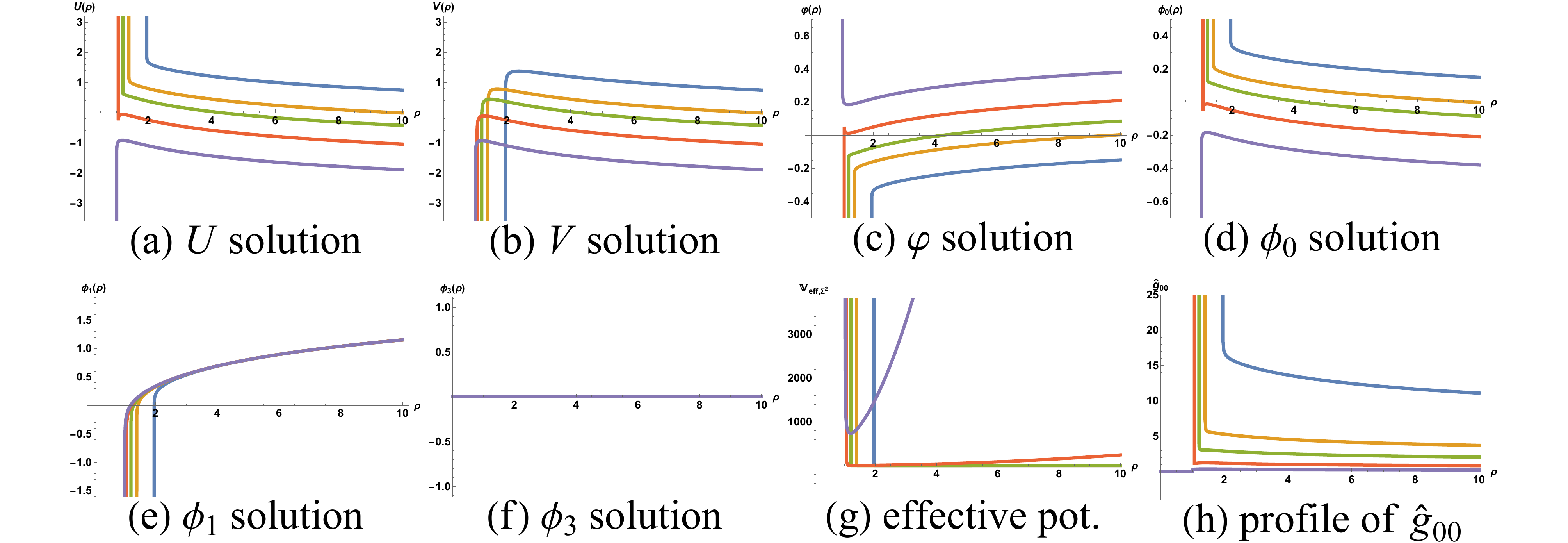}
\caption{Interpolating solutions between the locally $SO(2)$ flat domain wall as $\rho\rightarrow+\infty$ and $Mkw_3\times H^2$-sliced curved domain walls for $SO(2)$ twist in $SO(2,2)\ltimes \mathbb{R}^{4}$ gauge group. The blue, orange, green, red, and purple curves refer to $g=-0.30, -0.64, -0.97, -1.80, -4.22$, respectively.}
\label{40_H2_SO(2)_SO(22)gg_flows}
\end{figure}

\begin{figure}[h!]
  \centering
    \includegraphics[width=\linewidth]{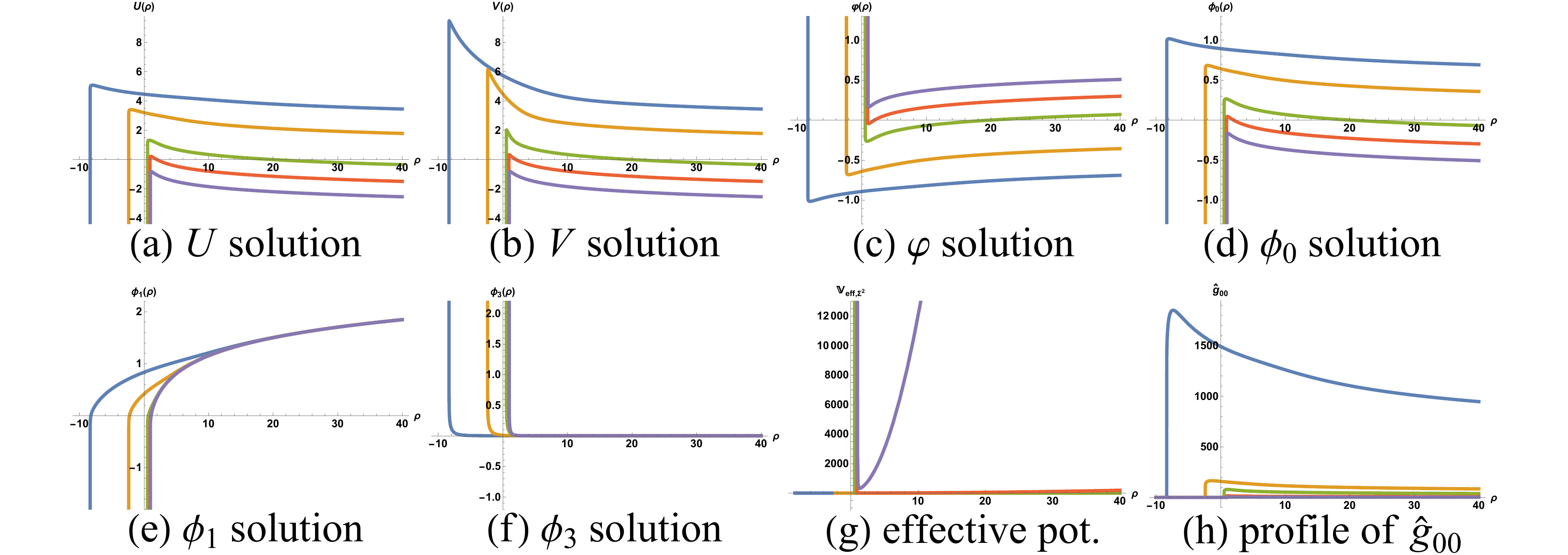}
\caption{Interpolating solutions between the locally $SO(2)$ flat domain wall as $\rho\rightarrow+\infty$ and $Mkw_3\times S^2$-sliced curved domain walls for $SO(2)$ twist in $CSO(2,1,1)\ltimes \mathbb{R}^{4}$ gauge group. The blue, orange, green, red, and purple curves refer to $g=-0.01, -0.05, -0.45, -1.40, -4$, respectively.}
\label{40_S2_SO(2)_CSO(211)gg_flows}
\end{figure}

\begin{figure}[h!]
  \centering
    \includegraphics[width=\linewidth]{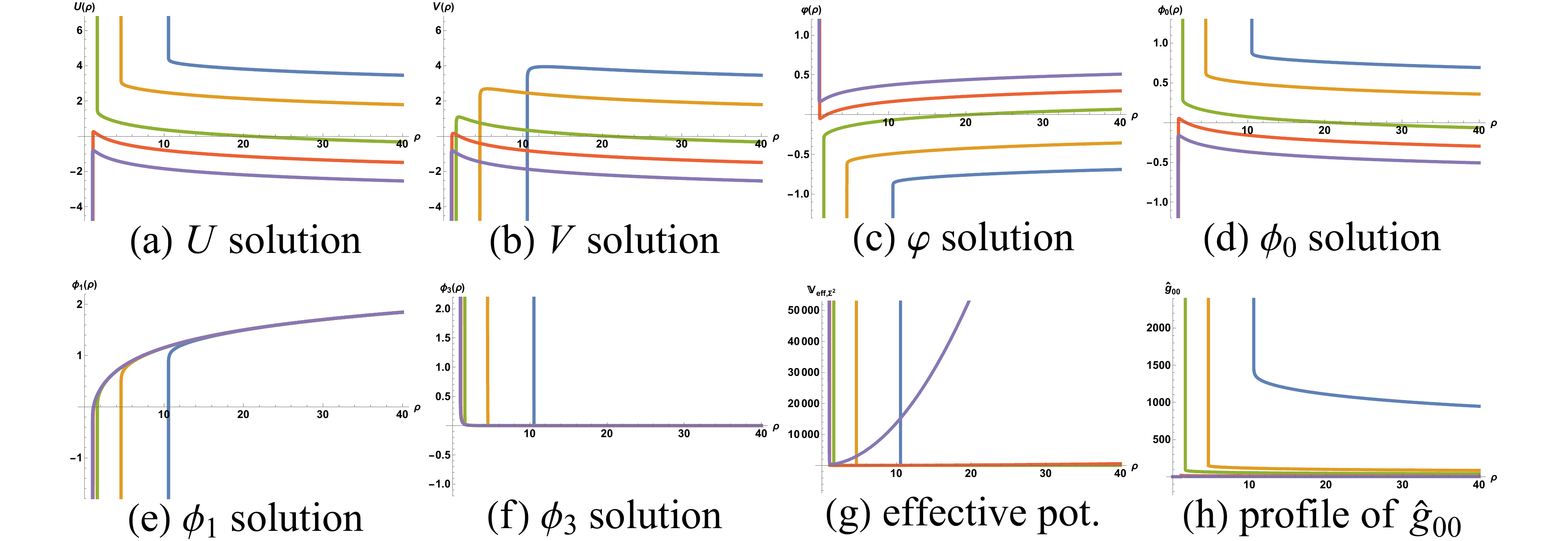}
\caption{Interpolating solutions between the locally $SO(2)$ flat domain wall as $\rho\rightarrow+\infty$ and $Mkw_3\times H^2$-sliced curved domain walls for $SO(2)$ twist in $CSO(2,1,1)\ltimes \mathbb{R}^{4}$ gauge group. The blue, orange, green, red, and purple curves refer to $g=-0.01, -0.05, -0.44, -1.40, -4$, respectively.}
\label{40_H2_SO(2)_CSO(211)gg_flows}
\end{figure}

We now move to another type of topological twists on $\Sigma^2$ achieved by turning on an $SO(2)$ gauge field. There are many gauge groups admitting an $SO(2)$ subgroup. They are collectively characterized by the following component of the embedding tensor
\begin{equation}\label{40_SO(2)w}
w^{ij}=\text{diag}(1,1,\kappa,\lambda)
\end{equation}
with $\kappa=0,\pm 1$ and $\lambda=0,\pm 1$.
\\
\indent Apart from the two dilatons, there are additional nine $SO(2)$ singlet scalars. Three of them are in the $SL(4)/SO(4)$ coset corresponding to non-compact generators
\begin{equation}
\overline{\mathcal{Y}}_1=\hat{\boldsymbol{t}}^+_{1\dot{1}}+\hat{\boldsymbol{t}}^+_{2\dot{2}}-\hat{\boldsymbol{t}}^+_{3\dot{3}}-\hat{\boldsymbol{t}}^+_{4\dot{4}},\qquad
\overline{\mathcal{Y}}_2=\hat{\boldsymbol{t}}^+_{3\dot{4}},\qquad
\overline{\mathcal{Y}}_3=\hat{\boldsymbol{t}}^+_{3\dot{3}}-\hat{\boldsymbol{t}}^+_{4\dot{4}}\, .
\end{equation}
The remaining ones consist of two nilpotent scalars associated with 
\begin{equation}
\overline{\mathcal{Y}}_4=\hat{\boldsymbol{t}}^+_{3\dot{5}},\qquad
\overline{\mathcal{Y}}_5=\hat{\boldsymbol{t}}^+_{4\dot{5}}
\end{equation}
and four shift scalars corresponding to
\begin{equation}
\overline{\mathcal{Y}}_6=\boldsymbol{s}_{12},\qquad
\overline{\mathcal{Y}}_7=\boldsymbol{s}_{35},\qquad
\overline{\mathcal{Y}}_8=\boldsymbol{s}_{45},\qquad
\overline{\mathcal{Y}}_9=\boldsymbol{s}_{34}\, .
\end{equation}
The analysis turns out to be highly complicated, so we will perform a subtruncation by setting the shift scalar corresponding to $\boldsymbol{s}_{12}$ and the two nilpotent scalars corresponding to $\overline{\mathcal{Y}}_4$ and $\overline{\mathcal{Y}}_5$ to zero. It is straightforward to verify that this is a consistent truncation and still gives interesting solutions. Furthermore, consistency of the resulting BPS equations requires the vanishing of the shift scalar corresponding to $\boldsymbol{s}_{35}$. We then end up with seven singlet scalars with the coset representative
\begin{equation}\label{40SO(2)FullCoset}
V=e^{\varphi\boldsymbol{d}+\phi_0\widetilde{\mathcal{Y}}_0+\phi_1\overline{\mathcal{Y}}_1
+\phi_2\overline{\mathcal{Y}}_2+\phi_3\overline{\mathcal{Y}}_3+\varsigma_1\overline{\mathcal{Y}}_8+\varsigma_2\overline{\mathcal{Y}}_9}\,.
\end{equation} 
This gives rise to the scalar potential of the form
\begin{eqnarray}
\mathbf{V}&=&-\frac{g^2}{32}e^{2\left(\varphi-4(\phi_0+\phi_1)\right)}\left[\kappa^2+10\kappa\lambda+\lambda^2-2(\kappa+\lambda)^2\cosh{4\phi_2}\cosh^2{4\phi_3}\phantom{\frac{1}{\sqrt{2}}}\right.\nonumber\\&&\left.\phantom{+\frac{1}{\sqrt{2}}}-(3\kappa^2-2\kappa\lambda+3\lambda^2)\cosh{8\phi_3}+16(\kappa-\lambda)e^{8\phi_1}\sinh{4\phi_3}\right.\nonumber\\&&\left.\phantom{+\frac{1}{\sqrt{2}}}+4(\kappa+\lambda)\cosh{2\phi_2}\left(4e^{8\phi_1}\cosh{4\phi_3}-(\kappa-\lambda)\sinh{8\phi_3}\right)\right].\label{ZSO(2)Pot}
\end{eqnarray}
\indent The ansatz for the $SO(2)$ gauge field is obtained from \eqref{40_SO(2)_SO(2)_SIg2_A} in the case of $SO(2)\times SO(2)$ twist by setting $p_2=0$ and $p_1=p$. Imposing the twist condition $gp=k$ and the projector \eqref{pureZProj} together with
\begin{equation}\label{40_SO(2)Projcon}
\hat{\gamma}_{\hat{\zeta}\hat{\theta}}\epsilon_{+\alpha}={(\gamma_{12})_\alpha}^\beta\epsilon_{+\beta} \qquad \textrm{and}\qquad 
\hat{\gamma}_{\hat{\zeta}\hat{\theta}}\epsilon_{-\dot{\alpha}}={(\gamma_{\dot{1}\dot{2}})_{\dot{\alpha}}}^{\dot{\beta}}\epsilon_{-\dot{\beta}}
\end{equation}
on the Killing spinors of the form
\begin{equation}\label{40_SO(2)_Killing_spinors}
\epsilon_+=e^{\frac{U(r)}{2}+W(r)\gamma_{34}}\epsilon_{+}^{0}\qquad\text{ and }\qquad\epsilon_-=e^{\frac{U(r)}{2}+W(r)\gamma_{\dot{3}\dot{4}}}\epsilon_{-}^{0},	
\end{equation}
we obtain the BPS equations
\begin{eqnarray}
U'&=&\frac{g}{8}e^{\varphi-4(\phi_0+\phi_1)}\left[2e^{8\phi_1}+(\kappa+\lambda)\cosh{2\phi_2}\cosh{4\phi_3}+(\kappa-\lambda)\sinh{4\phi_3}\right]\nonumber\\&&-\frac{p}{8}e^{-2V-\varphi+4(\phi_0-\phi_1)},\label{40SO(2)GenBPS1}\\
V'&=&\frac{g}{8}e^{\varphi-4(\phi_0+\phi_1)}\left[2e^{8\phi_1}+(\kappa+\lambda)\cosh{2\phi_2}\cosh{4\phi_3}+(\kappa-\lambda)\sinh{4\phi_3}\right]\nonumber\\&&+\frac{3p}{8}e^{-2V-\varphi+4(\phi_0-\phi_1)},\label{40SO(2)GenBPS2}\\
\varphi'&=&-\frac{g}{40}e^{\varphi-4(\phi_0+\phi_1)}\left[2e^{8\phi_1}+(\kappa+\lambda)\cosh{2\phi_2}\cosh{4\phi_3}+(\kappa-\lambda)\sinh{4\phi_3}\right]\nonumber\\&&+\frac{p}{40}e^{-2V-\varphi+4(\phi_0-\phi_1)},\qquad\ \label{40SO(2)GenBPS3}
%\\
\end{eqnarray}
\begin{eqnarray}
\phi'_0&=&\frac{g}{40}e^{\varphi-4(\phi_0+\phi_1)}\left[2e^{8\phi_1}+(\kappa+\lambda)\cosh{2\phi_2}\cosh{4\phi_3}+(\kappa-\lambda)\sinh{4\phi_3}\right]\nonumber\\&&-\frac{p}{40}e^{-2V-\varphi+4(\phi_0-\phi_1)},\label{40SO(2)GenBPS4}\\
\phi_1'&=&-\frac{g}{8}e^{\varphi-4(\phi_0+\phi_1)}\left[2e^{8\phi_1}-(\kappa+\lambda)\cosh{2\phi_2}\cosh{4\phi_3}-(\kappa-\lambda)\sinh{4\phi_3}\right]\nonumber\\&&+\frac{p}{8}e^{-2V-\varphi+4(\phi_0-\phi_1)},\label{40SO(2)GenBPS5}\\
\phi_2'&=&-\frac{g}{2}e^{\phi_0-4(\phi_0+\phi_1)}(\kappa+\lambda)\sinh{2\phi_2}\,\text{sech}\,{4\phi_3},\label{40SO(2)GenBPS6}\\
\phi_3'&=&-\frac{g}{4}e^{\phi_0-4(\phi_0+\phi_1)}\left((\kappa+\lambda)\cosh{2\phi_2}\sinh{4\phi_3}+(\kappa-\lambda)\cosh{4\phi_3}\right)\label{40SO(2)GenBPS7}
\end{eqnarray}
together with 
\begin{eqnarray}
\varsigma'_1&=&-\frac{ge^{\varphi+4\phi_3}}{2e^{4(\phi_0+\phi_1)}}\left[\varsigma_1\left(\kappa-\lambda+(\kappa+\lambda)\cosh{2\phi_2}\right)+\varsigma_2(\kappa+\lambda)\sinh{2\phi_2}\,\text{sech}\,{4\phi_3}\right],\nonumber\\ \label{40SO(2)GenBPS8}\\
\varsigma'_2&=&-\frac{ge^{\varphi-4\phi_3}}{2e^{4(\phi_0+\phi_1)}}\left[\varsigma_1(\kappa+\lambda)\sinh{2\phi_2}\,\text{sech}\,{4\phi_3}-\varsigma_2\left(\kappa-\lambda-(\kappa+\lambda)\cosh{2\phi_2}\right)\right],\nonumber\\\label{40SO(2)GenBPS9}\\
W'&=&-\frac{g}{4}e^{\phi_0-4(\phi_0+\phi_1)}(\kappa+\lambda)\sinh{2\phi_2}\tanh{4\phi_3}.\label{40SO(2)GenBPS10}
\end{eqnarray}
The solutions with $k=0$ will essentially imply $p=0$ leading to flat domain wall solutions. Therefore, we will not consider these solutions any further. We also note that in this case, the solutions with Killing spinors subject to the projectors \eqref{pureZProj} and \eqref{40_SO(2)Projcon} preserve eight supercharges or $N=4$ supersymmetry in three-dimensional field theories.
\\
\indent To solve these equations, we begin with a simple case of $CSO(2,0,2)\ltimes \mathbb{R}^{4}$ gauge group. In this case, we have $\kappa=\lambda=0$, and the BPS equations simplify considerably to
\begin{eqnarray}
U'&=&\frac{g}{4}e^{\varphi-4(\phi_0-\phi_1)}-\frac{p}{8}e^{-2V-\varphi+4(\phi_0-\phi_1)},\label{40_SO(2)_CSO202_BPS1}\\
V'&=&\frac{g}{4}e^{\varphi-4(\phi_0-\phi_1)}+\frac{3p}{8}e^{-2V-\varphi+4(\phi_0-\phi_1)},\label{40_SO(2)_CSO202_BPS2}\\
\varphi'&=&-\frac{g}{20}e^{\varphi-4(\phi_0-\phi_1)}+\frac{p}{40}e^{-2V-\varphi+4(\phi_0-\phi_1)},\label{40_SO(2)_CSO202_BPS3}\\
\phi'_0&=&\frac{g}{20}e^{\varphi-4(\phi_0-\phi_1)}-\frac{p}{40}e^{-2V-\varphi+4(\phi_0-\phi_1)},\label{40_SO(2)_CSO202_BPS4}\\
\phi_1'&=&-\frac{g}{4}e^{\varphi-4(\phi_0-\phi_1)}+\frac{p}{8}e^{-2V-\varphi+4(\phi_0-\phi_1)}\label{40_SO(2)_CSO202_BPS5}
\end{eqnarray}
together with $\phi_2'=\phi_3'=W'=\varsigma'_1=\varsigma'_2=0$. From these equations, we readily find the solutions for $U$, $\varphi$, and $\phi_0$ in terms of $\phi_1$
\begin{equation}
U=-\phi_1,\qquad \varphi=C+\frac{\phi_1}{5},\qquad \phi_0=C_0-\frac{\phi_1}{5}\, .\label{40_Sig2_SO(2)_CSO202_simsoln}
\end{equation}
Since $\phi_2$, $\phi_3$, $\varsigma_1$, $\varsigma_2$, and $W$ are constant, we will also choose $\phi_2=\phi_3=W=\varsigma_1=\varsigma_2=0$ for simplicity. We have also neglected an additive integration constant for the warp factor $U$. 
\\
\indent In order to find the other warp factor $V$, we consider a linear combination
\begin{equation}
V'+\phi'_1=\frac{p}{2}e^{-\varphi-2 V+4(\phi_0-\phi_1)}\, .
\end{equation}
Substituting \eqref{40_Sig2_SO(2)_CSO202_simsoln} into this equation and changing to a new radial coordinate $\rho$ defined by $\frac{d\rho}{dr}=e^{-3\phi_1}$, we find
\begin{equation}
V=\frac{1}{2}\ln \left[p\rho e^{4C_0}\right]-\phi_1-\frac{C}{2}\, .
\end{equation}
With all these results, we finally find the solution for $\phi_1$ of the form
\begin{equation}
\phi_1=\frac{C_0}{2}-\frac{1}{8}\ln \left[e^Cg \rho +\frac{C_1}{\rho }\right]
\end{equation}
in which $C_1$ is another integration constant. 
\\
\indent From the solution, we find that as $\rho\rightarrow +\infty$, the contribution of gauge fields in the BPS equations is small with 
\begin{equation}
U\sim V\qquad \textrm{and}\qquad \varphi\sim-\phi_0 \sim\frac{1}{40}\ln g\rho\, .
\end{equation}
This takes the form of a locally flat domain wall solution in the UV limit.
\\
\indent As $\rho\rightarrow 0$, we find the following asymptotic behavior, for $C_1\neq 0$,
\begin{equation}
U\sim-\phi_1\sim-\frac{1}{8}\ln \rho,\qquad V\sim\frac{3}{8}\ln \rho\qquad \varphi\sim-\phi_0\sim \frac{1}{40}\ln \rho\, .
\end{equation}
The effective potential \eqref{Sig2_effpot} is unbounded near the singularity. On the other hand, for $C_1=0$, we find 
\begin{equation}
U\sim-\phi_1\sim\frac{1}{8}\ln \rho,\qquad V\sim\frac{3}{8}\ln \rho\qquad \varphi\sim-\phi_0\sim -\frac{1}{40}\ln \rho\, .
\end{equation}
Near this singularity, the effective potential \eqref{Sig2_effpot} vanishes. 
\\
\indent Upon uplifting to type IIA theory, we find the metric component $\hat{g}_{00}$ near the IR singularity for $C_1\neq 0$,
\begin{equation}
\hat{g}_{00}\sim e^{-3\phi_0+2\varphi-\phi_1+2U}\rightarrow +\infty
\end{equation}
while for $C_1=0$, we find 
\begin{equation}
\hat{g}_{00}\sim e^{-3\phi_0+2\varphi+\phi_1+2U}\rightarrow 0\, . 
\end{equation}
Therefore, this IR singularity is physically acceptable for $C_1=0$, and the solution can be interpreted as holographic duals of an $N=4$ non-conformal field theory in three dimensions arising from a twisted compactification of five-dimensional field theory on a Riemann surface.
\\
\indent For other gauge groups, we will look for numerical solutions to the BPS equations. We first consider solutions with an asymptotic behavior given by a locally flat domain wall with $SO(4)$ symmetry of the form \begin{equation}\label{40_SO(4)_flat_DW_IR_asym}
U\sim V\sim-\ln \left[-\frac{g\rho}{2}\right]\qquad \textrm{and}\qquad \varphi\sim-\phi_0\sim \frac{1}{5}\ln \left[-\frac{g\rho}{2}\right]
\end{equation}
together with $\phi_2=\phi_3=\varsigma_1=\varsigma_2=0$. Solutions with this asymptotic geometry can only arise from $SO(4)\ltimes \mathbb{R}^4$ gauge group. As in the previous case, it turns out that all solutions in this gauge group lead to the uplifted ten-dimensional solutions with unphysical IR singularities, and we will not give numerical solutions for this case. 
\\
\indent Another class of solutions are given by those asymptotic to a locally $SO(2)$ symmetric flat domain wall of the form
\begin{eqnarray}\label{40_SO(2)_flat_DW_IR_asym}
& &U\sim V\sim\frac{1}{2}\ln \rho-\ln \left[-\frac{g\rho}{2}\right],\qquad  \phi_1\sim\frac{1}{2}\ln \rho\nonumber \\
& &\varphi\sim-\phi_0\sim\frac{1}{5}\ln \left[-\frac{g\rho}{2}\right]-\frac{1}{10}\ln \rho
\end{eqnarray}
with $\phi_2=\phi_3=\varsigma_1=\varsigma_2=0$. We find that all solutions interpolating between this locally flat domain wall and singular geometries in the IR have $\phi_2=\varsigma_1=\varsigma_2=0$ along the entire flow solutions. Furthermore, all solutions in $SO(4)\ltimes \mathbb{R}^4$ and $CSO(3,0,1)\ltimes \mathbb{R}^4$ gauge groups are uplifted to type IIA solutions with unphysical IR singularities, and we will omit these numerical solutions. For completeness, we also include these solutions in table \ref{tab10}.
\\
\indent For $SO(3,1)\ltimes \mathbb{R}^4$, $SO(2,2)\ltimes \mathbb{R}^4$, and $CSO(2,1,1)\ltimes \mathbb{R}^4$ gauge groups, there are solutions that lead to ten-dimensional solutions with physical singularities as shown in figures \ref{40_S2_SO(2)_SO(31)gg_flows} to \ref{40_H2_SO(2)_CSO(211)gg_flows}. In particular, all $Mkw_3\times S^2$-sliced solutions in these three gauge groups can be uplifted to solutions of type IIA theory with physical IR singularities. For $Mkw_3\times H^2$-sliced solutions, the solutions from $SO(3,1)\ltimes \mathbb{R}^4$ and $CSO(2,1,1)\ltimes \mathbb{R}^4$ gauge groups shown by the red and purple curves in figures \ref{40_H2_SO(2)_SO(31)gg_flows} and \ref{40_H2_SO(2)_CSO(211)gg_flows} together with the solution from $SO(2,2)\ltimes \mathbb{R}^4$ gauge group represented by the purple curve in figure \ref{40_H2_SO(2)_SO(22)gg_flows} also lead to ten-dimensional solutions with physical IR singularities. 
%%%%%%%%%%%%%%%%%%%%%%%%%%%%%%%%%%%%%%%%%
\subsection{D4-branes wrapped on a Riemannian three-manifold}\label{40_Sig3_Sec}
In this section, we consider solutions describing D4-branes wrapped on a Riemannian three-manifold $\Sigma^3$. The ansatz for the six-dimensional metric is given in \eqref{SO(3)6Dmetric}. We will consider two different twists by turning on $SO(3)$ and $SU(2)$ gauge fields. 
%%%%%%%%%%%%%%%%%%%%%%%%%%%%%%%%%%%%%%%%%%%%%%%%
\subsubsection{D4-branes wrapped on a Riemannian three-manifold with $SO(3)$ twist}\label{40_Sig3_SO(3)_section}
In the case of $SO(3)$ twist, we turn on the following gauge fields
\begin{equation}\label{40_Sig3_SO(3)_gaugeAnt}
{A_{\hat{\theta}}}^{12}=-\frac{e^{-V}p}{2\sqrt{2}k}\frac{f'_{k}(\psi)}{f_{k}(\psi)},\ \ \ {A_{\hat{\zeta}}}^{13}=-\frac{e^{-V}p}{2\sqrt{2}k}\frac{f'_{k}(\psi)}{f_{k}(\psi)}, \ \ \ {A_{\hat{\zeta}}}^{23}=-\frac{e^{-V}p}{2\sqrt{2}k}\frac{\cot\theta}{f_{k}(\psi)}\, .\
\end{equation}
There are three gauge groups with an $SO(3)$ subgroup. These are given by $SO(4)\ltimes \mathbb{R}^{4}$, $SO(3,1)\ltimes \mathbb{R}^{4}$, and $CSO(3,0,1)\ltimes \mathbb{R}^{4}$ which can be collectively characterized by the symmetric tensor
\begin{equation}\label{40_SO3_w}
w^{ij}=\textrm{diag}(1,1,1,\kappa)
\end{equation}
for $\kappa=1,-1,0$, respectively.
\vfil
\vfil
\vfil\pagebreak
\vfil
\vfil
\vfil
\clearpage\newpage

\begin{table}[h!]
\centering
\begin{tabular}{| c | c | c | c | c |}
\hline
UV flat&  \multirow{2}{*}{gauge group} & Riemann  & criterion \cite{Gubser_Sing}  & criterion \cite{Maldacena_nogo} \\
domain wall& & surface & ($\mathbf{V}_{\text{eff},\Sigma^3}\nrightarrow+\infty$) & ($\hat{g}_{00}\rightarrow0$) \\\hline
$SO(4)$ symmetric & $SO(4)\ltimes\mathbb{R}^4$& $S^2$ & any $g$ & $\times$ \\
&& $H^2$ & $g\leq-4.81$ & $\times$ \\\hline
$SO(2)$ symmetric &$SO(4)\ltimes\mathbb{R}^4$& $S^2$ & any $g$ & $\times$ \\
&& $H^2$ & $\times$ & $\times$ \\\cline{2-5}
&$SO(3,1)\ltimes\mathbb{R}^4$& $S^2$ & $g\leq-0.05$ & any $g$ \\
& & $H^2$ & $\times$ & $g\leq-1.23$ \\\cline{2-5}
&$SO(2,2)\ltimes\mathbb{R}^4$& $S^2$ & $\times$ & any $g$ \\
& & $H^2$ & $\times$ & $g\leq-4.22$ \\\cline{2-5}
&$CSO(3,0,1)\ltimes\mathbb{R}^4$& $S^2$ & any $g$ & $\times$ \\
& & $H^2$ & $\times$ & $\times$ \\\cline{2-5}
&$CSO(2,1,1)\ltimes\mathbb{R}^4$& $S^2$ & $g\geq-0.05$ & any $g$ \\
& & $H^2$ & $\times$ & $g\leq-1.4$ \\\cline{2-5}
&$CSO(2,0,2)$ & \multirow{2}{*}{$S^2$, $H^2$}& \multirow{2}{*}{$C_1=0$} & \multirow{2}{*}{$C_1=0$} \\
&(analytic) & &  &  \\\hline
\end{tabular}
\caption{Summary of satisfaction of criteria \cite{Gubser_Sing} and \cite{Maldacena_nogo} for the IR singularities of the $Mkw_3\times \Sigma^2$-sliced curved domain walls obtained from $SO(2)$ twist in $SO(4)\ltimes\mathbb{R}^4$, $SO(3,1)\ltimes\mathbb{R}^4$, $SO(2,2)\ltimes\mathbb{R}^4$, $CSO(3,0,1)\ltimes\mathbb{R}^4$, $CSO(2,1,1)\ltimes\mathbb{R}^4$, and $CSO(2,0,2)\ltimes\mathbb{R}^4$  gauge groups.}\label{tab10}
\end{table}

\begin{figure}[h!]
  \centering
    \includegraphics[width=\linewidth]{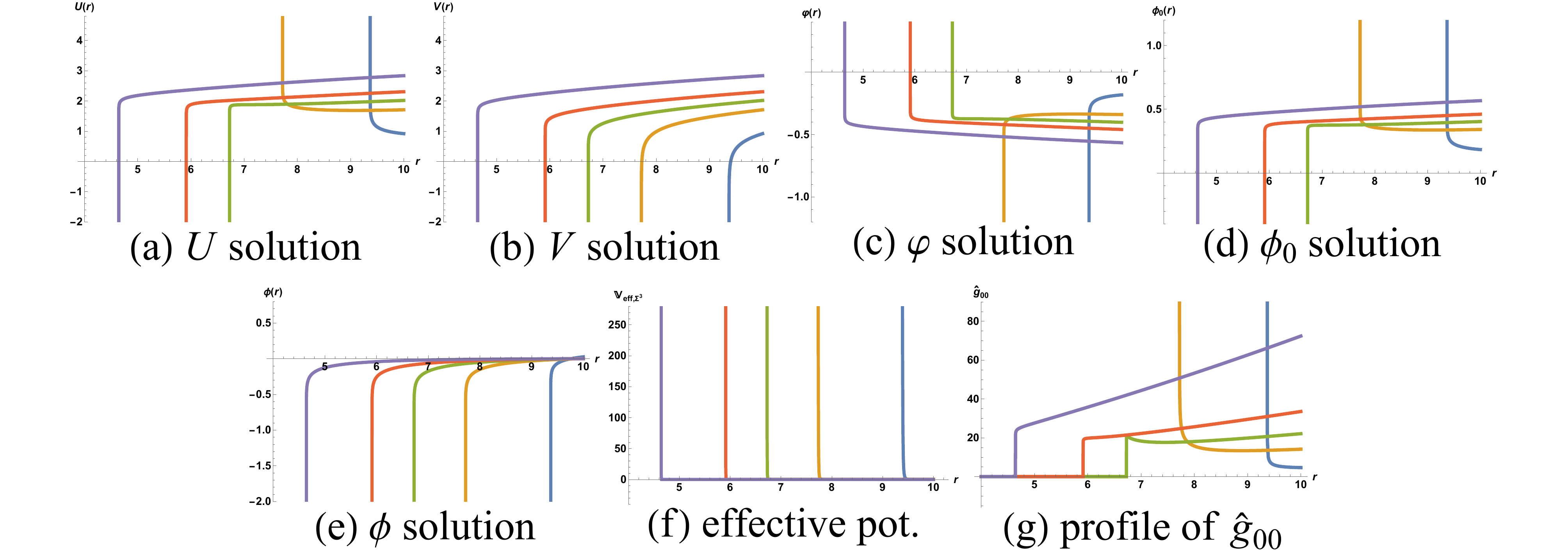}
\caption{Interpolating solutions between the locally $SO(4)$ flat domain wall as $\rho\rightarrow +\infty$ and $Mkw_2\times S^3$-sliced curved domain walls for $SO(3)$ twist in $SO(4)\ltimes \mathbb{R}^{4}$ gauge group. The blue, orange, green, red, and purple curves refer to $g=0.5, 1.1, 1.5, 2, 3.4$, respectively.}
\label{40_S3_UV_SO(3)_SO(4)gg_flows}
\end{figure}

\begin{figure}[h!]
  \centering
    \includegraphics[width=\linewidth]{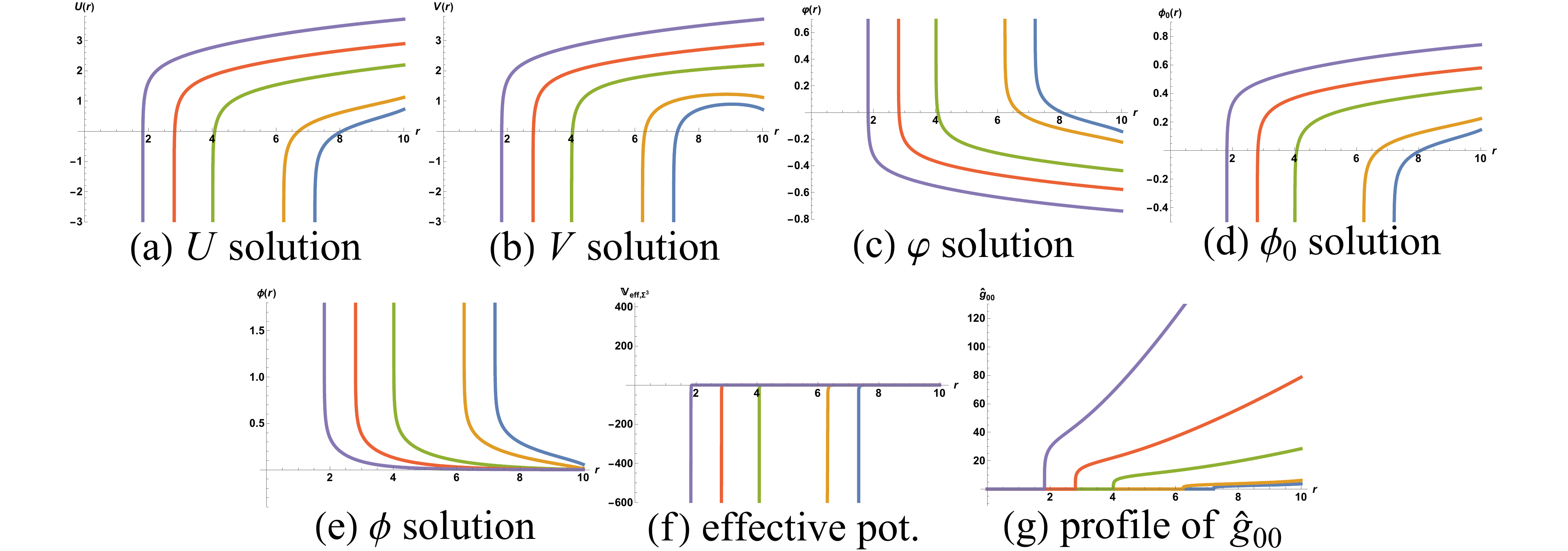}
\caption{Interpolating solutions between the locally $SO(4)$ flat domain wall as $\rho\rightarrow+\infty$ and $Mkw_2\times H^3$-sliced curved domain walls for $SO(3)$ twist in $SO(4)\ltimes \mathbb{R}^{4}$ gauge group. The blue, orange, green, red, and purple curves refer to $g=0.41, 0.61, 1.78, 3.60, 8.10$, respectively.}
\label{40_H3_UV_SO(3)_SO(4)gg_flows}
\end{figure}

\begin{figure}[h!]
  \centering
    \includegraphics[width=\linewidth]{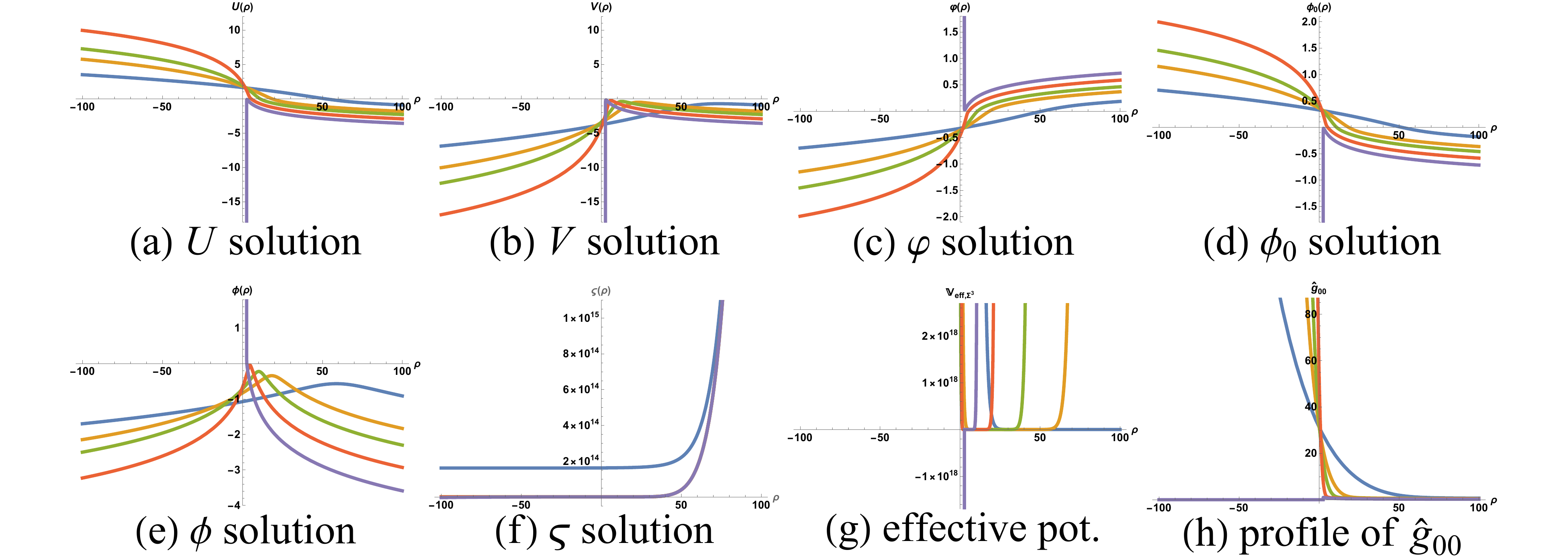}
\caption{Interpolating solutions between the locally $SO(3)$ flat domain wall as $\rho\rightarrow+\infty$ and $Mkw_2\times S^3$-sliced curved domain walls for $SO(3)$ twist in $SO(3,1)\ltimes \mathbb{R}^{4}$ gauge group. The blue, orange, green, red, and purple curves refer to $g=0.2, 0.5, 0.8, 1.5, 2.9$, respectively.}
\label{40_S3_SO(3)_SO(31)gg_flows}
\end{figure}

\begin{figure}[h!]
  \centering
    \includegraphics[width=\linewidth]{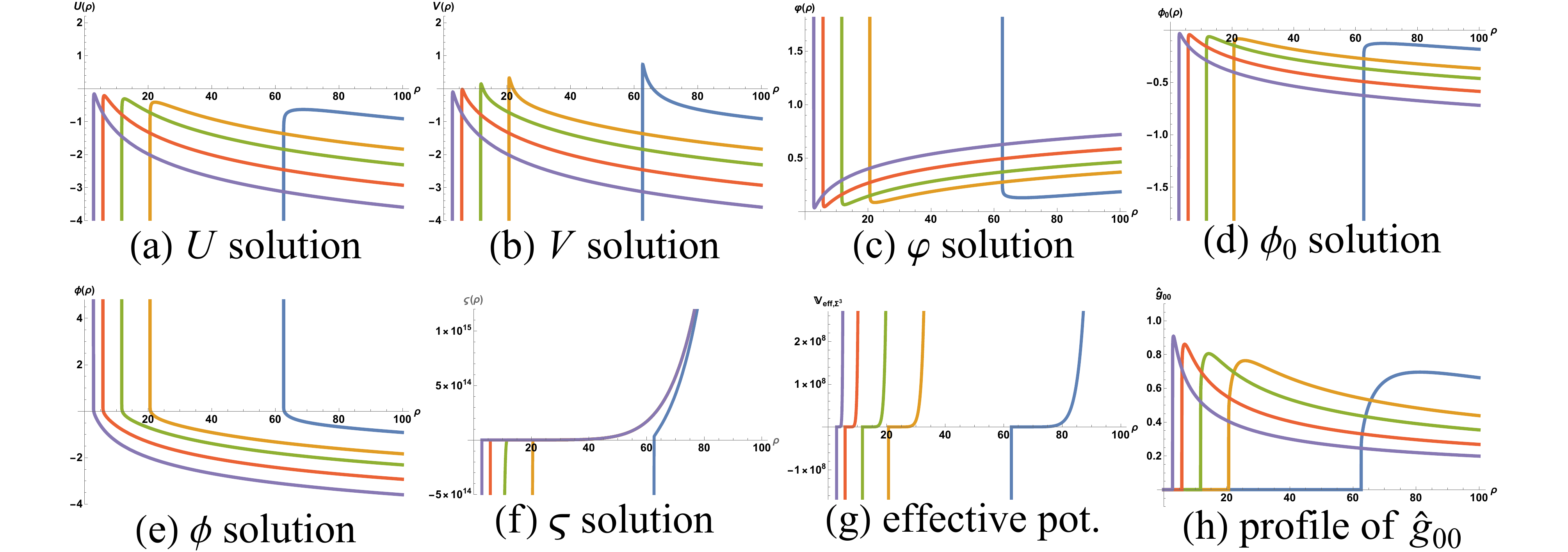}
\caption{Interpolating solutions between the locally $SO(3)$ flat domain wall as $\rho\rightarrow+\infty$ and $Mkw_2\times H^3$-sliced curved domain walls for $SO(3)$ twist in $SO(3,1)\ltimes \mathbb{R}^{4}$ gauge group. The blue, orange, green, red, and purple curves refer to $g=0.2, 0.5, 0.8, 1.5, 2.9$, respectively.}
\label{40_H3_SO(3)_SO(31)gg_flows}
\end{figure}

There are five $SO(3)$ singlet scalars consisting of the two dilatons and additional three scalars corresponding to the following non-compact generators 
\begin{equation}
\widetilde{\mathcal{Y}}_1=\hat{\boldsymbol{t}}^+_{1\dot{1}}+\hat{\boldsymbol{t}}^+_{2\dot{2}}+\hat{\boldsymbol{t}}^+_{3\dot{3}}-3\,\hat{\boldsymbol{t}}^+_{4\dot{4}},\qquad \widetilde{\mathcal{Y}}_2=\hat{\boldsymbol{t}}^+_{4\dot{5}},\qquad \widetilde{\mathcal{Y}}_3=\boldsymbol{s}_{45}\, .
\end{equation}
The coset representative then takes the form
\begin{equation}\label{40_SO(3)_coset}
V=e^{\varphi\boldsymbol{d}+\phi_0\widetilde{\mathcal{Y}}_0+\phi\widetilde{\mathcal{Y}}_1+b\widetilde{\mathcal{Y}}_2
+\varsigma\widetilde{\mathcal{Y}}_3}\, .
\end{equation}
It turns out that consistency of the BPS equations requires the axionic scalar $b$ to vanish. With $b=0$, the scalar potential takes the form
\begin{equation}\label{ZSO(3)Pot}
\mathbf{V}=-\frac{g^2}{16}e^{2(\varphi-4(\phi_0+3\phi))}\left(6\kappa e^{16\phi}+(9e^{32\phi}+\kappa^2)\right).
\end{equation}
\indent Imposing the projectors \eqref{SO(3)Projcon+}, \eqref{SO(3)Projcon-}, and \eqref{pureZProj} on the Killing spinors together with the twist condition $gp=k$, we find the following BPS equations
\begin{eqnarray}
U'&=&\frac{g}{8}e^{\varphi-4(\phi_0+3\phi)}(3e^{16\phi}+\kappa)-\frac{3p}{8}e^{-2V-\varphi+4(\phi_0-\phi)},\\
V'&=&\frac{g}{8}e^{\varphi-4(\phi_0+3\phi)}(3e^{16\phi}+\kappa)+\frac{5p}{8}e^{-2V-\varphi+4(\phi_0-\phi)},\\
\varphi'&=&-\frac{g}{40}e^{\varphi-4(\phi_0+3\phi)}(3e^{16\phi}+\kappa)+\frac{3p}{40}e^{-2V-\varphi+4(\phi_0-\phi)},\\
\phi'_0&=&\frac{g}{40}e^{\varphi-4(\phi_0+3\phi)}(3e^{16\phi}+\kappa)-\frac{3p}{40}e^{-2V-\varphi+4(\phi_0-\phi)},\\
\phi'&=&-\frac{g}{8}e^{\varphi-4(\phi_0+3\phi)}(e^{16\phi}-\kappa)+\frac{p}{8}e^{-2V-\varphi+4(\phi_0-\phi)},\\
\varsigma'&=&-g\kappa e^{\varphi-4(\phi_0+3\phi)}\varsigma\, .
\end{eqnarray}
As in the previous section, the solutions preserve $N=(2,2)$ supersymmetry in two dimensions.
\\
\indent 
For $CSO(3,0,1)\ltimes \mathbb{R}^{4}$ gauge group, we can analytically solve the BPS equations by the same procedure as in the case of $CSO(2,0,2)\ltimes \mathbb{R}^4$ gauge group. The resulting solution is given by
\begin{eqnarray}
U&=&-3\phi,\qquad\qquad V=\frac{1}{2}\ln\left[p\rho e^{4C_0}\right] -3\phi-\frac{C}{2},\qquad\quad\varphi=C+\frac{3\phi}{5},\nonumber\\
\qquad\ \phi_0&=&C_0-\frac{3\phi}{5},\qquad\ \phi=\frac{C_0}{2}-\frac{1}{8}\ln \left[\frac{2(e^{C} g \rho\sqrt{\rho} +C_1)}{3\sqrt{\rho}}\right],\quad\ \varsigma=0.\quad
\end{eqnarray}
Apart from some numerical factors, this solution is very similar to that of $CSO(2,0,2)\ltimes \mathbb{R}^4$ gauge group. As $\rho\rightarrow +\infty$, we again find a locally flat domain wall geometry
while as $\rho\rightarrow 0$, the solution becomes
\begin{equation}
U\sim -3\phi,\quad \varphi\sim -\phi_0\sim \frac{3}{5}\phi,\quad V\sim -3\phi+\frac{1}{2}\ln\rho,\quad 
\phi\sim \frac{1}{16}\ln\rho
\end{equation}
for $C_1\neq 0$. The effective potential \eqref{Sig3_effpot} is unbounded above, and the type IIA metric component gives
\begin{equation}
\hat{g}_{00}\sim e^{-3\phi_0-\phi+2\varphi+2U}\rightarrow +\infty\, .
\end{equation}
Therefore, the solution has an unphysical singularity.
\\
\indent For $C_1=0$, we find, as $\rho\rightarrow 0$,
\begin{equation}
U\sim -3\phi,\quad \varphi\sim -\phi_0\sim \frac{3}{5}\phi,\quad V\sim -3\phi+\frac{1}{2}\ln\rho,\quad 
\phi\sim -\frac{1}{8}\ln\rho
\end{equation}
In this case, the effective potential \eqref{Sig3_effpot} vanishes, and the type IIA metric component gives
\begin{equation}
\hat{g}_{00}\sim e^{-3\phi_0-\phi+2\varphi+2U}\rightarrow 0\, .
\end{equation}
Accordingly, the singularity at $\rho=0$ is physical for $C_1=0$, and the solution can be interpreted as a twisted compactification of five-dimensional field theory on $\Sigma^3$ to two-dimensional non-conformal field theory in the IR.
\\
\indent For other gauge groups, we will find numerical solutions to the BPS equations. For $SO(4)\ltimes \mathbb{R}^{4}$ gauge group with $\kappa=1$, there are solutions with an asymptotic geometry as $r\rightarrow +\infty$ given by a locally $SO(4)$ flat domain wall
\begin{equation}
U\sim V\sim\ln \left[\frac{g r}{2}\right],\qquad \varphi\sim-\phi_0\sim-\frac{1}{5}\ln \left[\frac{g r}{2}\right],\qquad \phi\sim\frac{16}{g^4r^4},\qquad \varsigma\sim 0\, .\label{40_SO(5)_fDW_asym_fromSO(3)}
\end{equation}
Examples for this type of solutions are shown in figures \ref{40_S3_UV_SO(3)_SO(4)gg_flows} and \ref{40_H3_UV_SO(3)_SO(4)gg_flows} for different values of $g$. In these figures, we have omitted the solutions for $\varsigma$ since in all cases $\varsigma$ identically vanishes along the flows. We find that all $Mkw_2\times H^3$-sliced solutions admit physical IR singularities in the frameworks of both six-dimensional gauged supergravity and type IIA theory. For $Mkw_2\times S^3$-sliced domain walls, the solutions represented by the red, green, and purple curves in figure \ref{40_S3_UV_SO(3)_SO(4)gg_flows} can be uplifted to physical solutions in type IIA theory. 
\vfil
\vfil
\vfil\pagebreak
Another type of solution is characterized by the following asymptotic geometry as $\rho\rightarrow +\infty$ 
\begin{equation}
U\sim V\sim\phi\sim -\ln \left[-\frac{ g\kappa \rho}{8}\right],\qquad \varphi\sim-\phi_0\sim\frac{1}{5} \ln \left[-\frac{ g\kappa \rho}{8}\right],\qquad\ \varsigma\sim\rho^8\label{40_SO(3)_flat_DW_asym}
\end{equation}
with the new radial coordinate $\rho$ defined by $\frac{d\rho}{dr}=e^{-V-5\phi}$. For $SO(4)\ltimes \mathbb{R}^4$ gauge group, all solutions lead to unphysical ten-dimensional solutions, so we will not give numerical solutions in this case. For $SO(3,1)\ltimes \mathbb{R}^4$ gauge group, examples of numerical solutions are shown in figures \ref{40_S3_SO(3)_SO(31)gg_flows} to \ref{40_H3_SO(3)_SO(31)gg_flows} for different values of $g$. In this case, all $Mkw_2\times H^3$-sliced domain walls can be uplifted to type IIA solutions with physical IR singularities. For $Mkw_2\times S^3$-sliced domain walls, only the solution represented by the purple curve in figure \ref{40_S3_SO(3)_SO(31)gg_flows} leads to a physical ten-dimensional solution, see also table \ref{tab11}.

\begin{table}[h!]
\centering
\begin{tabular}{| c | c | c | c | c |}
\hline
UV flat&  \multirow{2}{*}{gauge group} & Riemannian  & criterion \cite{Gubser_Sing}  & criterion \cite{Maldacena_nogo} \\
domain wall& & three-manifold  & ($\mathbf{V}_{\text{eff},\Sigma^3}\nrightarrow+\infty$) & ($\hat{g}_{00}\rightarrow0$) \\\hline
$SO(4)$ symmetric & $SO(4)\ltimes\mathbb{R}^4$& $S^3$ & $\times$ & $g\geq1.5$ \\
&& $H^3$ & any $g$ & any $g$ \\\hline
$SO(3)$ symmetric &$SO(4)\ltimes\mathbb{R}^4$& $S^3$ & any $g$ & $\times$ \\
&& $H^3$ & $\times$ & $\times$ \\\cline{2-5}
&$SO(3,1)\ltimes\mathbb{R}^4$ & $S^3$& $g\geq2.9$ & $g\geq2.9$ \\
& & $H^3$& any $g$ & any $g$ \\\cline{2-5}
&$CSO(3,0,1)$ &  \multirow{2}{*}{$S^3$, $H^3$} &  \multirow{2}{*}{$C_1=0$} &  \multirow{2}{*}{$C_1=0$} \\
&$\ltimes\mathbb{R}^4$ (analytic) & &  &  \\\hline
\end{tabular}
\caption{Summary of satisfaction of criteria \cite{Gubser_Sing} and \cite{Maldacena_nogo} for the $Mkw_2\times \Sigma^3$-sliced curved domain walls obtained from $SO(3)$ twist in $SO(4)\ltimes\mathbb{R}^4$, $SO(3,1)\ltimes\mathbb{R}^4$, and $CSO(3,0,1)\ltimes\mathbb{R}^4$ gauge groups.}\label{tab11}
\end{table}

%%%%%%%%%%%%%%%%%%%%%%
\subsubsection{D4-branes wrapped on a Riemannian three-manifold with $SU(2)$ twist}\label{40_Sig3_SO(3)sd_section}
As a final example, we consider another twist on $\Sigma^3$ by turning on the gauge fields of $SU(2)$ identified with the self-dual $SU(2)_+\subset SO(4)\sim SU(2)_+\times SU(2)_-$ gauge symmetry. Only the $SO(4)\ltimes \mathbb{R}^{4}$ gauge group with the embedding tensor given by $w^{ij}=\text{diag}(1,1,1,1)$ contains this $SU(2)_+$ subgroup.
\\
\indent There are only two $SU(2)_+$ singlet scalars given by the two dilatons parametrized by the coset representative 
\begin{equation}
V=e^{\varphi\boldsymbol{d}+\phi_0\widetilde{\mathcal{Y}}_0}
\end{equation}
which leads to the scalar potential
\begin{equation}\label{ZSO(4)Pot}
\mathbf{V}=-g^2e^{2\varphi-8\phi_0}\, .
\end{equation}
We perform a topological twist by turning on the following gauge fields
\begin{eqnarray}
{A_{\hat{\theta}}}^{12}={A_{\hat{\theta}}}^{34}&=& -\frac{e^{-V}p}{4\sqrt{2}k}\frac{f'_{k}(\psi)}{f_{k}(\psi)},\nonumber \\ {A_{\hat{\zeta}}}^{13}={A_{\hat{\zeta}}}^{42}&=& -\frac{e^{-V}p}{4\sqrt{2}k}\frac{f'_{k}(\psi)}{f_{k}(\psi)}, \nonumber \\ {A_{\hat{\zeta}}}^{23}={A_{\hat{\zeta}}}^{14}&=& -\frac{e^{-V}p}{4\sqrt{2}k}\frac{\cot\theta}{f_{k}(\psi)}\, .
\end{eqnarray}
With the projectors \eqref{SO(3)Projcon+}, \eqref{SO(3)Projcon-}, \eqref{GammaSDProj}, and \eqref{pureZProj} together with the twist condition $gp=k$, the resulting BPS equations read
\begin{eqnarray}
U'&=&\frac{g}{2}e^{\varphi-4\phi_0}-\frac{3p}{8}e^{-2V-\varphi+4\phi_0},\\ 
V'&=&\frac{g}{2}e^{\varphi-4\phi_0}+\frac{5p}{8}e^{-2V-\varphi+4\phi_0},\\ 
\varphi'&=&-\frac{g}{10}e^{\varphi-4\phi_0}+\frac{3p}{40}e^{-2V-\varphi+4\phi_0},\\ 
 \phi'_0&=& \frac{g}{10}e^{\varphi-4\phi_0}-\frac{3p}{40}e^{-2V-\varphi+4\phi_0}\, .
\end{eqnarray}
From these equations, we again find that 
\begin{equation}
U=5\phi_0\qquad \textrm{and}\qquad \varphi=C-\phi_0\, . 
\end{equation}
By taking a linear combination $V'-5\phi_0'$ and defining a new radial coordinate $\rho$ by $\frac{d\rho}{dr}=e^{-V}$, we obtain
\begin{equation}
V=5\phi_0-C+\ln \left[\frac{p\rho +C_1}{g}\right].
\end{equation}
The integration constant $C_1$ can be set to zero by shifting the coordinate $\rho$. Substituting all the results into $ \phi'_0$ equation, we find, after changing to the coordinate $\rho$,
\begin{equation}
\phi_0=\frac{1}{20}g p \rho^2-\frac{3}{40} \ln \rho+C_2\, .
\end{equation}
This solution has an asymptotic behavior for $\rho\rightarrow +\infty$ as a locally flat domain wall with
\begin{equation}
U\sim V\sim 5\rho^2,\qquad \varphi\sim -\phi_0,\qquad \phi_0\sim \rho^2\, .
\end{equation}
As $\rho\rightarrow 0$, we find
\begin{eqnarray}
& &\phi_0\sim -\frac{3}{40}\ln\rho,\qquad \varphi\sim -\phi_0\sim \frac{3}{40}\ln \rho,\nonumber \\
& &U\sim 5\phi_0\sim -\frac{3}{8}\ln \rho,\qquad V\sim \ln \rho-\frac{3}{8}\ln\rho\, .
\end{eqnarray}
This singularity is however unphysical as can be seen by considering the metric component $\hat{g}_{00}$ in type IIA theory
\begin{equation}
\hat{g}_{00}\sim e^{-3\phi_0+2\varphi+2U}\sim \rho^{-\frac{3}{8}}\rightarrow +\infty\, .
\end{equation}

%%%%%%%%%%%%%%%%%%%%%%%%%%%%%%%%%%%%%%%%%%%%%%%%%%%%%%%%%%%%%%%%%%%%%%%%%%%%%%%%%%%%%%%%%%%%%%%%%%%%%%%%%%%%%%%%%%%%%%%%%%%%%%%%%%%%%%%%%
\section{Conclusions and discussions}\label{conclusion_sec}
We have found a large number of supersymmetric solutions interpolating between locally flat domain walls and $Mkw_3\times \Sigma^2$/$Mkw_2\times \Sigma^3$-sliced domain walls from maximal gauged supergravity in six dimensions with $CSO(p,q,5-p-q)$ and $CSO(p,q,4-p-q)\ltimes \mathbb{R}^4$ gauged groups. Both of these gauge groups can be embedded in string/M-theory although the complete truncation ansatze have not been worked out to date. However, with some partial results from exceptional field theory (EFT), it is possible to study the behavior of the metric components $\hat{g}_{00}$ near the IR singularities. We have found that upon uplifted to type IIA theory, many solutions found in this paper admit physically acceptable singularities. Therefore, these solutions describe holographic RG flows across dimensions from five-dimensional field theories to non-conformal field theories in three/two dimensions and are expected to arise from D4-branes wrapped on $\Sigma^2$ and $\Sigma^3$. 
\\
\indent It is interesting to construct the complete truncation ansatze of type IIA theory on both $H^{p,q}\times \mathbb{R}^{5-p-q}$ and $H^{p,q}\times \mathbb{R}^{4-p-q}\times S^1$. This would allow to uplift the solutions found here to explicit solutions of type IIA theory in which the interpretation in terms of various brane configurations involving D4-branes wrapped on $\Sigma^2$ and $\Sigma^3$ is possible. Identifying the dual field theories in both five and three/two dimensions participating in these RG flows across dimensions could also be of particular interest. With some of the solutions given analytically, we hope the present results could provide a very useful tool to further study along these and related directions. Extending the present study to the case of D4-branes wrapped on spindles and topological disks as in the recent results \cite{non-confornal_brane_spindle,non-conformal_brane_Minwoo} is also worth considering.   
\begin{acknowledgments}
P. N. is supported by the faculty research grants from Faculty of Science, Ramkhamhaeng University.
\end{acknowledgments}

\appendix
\section{Useful formulae}\label{AppA}
For convenience, in this appendix, we collect some formulae and relations used in the main text; see \cite{6D_DW_I} for more details. With the $SO(5,5)\rightarrow GL(5)$ branching rule for an $SO(5,5)$ vector representation
\begin{equation}\label{VecDec}
\mathbf{10}\ \rightarrow\ \mathbf{5}^{+2}\,\oplus\,\overline{\mathbf{5}}^{-2},
\end{equation}
we can decompose the $SO(5,5)$ generators as 
\begin{equation}
\boldsymbol{t}_{MN}\ \rightarrow\ (\boldsymbol{t}_{mn},{\boldsymbol{t}^m}_n,\boldsymbol{t}^{mn})
\end{equation}
with ${\boldsymbol{t}_m}^n=-{\boldsymbol{t}^n}_m$. 
\\
\indent In vector representation, $SO(5,5)$ generators $\boldsymbol{t}_{MN}=\boldsymbol{t}_{[MN]}$ can be chosen as
\begin{equation}
{(\boldsymbol{t}_{MN})_P}^Q\ =\ 4\eta_{P[M}\delta^Q_{N]}
\end{equation}
satisfying the Lie algebra 
\begin{equation}\label{SO(5,5)algebra}
\left[\boldsymbol{t}_{MN},\boldsymbol{t}_{PQ}\right]\ =\ 4(\eta_{M[P}\boldsymbol{t}_{Q]N}-\eta_{N[P}\boldsymbol{t}_{Q]M}).
\end{equation}
With the generators of the shift and hidden symmetries respectively denoted by $\boldsymbol{s}_{mn}=\boldsymbol{t}_{mn}$ and $\boldsymbol{h}^{mn}=\boldsymbol{t}^{mn}$, the $SO(5,5)$ generators can be written as
\begin{equation}
{(\boldsymbol{t}_{MN})_P}^Q=\begin{pmatrix} {\boldsymbol{t}^m}_n & \boldsymbol{h}^{mn} \\
								\boldsymbol{s}_{mn} & -{\boldsymbol{t}^n}_m \end{pmatrix}.
\end{equation}
In this form, we can see that the $GL(5)$ subgroup generated by ${\boldsymbol{t}^m}_n$ is embedded diagonally. 
\\
\indent The branching rules for spinor and conjugate spinor representations of $SO(5,5)$
\begin{equation}
\mathbf{16}_s\ \rightarrow\ \overline{\mathbf{5}}^{+3}\,\oplus\,\mathbf{10}^{-1}\,\oplus\,\mathbf{1}^{-5}\qquad\text{and}\qquad\mathbf{16}_c\ \rightarrow\ \mathbf{5}^{-3}\,\oplus\,\overline{\mathbf{10}}^{+1}\,\oplus\,\mathbf{1}^{+5}
\end{equation}
are realized respectively by the following relations
\begin{equation}
\Psi_A\ =\ \mathbb{T}_{Am}\Psi^m+\mathbb{T}_{A}^{mn}\Psi_{mn}+\mathbb{T}_{A\ast}\Psi_\ast
\end{equation}
with $\Psi_{mn}=\Psi_{[mn]}$ and
\begin{equation}
\Psi^A\ =\ \mathbb{T}^{Am}\Psi_m+\mathbb{T}^{A}_{mn}\Psi^{mn}+\mathbb{T}^{A}_{\ast}\Psi_\ast\, .
\end{equation}
The transformation matrices are defined by
\begin{eqnarray}
\mathbb{T}_{Am}&=&\frac{1}{2\sqrt{2}}(\Gamma_m)_{AB}\boldsymbol{p}^B_{\alpha\beta}\Omega^{\alpha\beta},\label{TranMatTDef1}\\
\mathbb{T}_{A}^{mn}&=&\frac{1}{4\sqrt{2}}{(\Gamma^{mn})_A}^B\boldsymbol{p}_B^{\alpha\beta}\Omega_{\alpha\beta},\label{TranMatTDef2}\\
\mathbb{T}_{A\ast}&=&\frac{1}{10}{({\Gamma^m}_m)_A}^B\boldsymbol{p}_B^{\alpha\beta}\Omega_{\alpha\beta}\label{TranMatTDef3}
\end{eqnarray}
in which the inverse matrices of $\mathbb{T}_A$ are simply given by their complex conjugation $\mathbb{T}^A=(\mathbb{T}_A)^{-1}=(\mathbb{T}_A)^*$ satisfying the relations
\begin{eqnarray}
\mathbb{T}^{Am}\mathbb{T}_{An}\ =\ \delta^m_n,\qquad\quad\mathbb{T}^{A}_{mn}\mathbb{T}_{A}^{pq}& =& \delta^{[p}_{m}\delta^{q]}_{n},\qquad\mathbb{T}^{A}_\ast\mathbb{T}_{A\ast}\ =\ 1,\nonumber\\
\mathbb{T}^{Am}\mathbb{T}_{Anp}\ =\ 0,\qquad\quad\,\mathbb{T}^{Am}\mathbb{T}_{A\ast}& =& 0,\qquad\quad\mathbb{T}^{A}_{mn}\mathbb{T}_{A\ast}\ =\ 0\label{TT_rel}
\end{eqnarray}
together with
\begin{equation}
\mathbb{T}^{Am}\mathbb{T}_{Bm}+\mathbb{T}^{A}_{mn}\mathbb{T}_{B}^{mn}+\mathbb{T}^{A}_\ast\mathbb{T}_{B\ast}=\delta^A_B\, .
\end{equation}
\indent The transformation matrices $\boldsymbol{p}_A^{\alpha\beta}$ and inverse matrices $\boldsymbol{p}^A_{\alpha\beta}$ are given by
\begin{eqnarray}\label{thepmatrix}
\boldsymbol{p}_A^{\alpha\beta}&=&\delta_A^\alpha\delta_1^{\beta}+\delta_A^{\alpha+4}\delta_2^{\beta}+\delta_A^{\alpha+8}
\delta_3^{\beta}+\delta_A^{\alpha+12}\delta_4^{\beta},\nonumber\\
\boldsymbol{p}^A_{\alpha\beta}&=&\delta^A_\alpha\delta^1_{\beta}+\delta^A_{\alpha+4}\delta^2_{\beta}
+\delta^A_{\alpha+8}\delta^3_{\beta}+\delta^A_{\alpha+12}\delta^4_{\beta}\, .
\end{eqnarray}
These matrices satisfy the relations 
\begin{equation}
\boldsymbol{p}_A^{\alpha\beta}\boldsymbol{p}^B_{\alpha\beta}\ =\ \delta_A^B\qquad\text{and}\qquad\boldsymbol{p}_A^{\alpha\delta}\boldsymbol{p}^A_{\beta\gamma}\ =\ \delta^{\alpha}_{\beta}\delta^{\delta}_{\gamma}
\end{equation}
and can be used to express a spinor index of $SO(5,5)$, $A,B,\ldots$, in terms of a pair of $USp(4)$ fundamental or $SO(5)$ spinor indices $\alpha\beta$.
\\
\indent The matrices $(\Gamma_m)_{AB}$ appearing in the $SO(5,5)$ gamma matrices $\Gamma_M=\left(\Gamma_m,\Gamma^m\right)$ are related to $SO(5)\times SO(5)$ gamma matrices $\Gamma_{\ul{A}}=(\Gamma_a,\Gamma_{\dot{a}})$ as
\begin{equation}
(\Gamma_M)_{AB}={{\mathbb{M}}_M}^{\underline{A}}(\Gamma_{\underline{A}})_{AB}
\end{equation}
with
\begin{equation}
(\Gamma_{\underline{A}})_{AB}={(\Gamma_{\underline{A}})_{A}}^{C}\boldsymbol{c}_{CB}\, .
\end{equation}
The charge conjugation matrix $\boldsymbol{c}_{AB}$ is given by
\begin{equation}
\boldsymbol{c}_{AB}=\boldsymbol{p}_{A}^{\alpha\dot{\alpha}}\boldsymbol{p}_B^{\beta\dot{\beta}}
\Omega_{\alpha\beta}\Omega_{\dot{\alpha}\dot{\beta}}
\end{equation}
in which $\boldsymbol{p}_{A}^{\alpha\dot{\alpha}}$ and $\boldsymbol{p}^{A}_{\alpha\dot{\alpha}}$ are defined in parallel with \eqref{thepmatrix}. For convenience, we also note the transformation matrix $\mathbb{M}$ here
\begin{equation}\label{offDiagTrans}
\mathbb{M}\ =\ \frac{1}{\sqrt{2}}\begin{pmatrix} \mathds{1}_5 & \mathds{1}_5 \\ \mathds{1}_5 & -\mathds{1}_5 \end{pmatrix}.
\end{equation}
\indent The $SO(5)\times SO(5)$ gamma matrices can in turn be written as 
\begin{equation}
{(\Gamma_{\underline{A}})_{A}}^{B}=\left({(\gamma_a)_{A}}^{B},{(\gamma_{\dot{a}})_{A}}^{B}\right)
\end{equation}
in which ${(\gamma_a)_{A}}^{B}$ and ${(\gamma_{\dot{a}})_{A}}^{B}$ are explicitly given by
\begin{eqnarray}
{(\gamma_a)_{A}}^{B}&=&\boldsymbol{p}_A^{\alpha\dot{\alpha}}{(\gamma_a)_{\alpha\dot{\alpha}}}^{\beta\dot{\beta}}\boldsymbol{p}^B_{\beta\dot{\beta}}\ = \ \boldsymbol{p}_A^{\alpha\dot{\alpha}}{(\gamma_a)_\alpha}^\beta\delta_{\dot{\alpha}}^{\dot{\beta}}\boldsymbol{p}^B_{\beta\dot{\beta}},\\
{(\gamma_{\dot{a}})_{A}}^{B}&=&\boldsymbol{p}_A^{\alpha\dot{\alpha}}{(\gamma_{\dot{a}})_{\alpha\dot{\alpha}}}^{\beta\dot{\beta}}\boldsymbol{p}^B_{\beta\dot{\beta}}\ = \ \boldsymbol{p}_A^{\alpha\dot{\alpha}}\delta_\alpha^\beta{(\gamma_{\dot{a}})_{\dot{\alpha}}}^{\dot{\beta}}
\boldsymbol{p}^B_{\beta\dot{\beta}}\, .
\end{eqnarray}
With all these, the matrices $(\Gamma_m)_{AB}$ can be explicitly defined in terms of the $SO(5)$ gamma matrices as
\begin{equation}\label{GammEx}
(\Gamma_m)_{AB}=\frac{1}{\sqrt{2}}\boldsymbol{p}_A^{\alpha\dot{\alpha}}\left[{(\gamma_m)_\alpha}^\beta\delta_{\dot{\alpha}}^{\dot{\beta}}
+\delta_\alpha^\beta{(\gamma_m)_{\dot{\alpha}}}^{\dot{\beta}}\right]\Omega_{\beta\gamma}\Omega_{\dot{\beta}\dot{\gamma}}
\boldsymbol{p}_B^{\gamma\dot{\gamma}}\, .
\end{equation}
\indent Finally, ${(\Gamma^{mn})_A}^B$ and ${({\Gamma^m}_m)_A}^B$ matrices are given by
\begin{equation}
{(\Gamma^{mn})_A}^B=\frac{1}{2}\left[{(\Gamma^m)_A}^{C}{(\Gamma^n)_{C}}^B-{(\Gamma^n)_A}^{C}{(\Gamma^m)_{C}}^B\right]
\end{equation}
and
\begin{equation}\label{Gammamm}
{({\Gamma^m}_m)_A}^B=\frac{1}{2}\left[{(\Gamma^m)_A}^{C}{(\Gamma_m)_{C}}^B-{(\Gamma_m)_A}^{C}{(\Gamma^m)_{C}}^B\right]
\end{equation}
with
\begin{eqnarray}\label{GammUpDown}
{(\Gamma_m)_A}^B&=&\frac{1}{\sqrt{2}}\boldsymbol{p}_A^{\alpha\dot{\alpha}}\left[{(\gamma_m)_\alpha}^\beta\delta_{\dot{\alpha}}^{\dot{\beta}}
+\delta_\alpha^\beta{(\gamma_m)_{\dot{\alpha}}}^{\dot{\beta}}\right]\boldsymbol{p}^B_{\beta\dot{\beta}},\\
{(\Gamma^m)_A}^B&=&\frac{1}{\sqrt{2}}\boldsymbol{p}_A^{\alpha\dot{\alpha}}\left[{(\gamma_m)_\alpha}^\beta\delta_{\dot{\alpha}}^{\dot{\beta}}
-\delta_\alpha^\beta{(\gamma_m)_{\dot{\alpha}}}^{\dot{\beta}}\right]\boldsymbol{p}^B_{\beta\dot{\beta}}\, .
\end{eqnarray}
In this work, we use the following representation for $SO(5)$ gamma matrices
\begin{eqnarray}
\gamma_1&=&-\sigma_2\otimes\sigma_2,\qquad\, \gamma_2\ =\ \mathds{1}_2\otimes\sigma_1,\qquad
\gamma_3\ =\ \mathds{1}_2\otimes\sigma_3,\nonumber\\ 
\gamma_4&=&\sigma_1\otimes\sigma_2,\qquad \gamma_5\ =\ \sigma_3\otimes\sigma_2
\end{eqnarray}
in which $\sigma_1$, $\sigma_2$, and $\sigma_3$ are the usual Pauli matrices given by
\begin{equation}
\sigma_1\ = \ \begin{pmatrix} 0 & 1\\ 1 & 0 \end{pmatrix},\qquad \sigma_2\ = \ \begin{pmatrix} 0 & -i\\ i & 0 \end{pmatrix},\qquad\sigma_3\ = \ \begin{pmatrix} 1 & 0\\ 0 & -1 \end{pmatrix}\, .
\end{equation}

\section{Effective potential}\label{AppB}
In this appendix, we derive the explicit form of the four- and three-dimensional effective scalar potentials arising from the compactifications of the maximal gauged supergravity in six dimensions on $\Sigma^2$ and $\Sigma^3$. The analysis closely follows \cite{Maldacena_nogo}. For convenience, we repeat the relevant six-dimensional action here
\begin{equation}
S=\int d^6x\sqrt{-g}\left(\frac{1}{4}R-\frac{1}{16}{P_{\mu}}^{a\dot{a}}{P^\mu}_{a\dot{a}}-\mathbf{V}-\frac{1}{4}M_{AB}{\mathcal{H}_{\mu\nu}}^{A}\mathcal{H}^{\mu\nu, B}\right)
\end{equation}
where $\sqrt{-g}=e=\det(e^{\hat{\mu}}_\mu)$. 

\subsection{Compactification on $\Sigma^2$}
We first consider a compactification to four-dimensional gauged supergravity on a Riemann surface with the six-dimensional metric given by \eqref{Sig_2_6D_metric}. With the choice of vielbein \eqref{AdS5xSigma2bein}, we find
\begin{eqnarray}
\sqrt{-g}&=&e^{3U+2V}f_k(\theta),\\
R&=&2ke^{-2V}-2(3U''+2V''+6U'^2+6U'V'+3V'^2).
\end{eqnarray}
With all the bosonic fields' ansatz, we can write the above action in the following form
\begin{eqnarray}
S&=&\int d^6xe^{3U+2V}f_k(\theta)\left[ \frac{1}{4}\left\{2ke^{-2V}-2(3U''+2V''+6U'^2+6U'V'+3V'^2)\right\}\right.\nonumber\\&&\qquad\qquad\qquad\qquad\left.-\frac{1}{16}{P_{r}}^{a\dot{a}}{P^r}_{a\dot{a}}-\mathbf{V}-\frac{1}{4}M_{AB}{\mathcal{H}_{\theta\zeta}}^{A}\mathcal{H}^{\theta\zeta, B}\right]\nonumber\\
&\sim&\int d^4xe^{2V}\sqrt{-g_{\text{eff}}}\left[ \frac{1}{4}R(g_{\text{eff}})+\frac{k}{2}e^{-2V}-\frac{g^{rr}}{16}{P_{r}}^{a\dot{a}}{P_{r}}^{a\dot{a}}-\mathbf{V}-\frac{1}{4}M_{AB}{\mathcal{H}_{\theta\zeta}}^{A}\mathcal{H}^{\theta\zeta, B}\right].\nonumber\\
\end{eqnarray}
In the second line, we have neglected the integration on $\Sigma^2$ coordinates $(\theta,\zeta)$ leading to the volume of $\Sigma^2$, $\int f_k(\theta)d\theta d\zeta=Vol_{\Sigma^2}$. We have also defined the following quantities 
\begin{equation}
\sqrt{-g_{\text{eff}}}=e^{3U}\qquad\text{ and }\qquad R(g_{\text{eff}})=-2(3U''+2V''+6U'^2+6U'V'+3V'^2).
\end{equation}
\indent Finally, to get rid of the overall factor $e^{2V}$ and recover the Einstein-frame metric in four dimensions, we perform a Weyl rescaling of the form
\begin{equation}
g_{\bar{\mu}\bar{\nu}}=e^{-2V}\tilde{g}_{\bar{\mu}\bar{\nu}}\qquad\text{ and }\qquad g_{rr}=e^{-2V}\tilde{g}_{rr}.
\end{equation}
As a result, we obtain the effective action describing a flat domain wall in four-dimensional space-time
\begin{equation}
S\sim \int d^4x\sqrt{-\tilde{g}_{\text{eff}}}\left[ \frac{1}{4}R(\tilde{g}_{\text{eff}})-\frac{\tilde{g}^{rr}}{16}{P_{r}}^{a\dot{a}}{P_{r}}^{a\dot{a}}-\mathbf{V}_{\text{eff},\Sigma^2}\right]
\end{equation}
where the effective potential is given by
\begin{equation}\label{Sig2_effpot}
\mathbf{V}_{\text{eff},\Sigma^2}=e^{-2V}\left(\mathbf{V}-\frac{k}{2}e^{-2V}+\frac{1}{2}M_{AB}{\mathcal{H}_{\theta\zeta}}^{A}\mathcal{H}^{\theta\zeta,B}\right).
\end{equation}

\subsection{Compactification on $\Sigma^3$}
We now move to the case of a compactification to three-dimensional gauged supergravity on a Riemannian three-manifold. The six-dimensional metric is given by \eqref{SO(3)6Dmetric} together with the vielbein \eqref{AdS4xSigma3bein}. In this case, we find
\begin{eqnarray}
\sqrt{-g}&=&e^{2U+3V}f_k(\psi)^2\sin\theta,\\
R&=&6ke^{-2V}-2(2U''+3V''+3U'^2+6U'V'+6V'^2).
\end{eqnarray}
By repeating the same analysis as in the previous case, we find the effective action 
\begin{eqnarray}
S&=&\int d^6xe^{2U+3V}f_k(\psi)^2\sin\theta\left[ \frac{1}{4}\left\{6ke^{-2V}-2(2U''+3V''+3U'^2+6U'V'+6V'^2)\right\}\right.\nonumber\\&&\qquad\qquad\qquad\qquad\qquad\left.-\frac{1}{16}{P_{r}}^{a\dot{a}}{P^r}_{a\dot{a}}-\mathbf{V}-\frac{1}{4}M_{AB}{\mathcal{H}_{ij}}^{A}\mathcal{H}^{ij, B}\right]\nonumber\\
&\sim&\int d^3xe^{3V}\sqrt{-g_{\text{eff}}}\left[ \frac{1}{4}R(g_{\text{eff}})+\frac{3k}{2}e^{-2V}-\frac{g^{rr}}{16}{P_{r}}^{a\dot{a}}{P_{r}}^{a\dot{a}}-\mathbf{V}-\frac{1}{4}M_{AB}{\mathcal{H}_{ij}}^{A}\mathcal{H}^{ij, B}\right]\nonumber\\
\end{eqnarray}
in which we have neglected the integration on $\Sigma^3$ of the form $\int f_k(\psi)^2\sin\theta d\psi d\theta d\zeta=Vol_{\Sigma^3}$ and introduced 
\begin{equation}
\sqrt{-g_{\text{eff}}}=e^{2U}\qquad\text{ and }\qquad R(g_{\text{eff}})=-2(2U''+3V''+3U'^2+6U'V'+6V'^2).
\end{equation}
Finally, performing a Weyl rescaling of the form
\begin{equation}
g_{\bar{\mu}\bar{\nu}}=e^{-6V}\tilde{g}_{\bar{\mu}\bar{\nu}}\qquad\text{ and }\qquad g_{rr}=e^{-6V}\tilde{g}_{rr}
\end{equation}
leads to four-dimensional action in the Einstein frame
\begin{equation}
S\sim \int d^3x\sqrt{-\tilde{g}_{\text{eff}}}\left[ \frac{1}{4}R(\tilde{g}_{\text{eff}})-\frac{\tilde{g}^{rr}}{16}{P_{r}}^{a\dot{a}}{P_{r}}^{a\dot{a}}-\mathbf{V}_{\text{eff},\Sigma^3}\right]
\end{equation}
with the effective potential given by
\begin{equation}\label{Sig3_effpot}
\mathbf{V}_{\text{eff},\Sigma^3}=e^{-6V}\left(\mathbf{V}-\frac{3k}{2}e^{-2V}+\frac{1}{4}M_{AB}{\mathcal{H}_{ij}}^{A}\mathcal{H}^{ij,B}\right).
\end{equation}
%%%%%%%%%%%%%%%%%%%%%%%%%%%%%%%%%%%%%%%%%%%%%%%%%%%%%%%%%%%%%%%%%%%%%%%%%%%%%%%%%%%%%%%%%%%%%%%%%%%%%%%%%%%%%%%%%%%%%%%%%%%%%%%%%%%%%%%%%


\begin{thebibliography}{99}
\bibitem{maldacena} J. M. Maldacena, ``The large $N$ limit of
superconformal field theories and supergravity'', Adv. Theor. Math.
Phys. \textbf{2} (1998) 231-252, arXiv: hep-th/9711200.
\bibitem{Gubser_AdS_CFT} S. S. Gubser, I. R. Klebanov and A. M. Polyakov, ``Gauge theory
correlators from noncritical string theory'', Phys. Lett. \textbf{B428} (1998) 105-114, arXiv: hep-th/9802.109.
\bibitem{Witten_AdS_CFT} E. Witten, ``Anti De Sitter Space and holography'', Adv. Theor. Math. Phys. \textbf{2} (1998) 253-291, arXiv: 9802150.
\bibitem{Maldacena_nogo} J. Maldacena and C. Nunez, ``Supergravity description of field theories on curved manifolds and a no go theorem'', Int. J. Mod. Phys. \textbf{A16} (2001) 822, arXiv: hep-th/0007018.
\bibitem{Kim_AdS_factor} J. P. Gauntlett, N. Kim, S. Pakis and D. Waldram, ``M-Theory solutions with AdS factors'', Class. Quant. Grav. \textbf{19} (2002) 3927-3946, arXiv: hep-th/0202184.
\bibitem{Cucu_AdSD-2} S. Cucu, H. Lu and J. F. Vazquez-Poritz, ``Interpolating from $AdS_{(D-2)} \times  S^2$ to $AdS_D$,'' Nucl. Phys. \textbf{B677}, 181 (2004) arXiv: hep-th/0304022.
\bibitem{Gauntlett1} J. P. Gauntlett, N. Kim, and D. Waldram, ``M five-branes Wrapped on Supersymmetric Cycles'', Phys. Rev. \textbf{D63} (2001) 126001, arXiv: hep-th/0012195.
\bibitem{Gauntlett2} J. P. Gauntlett and N. Kim, ``M five-branes wrapped on supersymmetric cycles 2'', Phys. Rev. \textbf{D65} (2002) 086003, arXiv: hep-th/0109039.
\bibitem{4D_SCFT_from_M5} I. Bah, C. Beem, N. Bobev, and B. Wecht, ``Four-Dimensional SCFTs from M5-Branes'', JHEP 06 (2012) \textbf{005}, arXiv:1203.0303.
\bibitem{2D_Bobev} F. Benini and N. Bobev, ``Two-dimensional SCFTs from wrapped branes and c-extremization", JHEP 1306 (2013) \textbf{005}, arXiv: 1302.4451.
\bibitem{Wraped_M5} P. Karndumri and E. O Colgain, ``3D Supergravity from wrapped M5-branes'', JHEP 03 (2016) \textbf{188}, arXiv: 1508.00963.
\bibitem{Bobby_wrapped_M5} B. S. Acharya, J. P. Gauntlett, and N. Kim, Five-branes wrapped on associative three cycles, Phys.Rev. \textbf{D63} (2001) 106003, arXiv: hep-th/0011190.
\bibitem{7D_noncompact} P. Karndumri, ``Noncompact gauging of $N=2$ 7D supergravity and AdS/CFT holography'', JHEP 02 (2015) \textbf{034}, arXiv: 1411.4542.
\bibitem{7D_twist} P. Karndumri, ``RG flows from $(1,0)$ 6D SCFTs to $N = 1$ SCFTs in four and three dimensions'', JHEP 06 (2015) \textbf{027}, arXiv: 1503.04997.
\bibitem{AdS3_7D_N2} P. Karndumri and P. Nuchino, ``Two-dimensional SCFTs from mattercoupled 7D $N=2$ gauged supergravity'', Eur. Phys. J. \textbf{C79} (2019) 652, arXiv: 1905.13085.
\bibitem{7D_Max_twist} P. Karndumri and P. Nuchino, ``Twisted compactifications of 6D field theories from maximal 7D gauged supergravity," Eur. Phys. J. C \textbf{80} (2020) 201, arXiv: 1912.04807.
\bibitem{Wraped_D3} P. Karndumri and E. O Colgain, ``3D Supergravity from wrapped D3-branes'', JHEP 10 (2013) \textbf{094}, arXiv: 1307.2086.
\bibitem{3D_CFT_from_LS_point} N. Bobev, K. Pilch, and O. Vasilakis, ``$(0,2)$ SCFTs from the Leigh-Strassler fixed point'', JHEP 06 (2014) \textbf{094}, arXiv:1403.7131.
\bibitem{flow_acrossD_bobev} N. Bobev and P. M. Crichigno, ``Universal RG Flows Across Dimensions and
Holography'', JHEP 12 (2017) \textbf{065}, arXiv: 1708.05052.
\bibitem{BBC} F. Benini, N. Bobev and P. M. Crichigno, ``Two-dimensional SCFTs from D3-branes'', JHEP 07 (2016) \textbf{020}, arXiv: 1511.09462.
\bibitem{N3_AdS2} P. Karndumri, ``Holographic renormalization group flows in $N=3$ Chern-Simons-Matter theory from $N=3$ 4D gauged supergravity'', Phys. Rev. \textbf{D94} (2016) 045006, arXiv: 1601.05703.
\bibitem{flow_across_Betti} A. Amariti and C. Toldo, ``Betti multiplets, flows across dimensions and c-extremization'', JHEP 07 (2017) \textbf{040}, arXiv: 1610.08858.
\bibitem{AdS2_trisasakian} P. Karndumri, ``Supersymmetric $AdS_2\times \Sigma_2$ solutions from tri-sasakian truncation'', Eur. Phys. J. \textbf{C77} (2017) 689, arXiv: 1707.09633.
\bibitem{6D_twist} P. Karndumri, ``Twisted compactification of $N = 2$ 5D SCFTs to three and two dimensions from $F(4)$ gauged supergravity'', JHEP 09 (2015) \textbf{034}, arXiv: 1507.01515.
\bibitem{5D_N4_flow} H. L. Dao and P. Karndumri, ``Holographic RG flows and $AdS_5$ black strings from 5D half-maximal gauged supergravity'', Eur. Phys. J. \textbf{C79} (2019) 137, arXiv: 1811.01608.
\bibitem{5Dtwist} H. L. Dao and P. Karndumri, ``Supersymmetric $AdS_5$ black holes and strings from 5D $N=4$ gauged supergravity'', Eur. Phys. J. \textbf{C79} (2019) 247, arXiv: 1812.10122.
\bibitem{BH_microstate_6D2} M. Suh, ``D4-branes wrapped on supersymmetric four-cycles'', JHEP 1901 (2019) \textbf{035}, arXiv: 1809.03517.
\bibitem{Minwoo_6D_BH2} M. Suh, ``D4-branes wrapped on supersymmetric four-cycles from matter coupled F(4) gauged supergravity'', JHEP 1902 (2019) \textbf{108}, arXiv: 1810.00675.
\bibitem{Calos_6D_flow1}C. Nunez, I. Y. Park, M. Schvellinger and T. A. Tran, ``Supergravity duals of gauge theories from F(4) gauged supergravity in six dimensions”, JHEP 04 (2001) \textbf{025}, arXiv: hep-th/0103080. 
\bibitem{Kim_wrapped_F4} N. Kim and M. Shim, ``Wrapped Brane Solutions in Romans F(4) Gauged Supergravity'', arXiv: 1909.01534.
\bibitem{ISO3_defect} P. Karndumri, ``Twisted compactifications and conformal defects from ISO(3)$\times$U(1) $F(4)$ gauged supergravity'', Eur. Phys. J. \textbf{C85} (2025) 187, arXiv: 2410.04403.
\bibitem{DW_QFT1} H.J. Boonstra, K. Skenderis and P.K. Townsend, ``The domain-wall/QFT
correspondence'', JHEP 01 (1999) \textbf{003}, arXiv: hep-th/9807137.
\bibitem{DW_QFT2} T. Gherghetta and Y. Oz, ``Supergravity, Non-Conformal Field Theories and
Brane-Worlds'', Phys. Rev. \textbf{D65} (2002) 046001, arXiv: hep-th/0106255.
\bibitem{DW_QFT3} Ingmar Kanitscheider, Kostas Skenderis and Marika Taylor, ``Precision holography for
non-conformal branes'', JHEP 09 (2008) \textbf{094}, arXiv: 0807.3324.
\bibitem{pure_N1} J. M. Maldacena and C. Nunez, ``Towards the large $N$ limit of pure $N=1$ superYang-Mills'', Phys. Rev. Lett. \textbf{86} (2001) 588–591, arXiv: hep-th/0008001.
\bibitem{N2_SYM_Gauntlett} J. P. Gauntlett, N. Kim, D. Martelli, and D. Waldram, ``Wrapped five-branes and $N=2$ super Yang-Mills theory'', Phys. Rev. \textbf{D64} (2001) 106008, arXiv: hep-th/0106117.
\bibitem{N2_SYM_Zaffaroni} F. Bigazzi, A. L. Cotrone, and A. Zaffaroni, ``$N=2$ gauge theories from wrapped
five-branes'', Phys. Lett. \textbf{B519} (2001) 269–276, arXiv: hep-th/0106160.
\bibitem{fractional_brane} P. Di Vecchia, H. Enger, E. Imeroni, and E. Lozano-Tellechea, ``Gauge theories from wrapped and fractional branes'', Nucl. Phys. \textbf{B631} (2002) 95–127, arXiv: hep-th/0112126.
\bibitem{SQCD_like} R. Casero, C. Nunez, and A. Paredes, ``Towards the string dual of $N=1$ SQCD-like theories'', Phys. Rev. \textbf{D73} (2006) 086005, arXiv: hep-th/0602027.
\bibitem{D6_wrapped} J. D. Edelstein and C. Nunez, ``D6-branes and M theory geometrical transitions from gauged supergravity'', JHEP 04 (2001) \textbf{028}, arXiv: hep-th/0103167.
\bibitem{D6_wrapped_spindle} P. Ferrero, ``D6 branes wrapped on a spindle and $Y^{p,q}$ manifolds'', JHEP 05 (2024) \textbf{182}, arXiv: 2403.03988.
\bibitem{non-confornal_brane_spindle} M. Boisvert and P. Ferrero, ``A story of non-conformal branes: spindles, disks, circles and black holes'', JHEP 06 (2024) \textbf{013}, arXiv: 2403.03989.
\bibitem{non-conformal_brane_Minwoo} M. Suh, ``Non-conformal branes wrapped on a disk'', arXiv: 2507.22991.
\bibitem{6D_Max_Gauging} E. Bergshoeff, H. Samtleben, and E. Sezgin, ``The Gauging of Maximal D=6 Supergravity", JHEP 05 (2020) \textbf{015}, arXiv: 0712.4277.
\bibitem{6D_DW_I} P. Karndumri and P. Nuchino,``Supersymmetric domain walls in maximal 6D gauged supergravity I'', Eur. Phys. J. C \textbf{81} (2021): 764, arXiv: 2102.11185.
\bibitem{6D_DW_II} P. Karndumri and P. Nuchino,``Supersymmetric domain walls in maximal 6D gauged supergravity II'', Phys. Rev. \textbf{D104} (2021): 106008, arXiv: 2108.08260.
\bibitem{6D_DW_III} P. Karndumri and P. Nuchino,``Supersymmetric domain walls in maximal 6D gauged supergravity IIII'', Eur. Phys. J. C \textbf{84} (2024): 333, arXiv: 2312.15777.
\bibitem{7D_Max_Gauging} H. Samtleben and M. Weidner, ``The maximal D=7 supergravities'', Nucl. Phys. \textbf{725} (2005) 383-419, arXiv: hep-th/0506237.
\bibitem{Henning_KK} O. Hohm and H. Samtleben, ``Consistent Kaluza-Klein Truncations via Exceptional Field Theory'', JHEP 01 (2015) \textbf{131}, arXiv: 1410.8145.
\bibitem{Pope_typeII_S3_S4} M. Cvetic, H. Lu, C. N. Pope, A. Sadrzadeh, and T. A. Tran, ``$S^3$ and $S^4$ Reductions of Type IIA Supergravity'', Nucl. Phys. \textbf{B590} (2000) 233-251, arXiv: hep-th/0005137.
\bibitem{Malek_IIA_IIB} E. Malek and H. Samtleben, ``Dualising consistent IIA /IIB truncations'', JHEP 12 (2015) \textbf{029}, arXiv: 1510.03433.
\bibitem{Gubser_Sing} Steven S. Gubser, ``Curvature singularities: the good, the bad and the naked,” Adv. Theor. Math.
Phys. \textbf{4} (2000), 679-745, arXiv: hep-th/0002160.
\end{thebibliography}
\end{document}